\documentclass[12pt]{article}
\usepackage{a4wide}
\usepackage{latexsym}
\usepackage{amsmath}
\usepackage{amsfonts}
\usepackage{amscd}
\usepackage{cite}
\usepackage{graphicx}
\usepackage{float}

\usepackage{pslatex}
\usepackage[latin1]{inputenc}
\usepackage[T1]{fontenc}

\def\bq{\begin{eqnarray}}
\def\eq{\end{eqnarray}}
\def\l{\langle}
\def\r{\rangle}

\begin{document}

\thispagestyle{empty}

\begin{flushright}
  MZ-TH/09-13
\end{flushright}

\vspace{1.5cm}

\begin{center}
  {\Large\bf Event shapes and jet rates in electron-positron annihilation at NNLO\\
  }
  \vspace{1cm}
  {\large Stefan Weinzierl\\
\vspace{2mm}
      {\small \em Institut f{\"u}r Physik, Universit{\"a}t Mainz,}\\
      {\small \em D - 55099 Mainz, Germany}\\
  } 
\end{center}

\vspace{2cm}

\begin{abstract}\noindent
  {
This article gives the perturbative NNLO results for the
most commonly used event shape variables associated to three-jet events
in electron-positron annihilation: Thrust, heavy jet mass, wide jet broadening,
total jet broadening, C parameter and the Durham three-to-two jet transition variable.
In addition the NNLO results for the jet rates corresponding to the
Durham, Geneva, Jade-E0 and Cambridge jet algorithms are presented.
   }
\end{abstract}

\vspace*{\fill}

\newpage

\section{Introduction}
\label{sec:intro}

The experiments at LEP (CERN), SLC (SLAC) and PETRA (DESY) have collected a wealth of data from electron-positron annihilation
with hadronic final state over a wide range of energies.
Of particular interest are three-jet events, which can be used to extract the value of the strong coupling constant.
Three-jet events are well suited for this task because the leading term in a perturbative calculation 
of three-jet observables is already proportional to the strong coupling.
In comparing experiments to theory it is important to restrict oneself to infra-red safe observables.
For three-jet events in electron-positron annihilation there is a well established set of infra-red safe observables, 
which is widely used. These are the event shape variables consisting of thrust, heavy jet mass, wide jet broadening,
total jet broadening and the $C$-parameter.
In addition there are observables related to a specific jet definition.
First of all these are the jet rates associated to the different jet definitions. 
In this paper the following jet algorithms are considered: Durham, Geneva, Jade-E0 and Cambridge.
As the Durham jet algorithm is the most popular one, for this jet algorithm also the
three-to-two jet transition variable is studied.

All these observables can be calculated in perturbation theory.
In this article I present the next-to-next-to-leading order (NNLO) results for these observables.
For completeness I also include the next-to-leading order (NLO) and leading order (LO) results.

Another group published results for these observables earlier on in \cite{GehrmannDeRidder:2007hr,GehrmannDeRidder:2008ug,Ridder:2009dp},
but omitted certain subtraction terms related to soft gluons.
The present calculation is based on the numerical Monte-Carlo program reported in \cite{Weinzierl:2008iv,Weinzierl:1999yf}.
The additional soft gluon subtraction terms modify the distributions in the peak region.
Meanwhile the other group has added the missing subtraction terms and the two programs are now for the main part of the
observables in good agreement.
A few remaining differences are discussed in the numerical section.
A detailed account of the subtraction terms used in this calculation is given in a companion paper \cite{Weinzierl:2009aaa}.

For the thrust distribution there exists an independent calculation of the logarithmic terms based on
soft-collinear effective theory \cite{Becher:2008cf}. 
This calculation should agree with the NNLO result for small values of $(1-T)$.
For large values of $(1-T)$ the logarithmic terms alone are not sufficient to give an accurate result.
On the other hand the perturbative calculation of this paper gives the correct NNLO result
for large and intermediate values of $(1-T)$, but has its limitations due to the Monte-Carlo integration
method for very small values of $(1-T)$.
This gives a region of overlap where the perturbative NNLO result and the one obtained from SCET should agree,
and indeed they do. 
 
This article reports the pure perturbative results for the event shapes.
Not included are soft-gluon resummations \cite{Catani:1992ua,Catani:1998sf,Becher:2008cf,Gehrmann:2008kh}
nor power corrections \cite{Dokshitzer:1997ew,Davison:2008vx}.

It should be mentioned that the results of this paper rely heavily on research carried out in the past years:
Integration techniques for two-loop amplitudes \cite{Gehrmann:1999as,Gehrmann:2000zt,Gehrmann:2001ck,Moch:2001zr},
the calculation of the relevant tree-, one- and two-loop-amplitudes \cite{Berends:1989yn,Hagiwara:1989pp,Falck:1989uz,Schuler:1987ej,Korner:1990sj,Giele:1992vf,Bern:1997ka,Bern:1997sc,Campbell:1997tv,Glover:1997eh,Garland:2001tf,Garland:2002ak,Moch:2002hm},
routines for the numerical evaluation of polylogarithms \cite{Gehrmann:2001pz,Gehrmann:2001jv,Vollinga:2004sn},
methods to handle infrared singularities \cite{Kosower:2002su,Kosower:2003cz,Weinzierl:2003fx,Weinzierl:2003ra,Kilgore:2004ty,Frixione:2004is,Gehrmann-DeRidder:2003bm,Gehrmann-DeRidder:2004tv,Gehrmann-DeRidder:2005hi,Gehrmann-DeRidder:2005aw,Gehrmann-DeRidder:2005cm,GehrmannDeRidder:2007jk,Somogyi:2005xz,Somogyi:2006da,Somogyi:2006db,Catani:2007vq,Somogyi:2008fc,Somogyi:2009ri,Aglietti:2008fe}
and experience from the NNLO calculations of
$e^+ e^- \rightarrow \mbox{2 jets}$ and other processes
\cite{Anastasiou:2004qd,Gehrmann-DeRidder:2004tv,Weinzierl:2006ij,Weinzierl:2006yt,Anastasiou:2002yz,Anastasiou:2003yy,Anastasiou:2003ds,Anastasiou:2004xq,Anastasiou:2005qj,Anastasiou:2007mz,Melnikov:2006di,Anastasiou:2005pn,Catani:2001ic,Catani:2001cr,Grazzini:2008tf,Harlander:2001is,Harlander:2002wh,Harlander:2003ai}.

This paper is organised as follows: Section~\ref{sec:def} gives the definition of the observables
related to three-jet events in electron-positron annihilation.
Section~\ref{sec:perturbative} describes the perturbative calculation of these observables.
Section~\ref{sec:num} gives the numerical results.
Finally, section~\ref{sec:conclusions} contains a summary.
In an appendix a useful algorithm for the determination of the thrust axis is described.

\section{Definition of the observables}
\label{sec:def}

The event shape variable thrust \cite{Brandt:1964sa,Farhi:1977sg}
is defined by
\bq
 T & = & 
 \frac{\max\limits_{\vec{n}} \sum\limits_j \left| \vec{p}_j \cdot \vec{n} \right|}{\sum\limits_j \left| \vec{p}_j \right|},
\eq
where $\vec{p}_j$ denotes the three-momentum of particle $j$ and the sum runs over all particles in the 
final state. The thrust variable maximises the total longitudinal momentum along the unit vector $\vec{n}$.
The value of $\vec{n}$, for which the maximum is attained is called the thrust axis
and denoted by $\vec{n}_T$. 
The value of thrust ranges between $1/2$ and $1$, where $T=1$ corresponds to an ideal collinear two-jet event
and $T=1/2$ corresponds to a perfectly spherical event.
Usually one considers instead of thrust $T$ the variable $(1-T)$, such that
the two-jet region corresponds to $(1-T) \rightarrow 0$.
For three-parton events we have $(1-T) \le 1/3$.

The plane orthogonal to the thrust axis divides the space into two hemispheres $H_1$ and $H_2$. These are used
to define the following event shape variables: 
The hemisphere masses \cite{Clavelli:1981yh} are defined by
\bq
 M_i^2 & = & \left( \sum\limits_{j\in H_i} p_j \right)^2, \;\;\;i=1,2,
\eq
where $p_j$ denotes the four-momentum of particle $j$.
The heavy hemisphere mass $M_H$ and the light hemisphere mass $M_L$ are then defined by
\bq
 M_H^2 = \max\left(M_1^2,M_2^2\right),
 & &
 M_L^2 = \min\left(M_1^2,M_2^2\right).
\eq
The light hemisphere mass is a four-jet observable and vanishes for three partons.
For the heavy jet mass it is convenient to introduce the dimensionless quantity
\bq
 \rho & = & \frac{M_H^2}{Q^2},
\eq
where $Q$ is the centre-of-mass energy.
For three-parton events we have $\rho \le 1/3$.
In leading order the distribution of the heavy jet mass is identical to the distribution of $(1-T)$.

The hemisphere broadenings \cite{Rakow:1981qn,Catani:1992jc} are defined by
\bq
 B_i & = & 
 \frac{\sum\limits_{j\in H_i} \left| \vec{p}_j \times \vec{n}_T \right|}
      {2 \sum\limits_k \left| \vec{p}_k \right|},
 \;\;\;i=1,2,
\eq
where the sum over $j$ runs over all particles in one of the hemispheres, whereas the sum over $k$ is over all
particles in the final state.
The wide jet broadening $B_W$, the narrow jet broadening $B_N$ and the total jet broadening $B_T$ are defined
by
\bq
 B_W = \max\left(B_1,B_2\right), 
 \;\;\;
 B_N = \min\left(B_1,B_2\right), 
 \;\;\;
 B_T = B_1 + B_2.
\eq
The narrow jet broadening is a four-jet observable and vanishes for three partons.
For three-parton events we have $B_W, B_T \le 1/(2\sqrt{3})$.

The $C$- and $D$-parameters \cite{Parisi:1978eg,Donoghue:1979vi} are obtained from the
linearised momentum tensor
\bq
\theta^{ij} & = & 
 \frac{1}{\sum\limits_l \left|\vec{p}_l\right|} \sum\limits_k \frac{p_k^i p_k^j}{\left|\vec{p}_k\right|},
 \;\;\; i,j=1,2,3,
\eq
where the sum runs over all final state particles and $p_k^i$ is the $i$-th component of the three-momentum
$\vec{p}_k$ of particle $k$ in the c.m. system. 
The tensor $\theta$ is normalised to have unit trace.
In terms of the eigenvalues of the $\theta$ tensor,
$\lambda_1$, $\lambda_2$, $\lambda_3$, with
$\lambda_1 + \lambda_2 + \lambda_3 = 1$, one defines
\bq
C & = & 3 \left( \lambda_1 \lambda_2 + \lambda_2 \lambda_3 + \lambda_3 \lambda_1 \right),
 \nonumber \\
D & = & 27 \lambda_1 \lambda_2 \lambda_3.
\eq
The range of values is $0 \leq C,D \leq 1$. 
The $D$-parameter is again a four-jet observable and vanishes for three partons.
For three-parton events we have $C \le 3/4$.
The $C$-parameter exhibits in perturbation theory a singularity at the 
three-parton boundary $C=3/4$ \cite{Catani:1997xc}.

The production rate for three-jet events is defined by 
the ratio of the cross section for three-jet events by the total hadronic cross section
\bq
 R_3 & = &  \frac{\sigma_{3-jet}}{\sigma_{tot}}.
\eq
This depends on the definition of the jet algorithm, usually specified through a resolution variable and a
recombination prescription.
The clustering procedure of a jet algorithm in electron-positron annihilation 
is in most cases defined through the following steps:
\begin{enumerate}
\item Define a resolution parameter $y_{cut}$
\item For every pair $(p_k, p_l)$ of final-state particles compute the corresponding
resolution variable $y_{kl}$.
\item If $y_{ij}$ is the smallest value of $y_{kl}$ computed above and
$y_{ij} < y_{cut}$ then combine $(p_i, p_j)$ into a single jet ('pseudo-particle')
with momentum $p_{ij}$ according to a recombination prescription.
\item Repeat until all pairs of objects (particles and/or pseudo-particles)
have $y_{kl} > y_{cut}$.
\end{enumerate}
The various jet algorithms differ in the precise definition of the resolution variable
and the recombination prescription.
The various recombination prescriptions are:
\begin{enumerate}
\item E-scheme:
\bq
E_{ij} = E_i + E_j,
 & &
\vec{p}_{ij} = \vec{p}_i + \vec{p}_j.
\eq
The E-scheme conserves energy and momentum, but for massless particles $i$ and $j$ the recombined four-momentum is not massless.
\item E0-scheme:
\bq
E_{ij} = E_i + E_j,
 & &
\vec{p}_{ij} = \frac{E_i + E_j}{\left| \vec{p}_i + \vec{p}_j \right|} \left( \vec{p}_i + \vec{p}_j \right).
\eq
The E0-scheme conserves energy, but not momentum. For massless particles $i$ and $j$ is the recombined four-momentum again massless. 
\item P-scheme:
\bq
E_{ij} = \frac{\left| \vec{p}_i + \vec{p}_j \right|}{E_i + E_j} \left( E_i + E_j \right) = 
\left| \vec{p}_i + \vec{p}_j \right|,
 & &
\vec{p}_{ij} = \vec{p}_i + \vec{p}_j.
\eq
The P-scheme conserves momentum, but not energy. For massless particles $i$ and $j$ is the recombined four-momentum again massless, 
as in the $E0$-scheme.
\end{enumerate}
For the Durham \cite{Stirling:1991ds}, Geneva \cite{Bethke:1991wk} and Jade-E0 \cite{Bartel:1986ua} jet algorithms the 
resolution variables and the recombination prescriptions are defined as follows:
\bq
\mbox{Durham:} & & 
y_{ij} = \frac{2 \min(E_i^2, E_j^2) \left( 1 - \cos \theta_{ij} \right)}{Q^2},
 \;\;\;\mbox{E-scheme},
 \nonumber \\
\mbox{Geneva:} & &
y_{ij} = \frac{8}{9} \; \cdot \; \frac{2 E_i E_j \left( 1 - \cos \theta_{ij} \right)}{\left(E_i + E_j \right)^2},
 \;\;\;\mbox{E-scheme},
 \nonumber \\
\mbox{Jade-E0:} & &
y_{ij} = \frac{2 E_i E_j \left( 1 - \cos \theta_{ij} \right)}{Q^2},
 \;\;\;\mbox{E0-scheme},
\eq
where $E_i$ and $E_j$ are the energies of particles $i$ and $j$, and $\theta_{ij}$ is the angle
between $\vec{p}_i$ and $\vec{p}_j$.
$Q$ is the centre-of-mass energy.
The jet transition variable $y_{23}$ is the value of the jet resolution parameter
$y_{cut}$, for which the event changes from a three-jet to a two-jet configuration.
Similar, $y_{34}$ is defined as the value of the resolution parameter, where the event changes from a four-jet to a
three-jet configuration. 
The three-jet rate is related to the jet transition variables $y_{23}$ and $y_{34}$:
\bq
R_3(y_{cut}) & = &
 \int\limits_{y_{cut}}^1 dy_{23} \frac{d\sigma}{dy_{23}}
 -
 \int\limits_{y_{cut}}^1 dy_{34} \frac{d\sigma}{dy_{34}}.
\eq
The Cambridge algorithm \cite{Dokshitzer:1997in} distinguishes an ordering variable $v_{ij}$ and a resolution variable $y_{ij}$.
The Cambridge algorithm is defined as follows:
\begin{enumerate}
\item Select a pair of objects $(p_i,p_j)$ with the minimal value of the ordering
variable $v_{ij}$.
\item If $y_{ij} < y_{cut}$ they are combined, one recomputes the relevant
values of the ordering variable and goes back to the first step.
\item If $y_{ij} \geq y_{cut}$ and $E_i < E_j$ then $i$ is defined as
a resolved jet and deleted from the table.
\item Repeat until only one object is left in the table. This object is
also defined as a jet and clustering is finished.
\end{enumerate} 
As ordering variable
\bq
v_{ij} = 2\left( 1 - \cos \theta_{ij} \right)
\eq
is used. The resolution variable is as in the Durham algorithm
\bq
y_{ij} & = & \frac{2 \min(E_i^2, E_j^2) \left( 1 - \cos \theta_{ij} \right)}{Q^2}
\eq
and the E-scheme is used as recombination prescription.

\section{Perturbative expansion}
\label{sec:perturbative}

The perturbative expansion of a differential distribution 
for any infrared-safe observable $\cal O$ for the process
$e^+ e^- \rightarrow \mbox{3 jets}$
can be written up to NNLO as
\bq
\frac{{\cal O}^n}{\sigma_{tot}(\mu)} \frac{d\sigma(\mu)}{d{\cal O}} & = & 
 \frac{\alpha_s(\mu)}{2\pi} \frac{{\cal O}^n d\bar{A}_{\cal O}(\mu)}{d{\cal O}}
 +
 \left( \frac{\alpha_s(\mu)}{2\pi} \right)^2 \frac{{\cal O}^n d\bar{B}_{\cal O}(\mu)}{d{\cal O}}
 +
 \left( \frac{\alpha_s(\mu)}{2\pi} \right)^3 \frac{{\cal O}^n d\bar{C}_{\cal O}(\mu)}{d{\cal O}}.
\eq
$\bar{A}_{\cal O}$ gives the LO result, $\bar{B}_{\cal O}$ the NLO correction and $\bar{C}_{\cal O}$ the NNLO correction.
The variable $n$ denotes the moment of the distribution.
Unless stated otherwise, the value $n=1$ is used as default.
$\sigma_{tot}$ denotes the total hadronic cross section calculated up to the relevant order.
The arbitrary renormalisation scale is denoted by $\mu$.
The $n$-th moment is given by
\bq 
 \left\langle {\cal O}^n \right\rangle
 & = &
 \frac{1}{\sigma_{tot}} \int {\cal O}^n \frac{d\sigma}{d{\cal O}} d{\cal O}.
\eq
In practise the numerical program computes the distribution
\bq
\frac{{\cal O}^n}{\sigma_{0}(\mu)} \frac{d\sigma(\mu)}{d{\cal O}} & = & 
 \frac{\alpha_s(\mu)}{2\pi} \frac{{\cal O}^n dA_{\cal O}(\mu)}{d{\cal O}}
 +
 \left( \frac{\alpha_s(\mu)}{2\pi} \right)^2 \frac{{\cal O}^n dB_{\cal O}(\mu)}{d{\cal O}}
 +
 \left( \frac{\alpha_s(\mu)}{2\pi} \right)^3 \frac{{\cal O}^n dC_{\cal O}(\mu)}{d{\cal O}},
\eq
normalised to $\sigma_0$, which is the LO cross section for $e^+ e^- \rightarrow \mbox{hadrons}$,
instead of the normalisation to $\sigma_{tot}$.
There is a simple relation between the two distributions:
The functions $A_{\cal O}$, $B_{\cal O}$ and $C_{\cal O}$ are related to the functions
$\bar{A}_{\cal O}$, $\bar{B}_{\cal O}$ and $\bar{C}_{\cal O}$ by
\bq
 \bar{A}_{\cal O} & = & A_{\cal O},
 \nonumber \\
 \bar{B}_{\cal O} & = & B_{\cal O} - A_{tot} A_{\cal O} ,
 \nonumber \\
 \bar{C}_{\cal O} & = & C_{\cal O} - A_{tot} B_{\cal O} - \left( B_{tot} - A_{tot}^2 \right) A_{\cal O} ,
 \nonumber 
\eq
where
\bq
A_{tot} & = & \frac{3(N_c^2-1)}{4 N_c},
 \nonumber \\
 B_{tot} & = &   
 \frac{N_c^2-1}{8 N_c} \left[ \left( \frac{243}{4} - 44 \zeta_3 \right) N_c + \frac{3}{4 N_c} 
                             + \left( 8 \zeta_3 - 11 \right) N_f \right]. 
 \nonumber 
\eq
$N_c$ denotes the number of colours and $N_f$ the number of light quark flavours.
$A_{tot}$ and $B_{tot}$ are obtained from the perturbative expansion of $\sigma_{tot}$ \cite{Dine:1979qh,Chetyrkin:1979bj,Celmaster:1979xr}:
\bq
 \sigma_{tot} & = & 
 \sigma_0 \left( 1 + \frac{\alpha_s}{2\pi} A_{tot}
                   + \left( \frac{\alpha_s}{2\pi} \right)^2 B_{tot}
                   + {\cal O}(\alpha_s^3) \right).
\eq
The perturbative calculation of the inclusive hadronic cross section $\l \sigma \r^{(tot)}$ is actually known
to ${\cal O}(\alpha_s^3)$ \cite{Gorishnii:1991vf,Surguladze:1990tg}, although we need here only the coefficients up to
${\cal O}(\alpha_s^2)$. 

It is sufficient to calculate the functions $\bar{A}_{\cal O}$, $\bar{B}_{\cal O}$ and $\bar{C}_{\cal O}$
for a fixed renormalisation scale $\mu_0$, which can be taken conveniently to be equal to the
centre-of-mass energy: $\mu_0=Q$.
The scale variation can be restored from the renormalisation group equation
\bq
\label{RGE_alpha_s}
 \mu^2 \frac{d}{d\mu^2} \left( \frac{\alpha_S}{2\pi} \right)
 & = & 
 - \frac{1}{2} \beta_0 \left( \frac{\alpha_S}{2\pi} \right)^2
 - \frac{1}{4} \beta_1 \left( \frac{\alpha_S}{2\pi} \right)^3
 - \frac{1}{8} \beta_2 \left( \frac{\alpha_S}{2\pi} \right)^4
 + {\cal O}(\alpha_s^5),
 \\
 \beta_0 & = & \frac{11}{3} C_A - \frac{4}{3} T_R N_f,
 \nonumber \\
 \beta_1 & = & \frac{34}{3} C_A^2 - 4 \left( \frac{5}{3} C_A + C_F \right) T_R N_f,
 \nonumber \\
 \beta_2 & = & \frac{2857}{54} C_A^3 - \left( \frac{1415}{27} C_A^2 + \frac{205}{9} C_A C_F - 2 C_F^2 \right) T_R N_f
             + \left( \frac{158}{27} C_A + \frac{44}{9} C_F \right) T_R^2 N_f^2.
 \nonumber
\eq
The colour factors are defined as usual by
\bq
 C_A = N_c,
 \;\;\;
 C_F = \frac{N_c^2-1}{2 N_c},
 \;\;\;
 T_R = \frac{1}{2}.
\eq
The values of the functions $\bar{A}_{\cal O}$, $\bar{B}_{\cal O}$ and $\bar{C}_{\cal O}$
at a scale $\mu$ are then obtained from the ones at the scale $\mu_0$ by
\bq
 \bar{A}_{\cal O}(\mu) & = & \bar{A}_{\cal O}(\mu_0),
 \nonumber \\
 \bar{B}_{\cal O}(\mu) & = & \bar{B}_{\cal O}(\mu_0) + \frac{1}{2} \beta_0 \ln\left(\frac{\mu^2}{\mu_0^2}\right) \bar{A}_{\cal O}(\mu_0),
 \nonumber \\
 \bar{C}_{\cal O}(\mu) & = & \bar{C}_{\cal O}(\mu_0) + \beta_0 \ln\left(\frac{\mu^2}{\mu_0^2}\right) \bar{B}_{\cal O}(\mu_0)
                           + \frac{1}{4} \left[ \beta_1 + \beta_0^2 \ln\left(\frac{\mu^2}{\mu_0^2}\right) \right] \ln\left(\frac{\mu^2}{\mu_0^2}\right) \bar{A}_{\cal O}(\mu_0).
\eq
Finally, an approximate solution of eq.~(\ref{RGE_alpha_s}) for $\alpha_s$ is given by
\bq
 \frac{\alpha_s(\mu)}{2\pi}
 & = &
 \frac{2}{\beta_0 L}
 \left\{
         1 - \frac{\beta_1}{\beta_0^2} \frac{\ln L}{L}
         + \frac{\beta_1^2}{\beta_0^4 L^2}
           \left[ \left( \frac{1}{2} - \ln L \right)^2
                  + \frac{\beta_0 \beta_2}{\beta_1^2}
                  - \frac{5}{4}
           \right]
 \right\},
\eq
where $L=\ln(\mu^2/\Lambda^2)$.

The function $C_{\cal O}$ can be decomposed into six colour pieces
\bq
 C_{\cal O} & = &
 \frac{\left( N_c^2-1 \right)}{8 N_c} 
 \left[ 
        N_c^2 C_{\cal O}^{lc}
      + C_{\cal O}^{sc}
      + \frac{1}{N_c^{2}} C_{\cal O}^{ssc}
      + N_f N_c C_{\cal O}^{nf}
      + \frac{N_f}{N_c} C_{\cal O}^{nfsc}
      + N_f^2 C_{\cal O}^{nfnf}
 \right].
\eq
In addition, there are singlet contributions, which arise from interference terms of amplitudes, where
the electro-weak boson couples to two different fermion lines.
These singlet contributions are expected to be numerically 
small \cite{Dixon:1997th,vanderBij:1988ac,Garland:2002ak}
and neglected in the present calculation.
We define
\bq
\label{def_colour_factors}
 & &
  \left. C_{\cal O} \right|_{lc} = \frac{\left( N_c^2-1 \right)}{8 N_c} N_c^2 C_{\cal O}^{lc},
 \;\;\;
  \left. C_{\cal O} \right|_{sc} = \frac{\left( N_c^2-1 \right)}{8 N_c} C_{\cal O}^{sc},
 \;\;\;
  \left. C_{\cal O} \right|_{ssc} = \frac{\left( N_c^2-1 \right)}{8 N_c} \frac{1}{N_c^{2}} C_{\cal O}^{ssc},
 \\
 & &
  \left. C_{\cal O} \right|_{nf} = \frac{\left( N_c^2-1 \right)}{8 N_c} N_f N_c C_{\cal O}^{nf},
 \;\;\;
  \left. C_{\cal O} \right|_{nfsc} = \frac{\left( N_c^2-1 \right)}{8 N_c} \frac{N_f}{N_c} C_{\cal O}^{nfsc},
 \;\;\;
  \left. C_{\cal O} \right|_{nfnf} = \frac{\left( N_c^2-1 \right)}{8 N_c} N_f^2 C_{\cal O}^{nfnf},
 \nonumber 
\eq
e.g. the function $\left. C_{\cal O} \right|_{lc}$ includes the colour factors.

The functions $A_{\cal O}$, $B_{\cal O}$ and $C_{\cal O}$ are calculated for a fixed renormalisation scale equal
to the centre-of-mass energy: $\mu_0=Q$.
They depend only on the value of the observable ${\cal O}$. 
Since only QCD corrections with non-singlet quark couplings are taken into account and 
singlet contributions to $C_{\cal O}$ are neglected, the functions 
$A_{\cal O}$, $B_{\cal O}$ and $C_{\cal O}$ do not depend on electro-weak couplings.

\section{Numerical results}
\label{sec:num}

\begin{figure}[p]
\begin{center}
\includegraphics[bb= 125 460 490 710,width=0.9\textwidth]{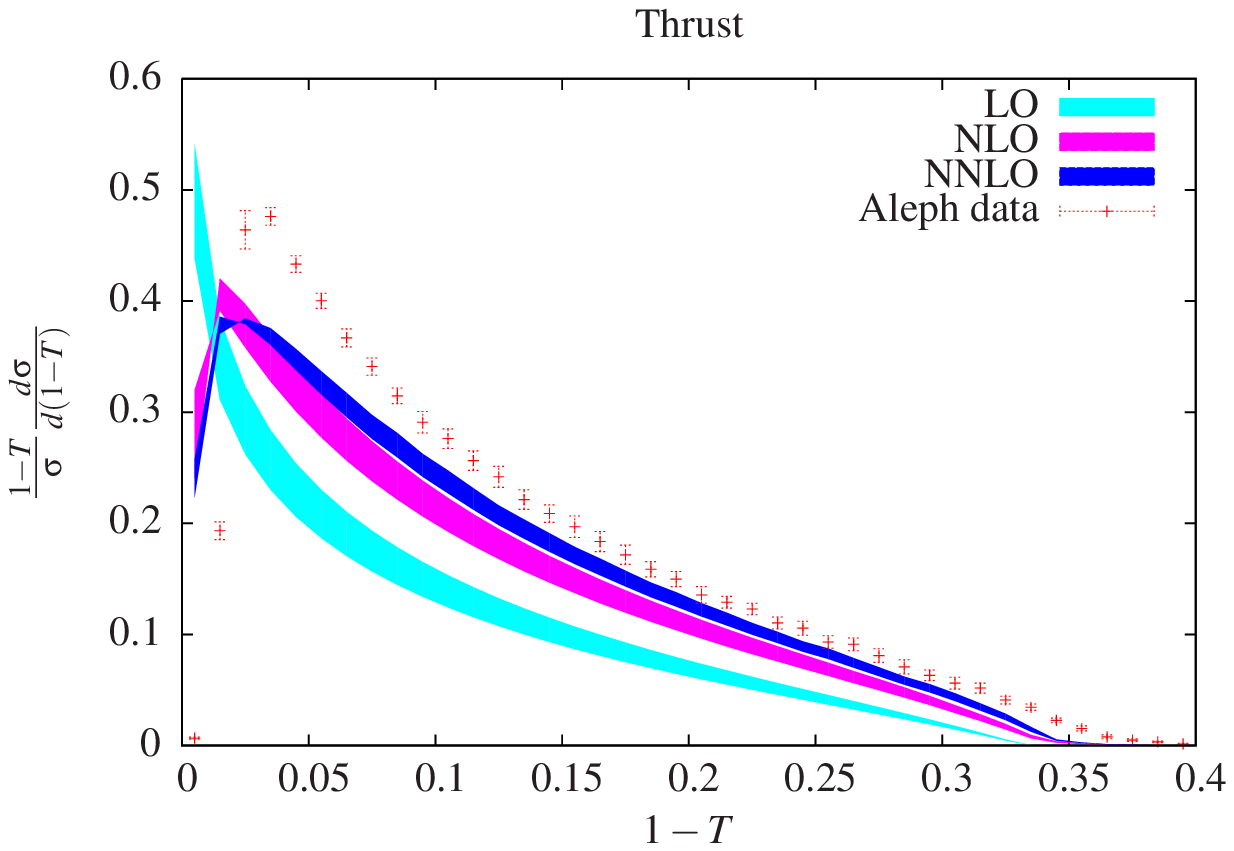}
\end{center}
\caption{The thrust distribution at LO, NLO and NNLO at $\sqrt{Q^2}=m_Z$ with $\alpha_s(m_Z)=0.118$.
The bands give the range for the theoretical prediction obtained from varying the renormalisation scale
from $\mu=m_Z/2$ to $\mu=2 m_Z$.
In addition the experimental data points from the Aleph experiment are shown.
}
\label{fig_thrust}
\end{figure}

\begin{figure}[p]
\begin{center}
\includegraphics[bb= 125 460 490 710,width=0.9\textwidth]{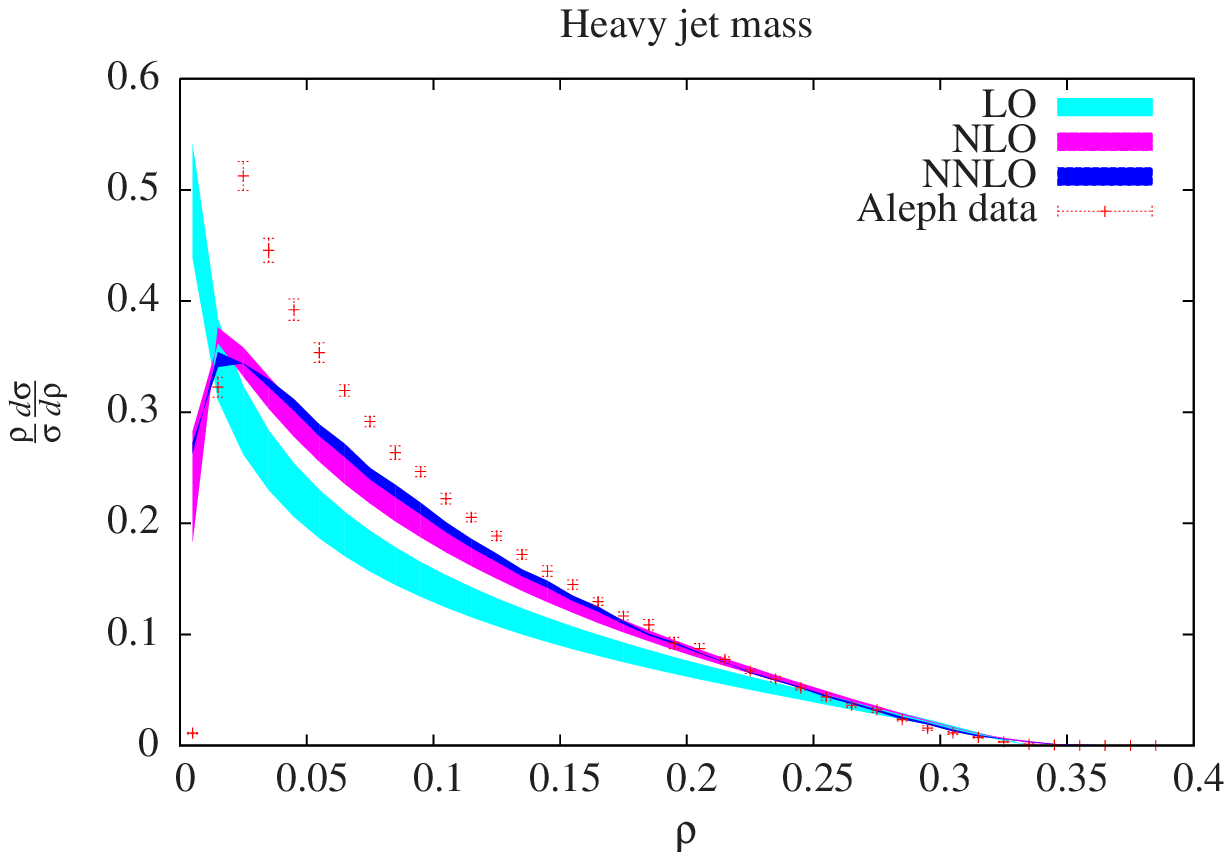}
\end{center}
\caption{The distribution of the heavy jet mass at LO, NLO and NNLO at $\sqrt{Q^2}=m_Z$ with $\alpha_s(m_Z)=0.118$.
The bands give the range for the theoretical prediction obtained from varying the renormalisation scale
from $\mu=m_Z/2$ to $\mu=2 m_Z$.
In addition the experimental data points from the Aleph experiment are shown.
}
\label{fig_heavyjetmass}
\end{figure}

\begin{figure}[p]
\begin{center}
\includegraphics[bb= 125 460 490 710,width=0.9\textwidth]{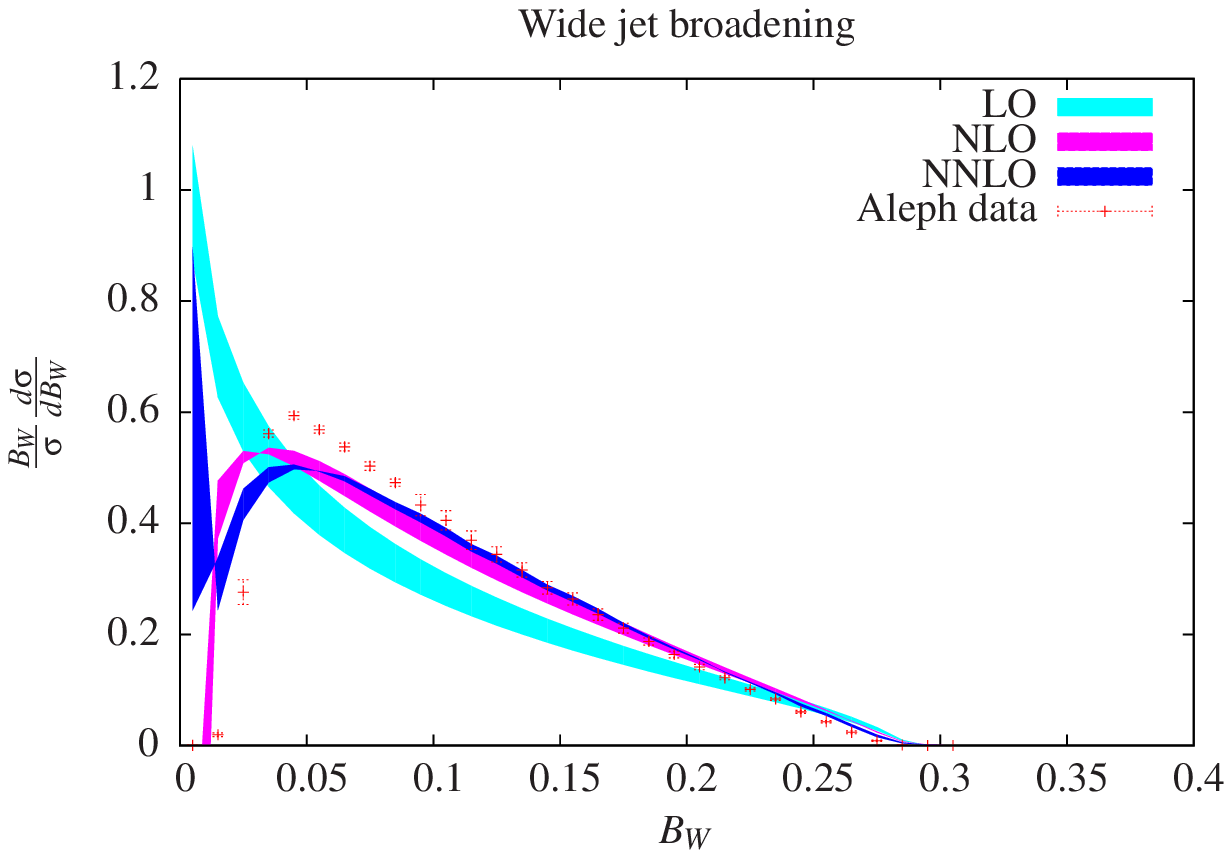}
\end{center}
\caption{The wide jet broadening distribution at LO, NLO and NNLO at $\sqrt{Q^2}=m_Z$ with $\alpha_s(m_Z)=0.118$.
The bands give the range for the theoretical prediction obtained from varying the renormalisation scale
from $\mu=m_Z/2$ to $\mu=2 m_Z$.
In addition the experimental data points from the Aleph experiment are shown.
}
\label{fig_widejetbroadening}
\end{figure}

\begin{figure}[p]
\begin{center}
\includegraphics[bb= 125 460 490 710,width=0.9\textwidth]{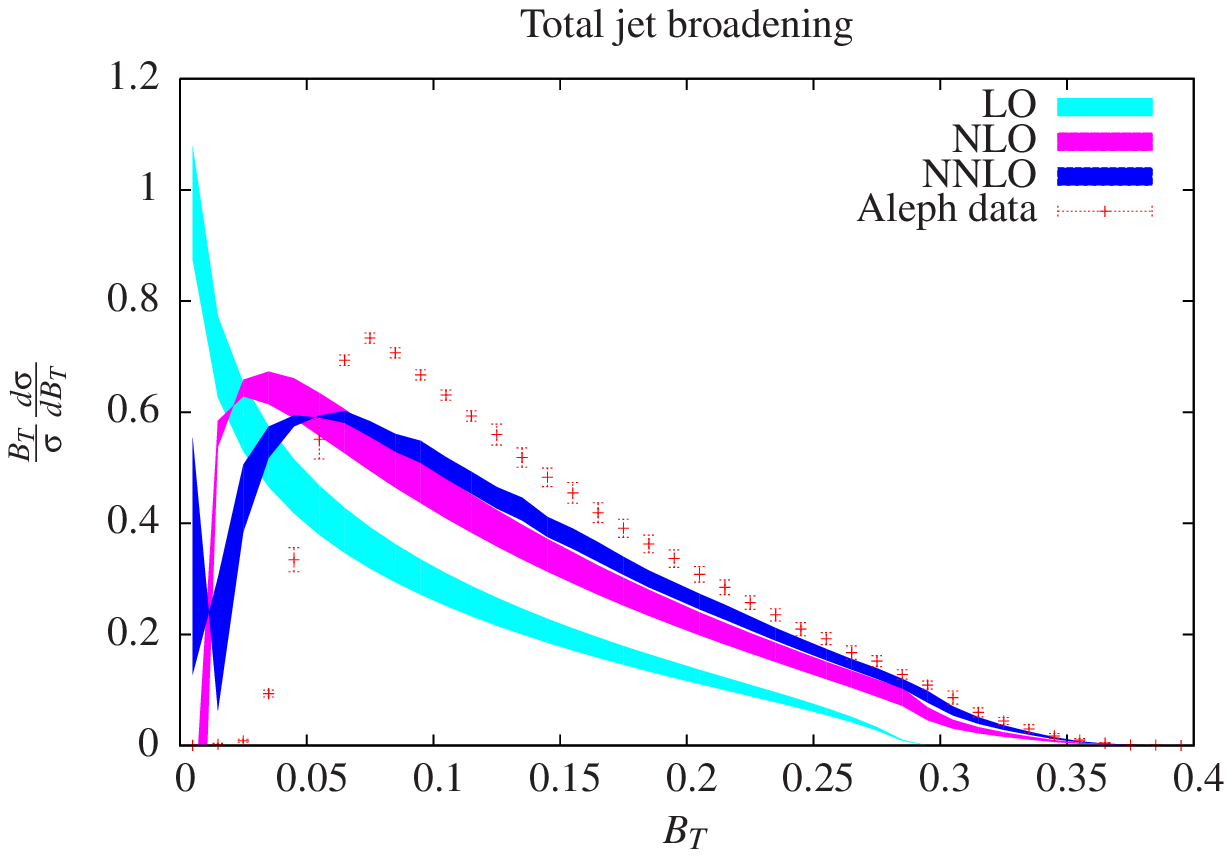}
\end{center}
\caption{The total jet broadening distribution at LO, NLO and NNLO at $\sqrt{Q^2}=m_Z$ with $\alpha_s(m_Z)=0.118$.
The bands give the range for the theoretical prediction obtained from varying the renormalisation scale
from $\mu=m_Z/2$ to $\mu=2 m_Z$.
In addition the experimental data points from the Aleph experiment are shown.
}
\label{fig_totaljetbroadening}
\end{figure}

\begin{figure}[p]
\begin{center}
\includegraphics[bb= 125 460 490 710,width=0.9\textwidth]{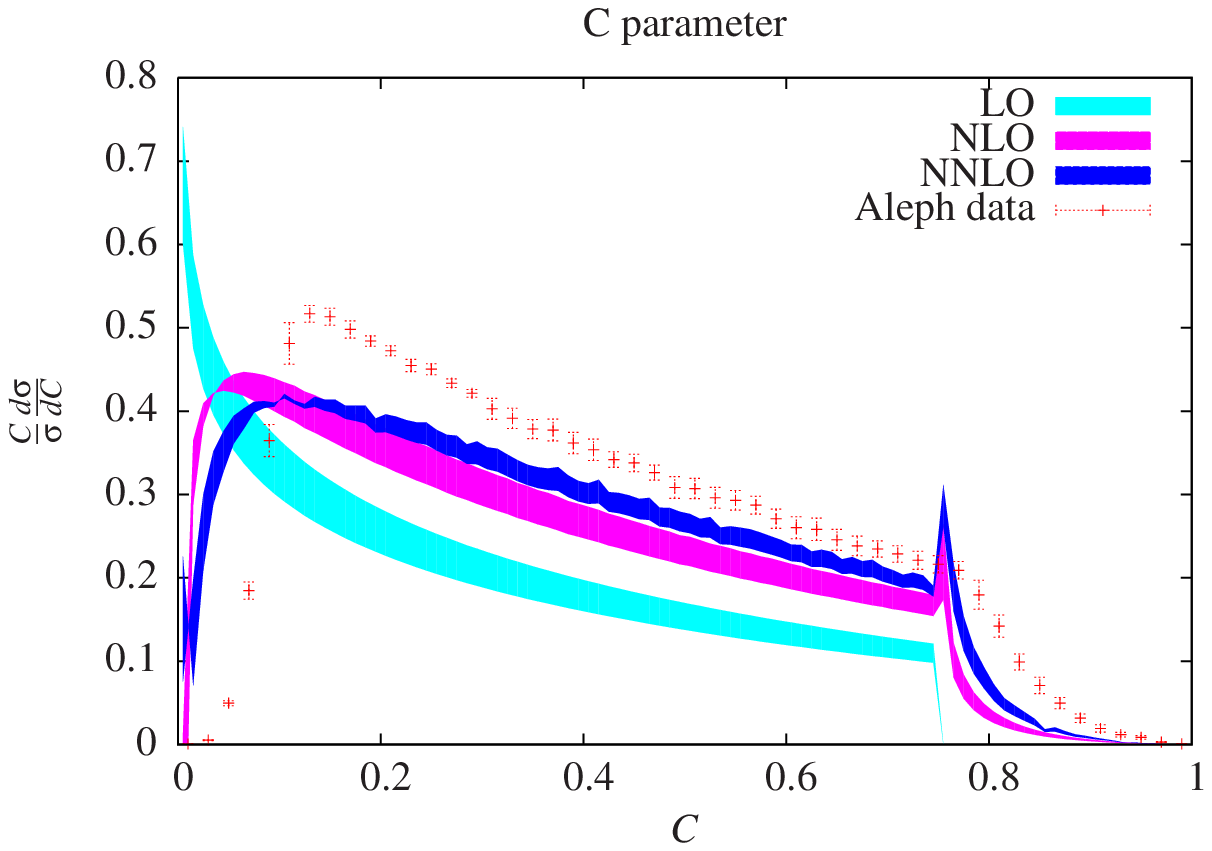}
\end{center}
\caption{The $C$-parameter distribution at LO, NLO and NNLO at $\sqrt{Q^2}=m_Z$ with $\alpha_s(m_Z)=0.118$.
The bands give the range for the theoretical prediction obtained from varying the renormalisation scale
from $\mu=m_Z/2$ to $\mu=2 m_Z$.
In addition the experimental data points from the Aleph experiment are shown.
}
\label{fig_Cparameter}
\end{figure}

\begin{figure}[p]
\begin{center}
\includegraphics[bb= 125 460 490 710,width=0.9\textwidth]{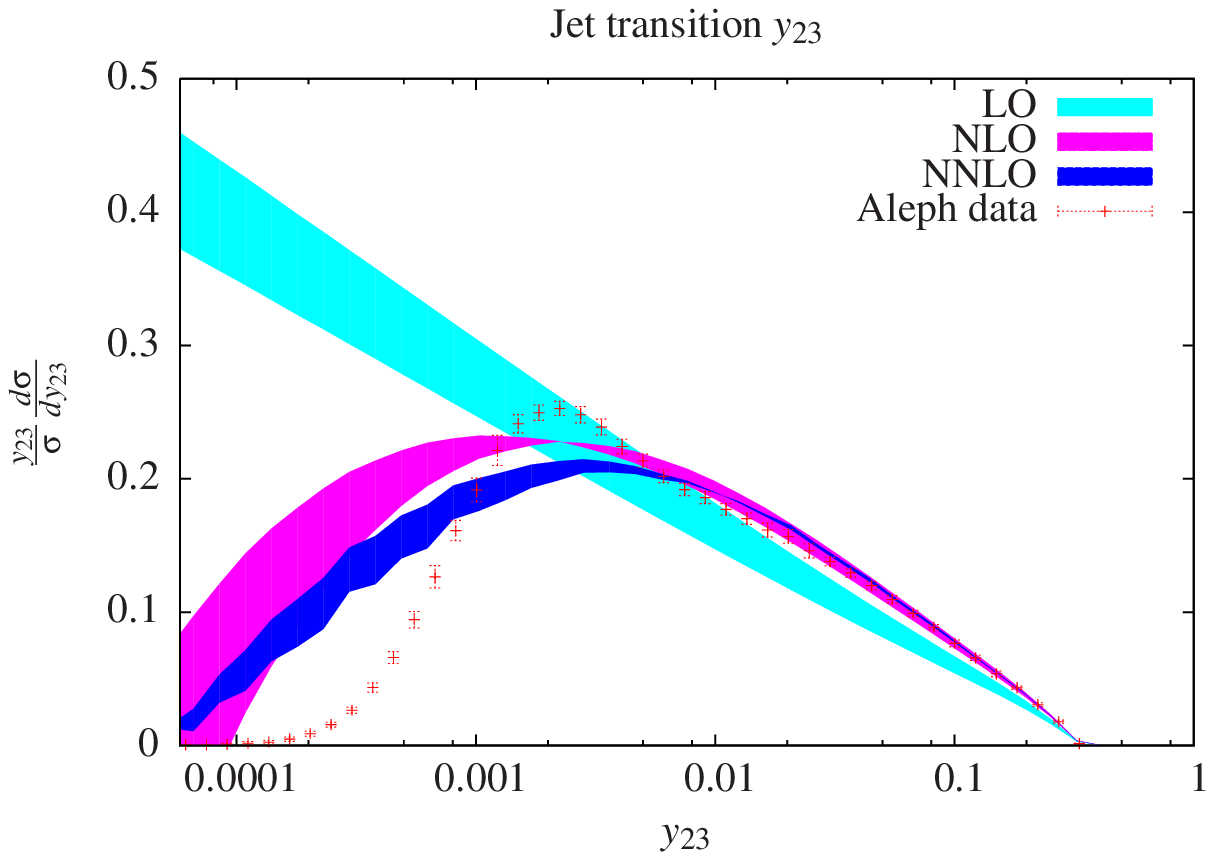}
\end{center}
\caption{The three-to-two jet transition distribution at LO, NLO and NNLO at $\sqrt{Q^2}=m_Z$ with $\alpha_s(m_Z)=0.118$.
The bands give the range for the theoretical prediction obtained from varying the renormalisation scale
from $\mu=m_Z/2$ to $\mu=2 m_Z$.
In addition the experimental data points from the Aleph experiment are shown.
}
\label{fig_y23}
\end{figure}

\begin{figure}[p]
\begin{center}
\includegraphics[bb= 125 460 490 710,width=0.9\textwidth]{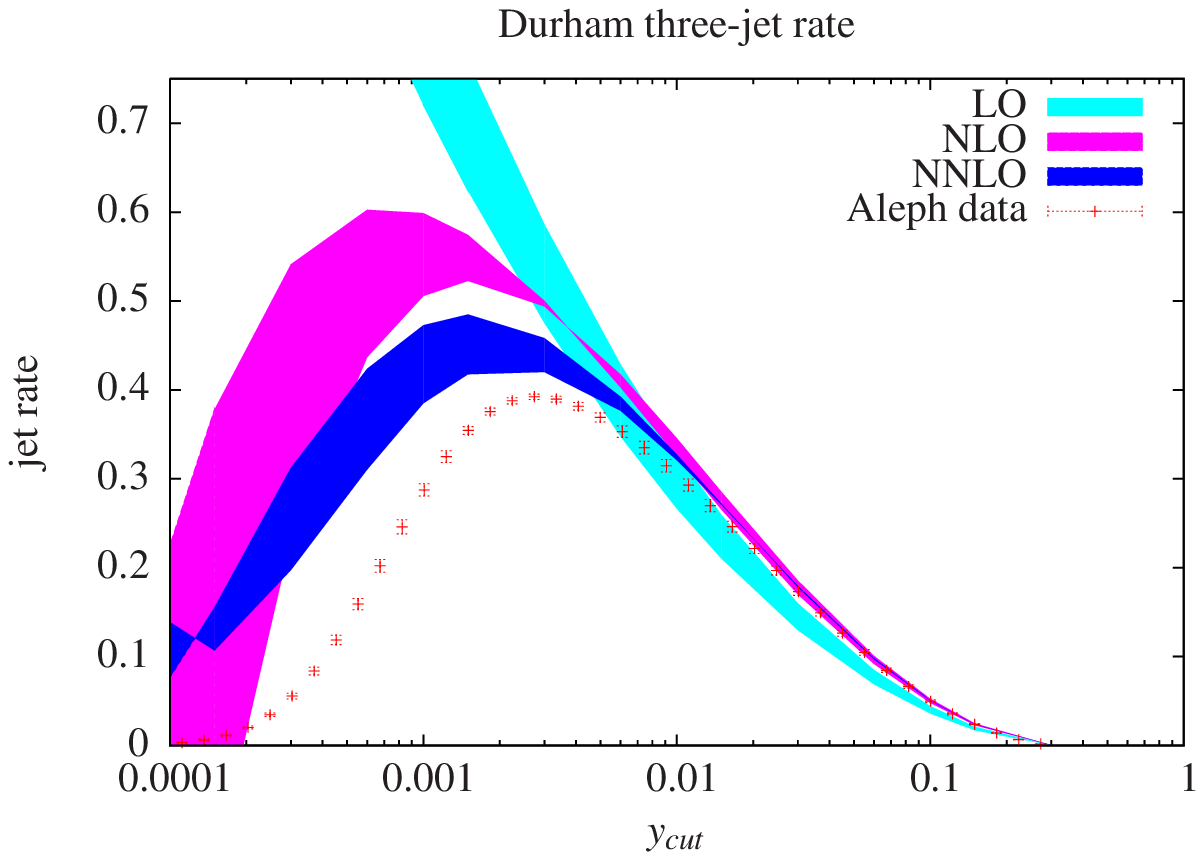}
\end{center}
\caption{The three jet rate with the Durham jet algorithm at LO, NLO and NNLO at $\sqrt{Q^2}=m_Z$ with $\alpha_s(m_Z)=0.118$.
The bands give the range for the theoretical prediction obtained from varying the renormalisation scale
from $\mu=m_Z/2$ to $\mu=2 m_Z$.
In addition the experimental data points from the Aleph experiment are shown.
}
\label{fig_jet_rate_durham}
\end{figure}

\begin{figure}[p]
\begin{center}
\includegraphics[bb= 125 460 490 710,width=0.9\textwidth]{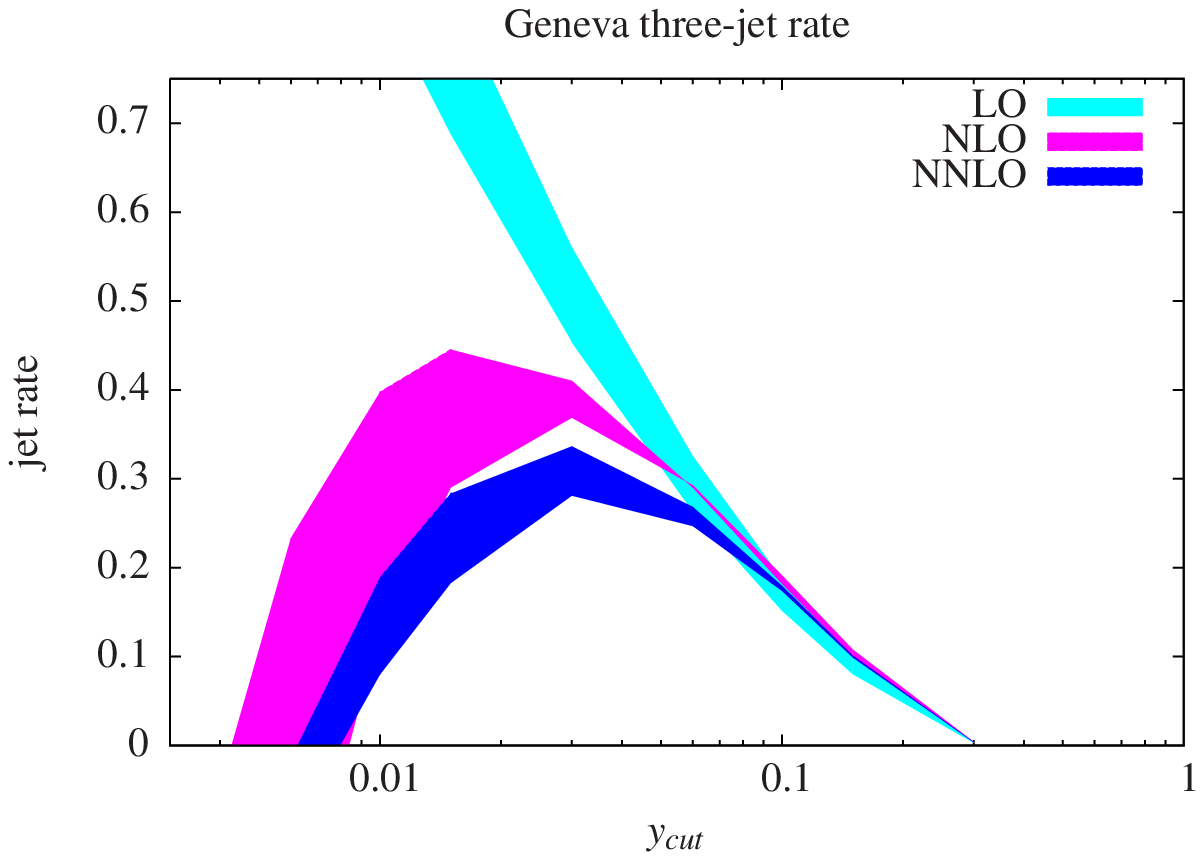}
\end{center}
\caption{The three jet rate with the Geneva jet algorithm at LO, NLO and NNLO at $\sqrt{Q^2}=m_Z$ with $\alpha_s(m_Z)=0.118$.
The bands give the range for the theoretical prediction obtained from varying the renormalisation scale
from $\mu=m_Z/2$ to $\mu=2 m_Z$.
}
\label{fig_jet_rate_geneva}
\end{figure}

\begin{figure}[p]
\begin{center}
\includegraphics[bb= 125 460 490 710,width=0.9\textwidth]{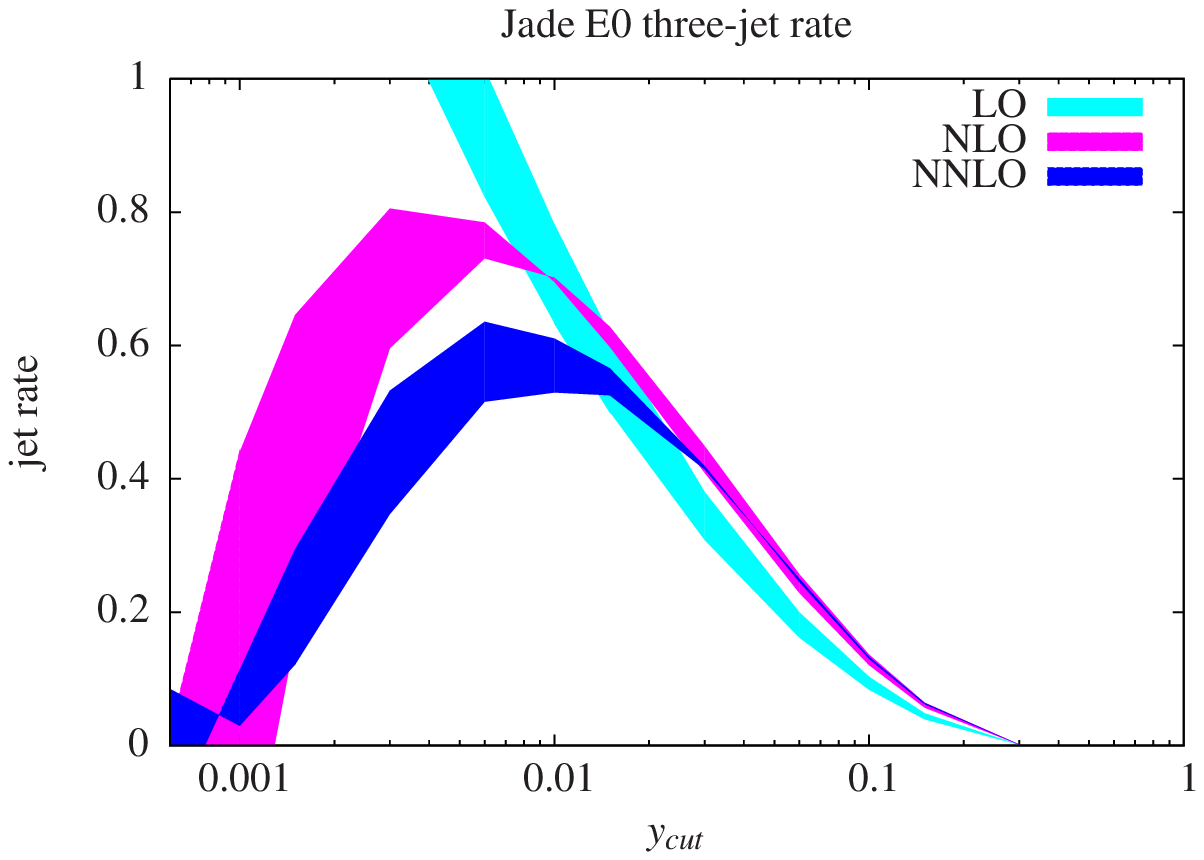}
\end{center}
\caption{The three jet rate with the Jade E0 jet algorithm at LO, NLO and NNLO at $\sqrt{Q^2}=m_Z$ with $\alpha_s(m_Z)=0.118$.
The bands give the range for the theoretical prediction obtained from varying the renormalisation scale
from $\mu=m_Z/2$ to $\mu=2 m_Z$.
}
\label{fig_jet_rate_jadeE0}
\end{figure}

\begin{figure}[p]
\begin{center}
\includegraphics[bb= 125 460 490 710,width=0.9\textwidth]{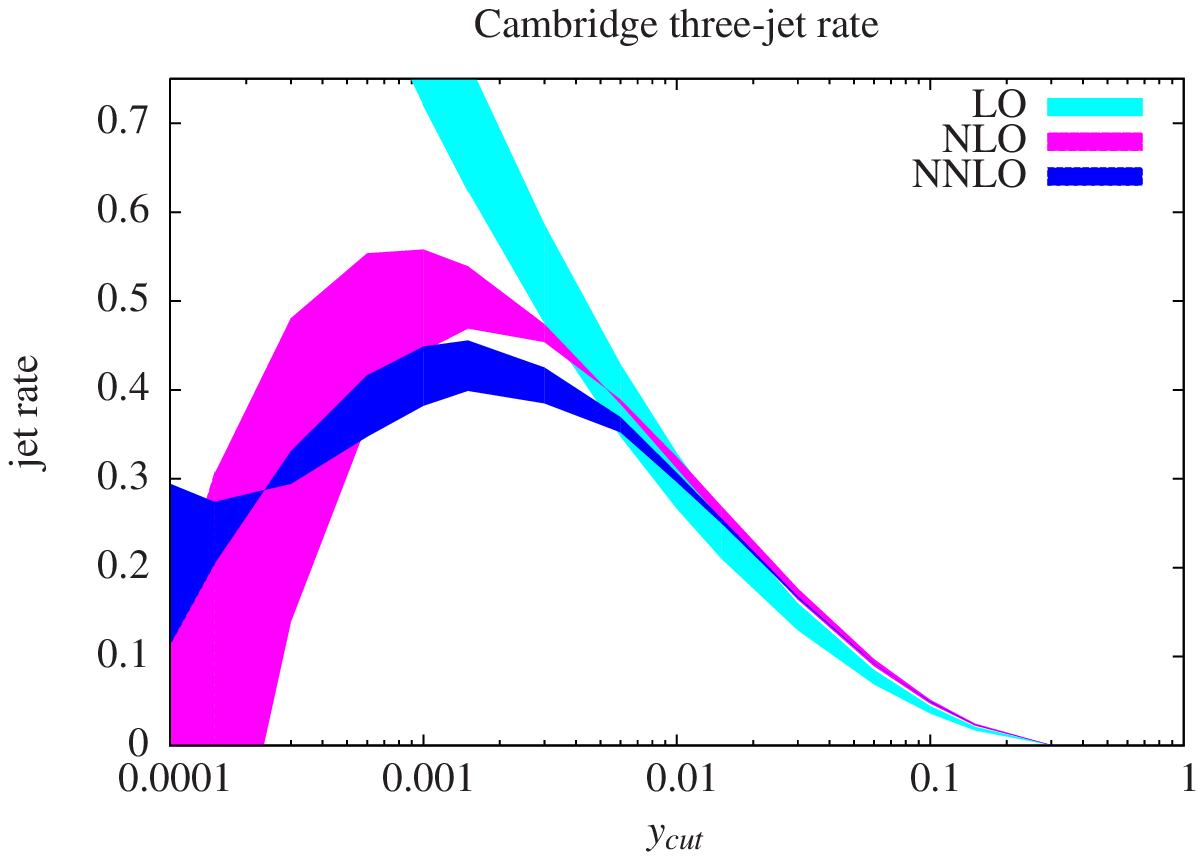}
\end{center}
\caption{The three jet rate with the Cambridge jet algorithm at LO, NLO and NNLO at $\sqrt{Q^2}=m_Z$ with $\alpha_s(m_Z)=0.118$.
The bands give the range for the theoretical prediction obtained from varying the renormalisation scale
from $\mu=m_Z/2$ to $\mu=2 m_Z$.
}
\label{fig_jet_rate_cambridge}
\end{figure}

In this section I present the numerical results for the event shape variables and jet rates at 
next-to-next-to-leading order.
All results have been obtained by numerical Monte-Carlo integration.
The Monte-Carlo integration introduces a statistical error, which will be quoted with all results.
For the infrared singularities a hybrid method between subtraction and slicing has been used.
Unless stated otherwise all results have been obtained with the slicing parameter
\bq
 \eta & = & \frac{s_{min}}{Q^2} = 10^{-5}.
\eq
It should be noted that the slicing procedure introduces in addition a systematic error.
The size of this error can be estimated by varying the slicing parameter $\eta$.
However, lowering the slicing parameter will increase the statistical error.
A practical criteria is to require that the variation due to the slicing parameter is smaller than
the statistical error, with the possible exception for the boundaries of the distributions.
Imposing this criteria $\eta=10^{-5}$ turns out to be a good compromise between accuracy and efficiency.
The boundaries of the distributions close to the two-jet region deserve special attention:
There the value of the observable is comparable to $\eta$ and one expects sizable corrections.
I will discuss the dependence on the slicing parameter in the close-to-two-jet region
in detail for a few examples. 
\\
\\
In a first series of plots I show the comparison of the NNLO results with the NLO and LO results
for the LEP I centre-of-mass energy $\sqrt{Q^2}=m_Z$ with $\alpha_S=0.118$.
The results for the event shape variables thrust, heavy jet mass, wide jet broadening, total jet broadening, C parameter
and three-to-two jet transition variable $y_{23}$ are shown in figs.~\ref{fig_thrust} to \ref{fig_y23}.
For each of these observable the distribution weighted by the observable and normalised to the total
hadronic cross section is shown, e.g. for thrust the distribution
\bq
 \frac{1-T}{\sigma} \frac{d\sigma}{d(1-T)}
\eq
is shown.
The corresponding plots for the jet rates for the jet algorithms of Durham, Geneva, Jade-E0 and Cambridge
are shown in figs.~\ref{fig_jet_rate_durham} to \ref{fig_jet_rate_cambridge}.
In all these plots is
the leading-order prediction shown in light blue, the next-to-leading-order prediction is shown in pink and the
next-to-next-to-leading order prediction is shown in dark blue.
The bands give the range for the theoretical prediction obtained from varying the renormalisation scale
from $\mu=Q/2$ to $\mu=2 Q$.
In addition the experimental data points \cite{Heister:2003aj} from the Aleph experiment (where available) are also
shown in these plot.
Note that the theory predictions in these plots are the pure perturbative predictions. Power corrections or soft gluon resummation
effects are not included in these results.

\begin{figure}[p]
\begin{center}
\includegraphics[bb= 125 460 490 710,width=0.45\textwidth]{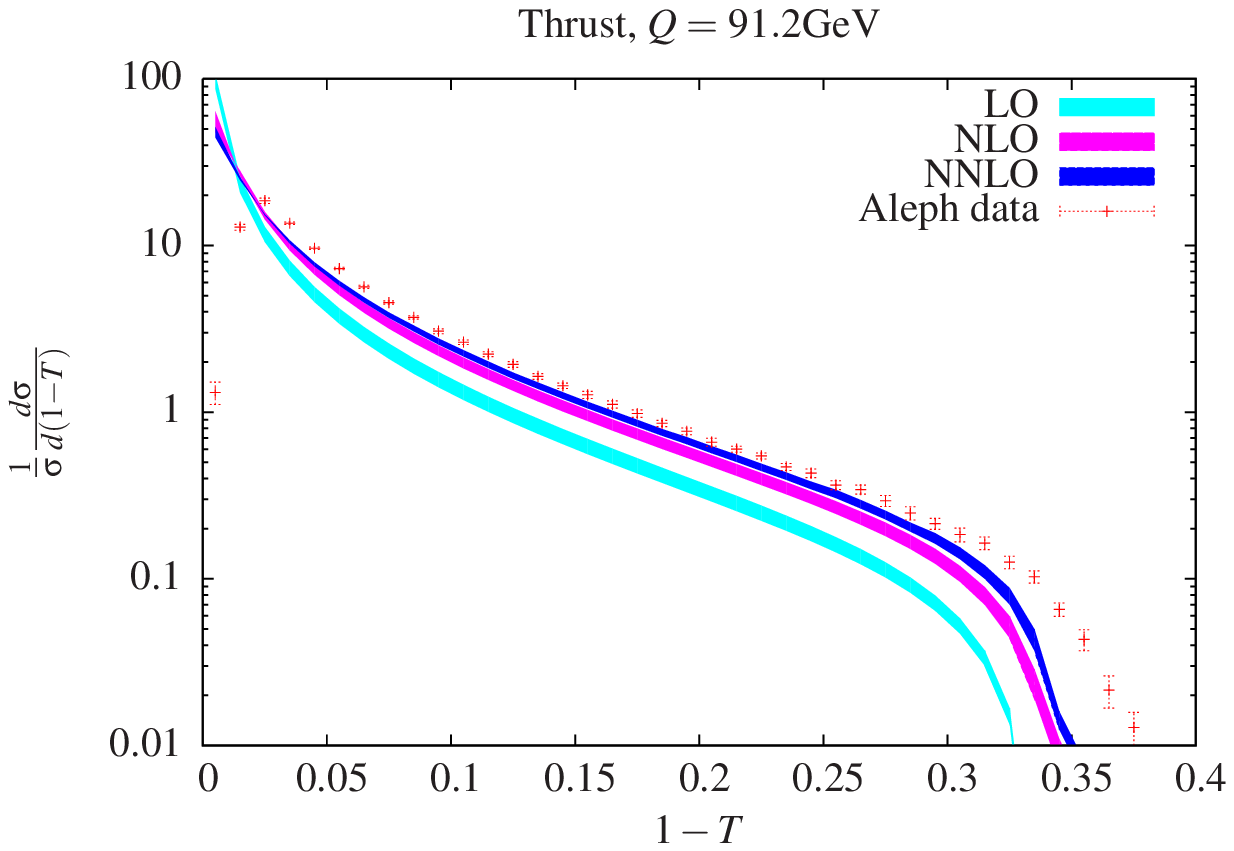}
\includegraphics[bb= 125 460 490 710,width=0.45\textwidth]{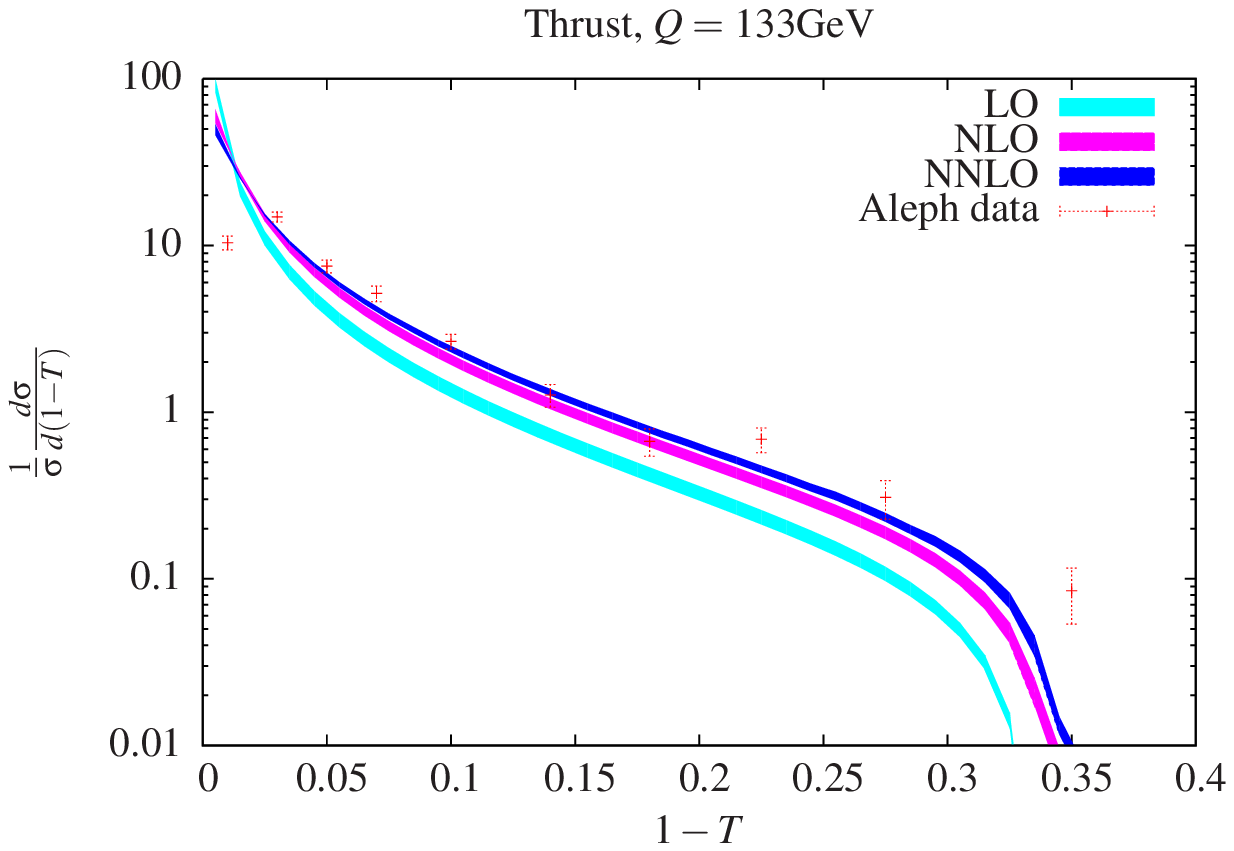}
\\
\includegraphics[bb= 125 460 490 710,width=0.45\textwidth]{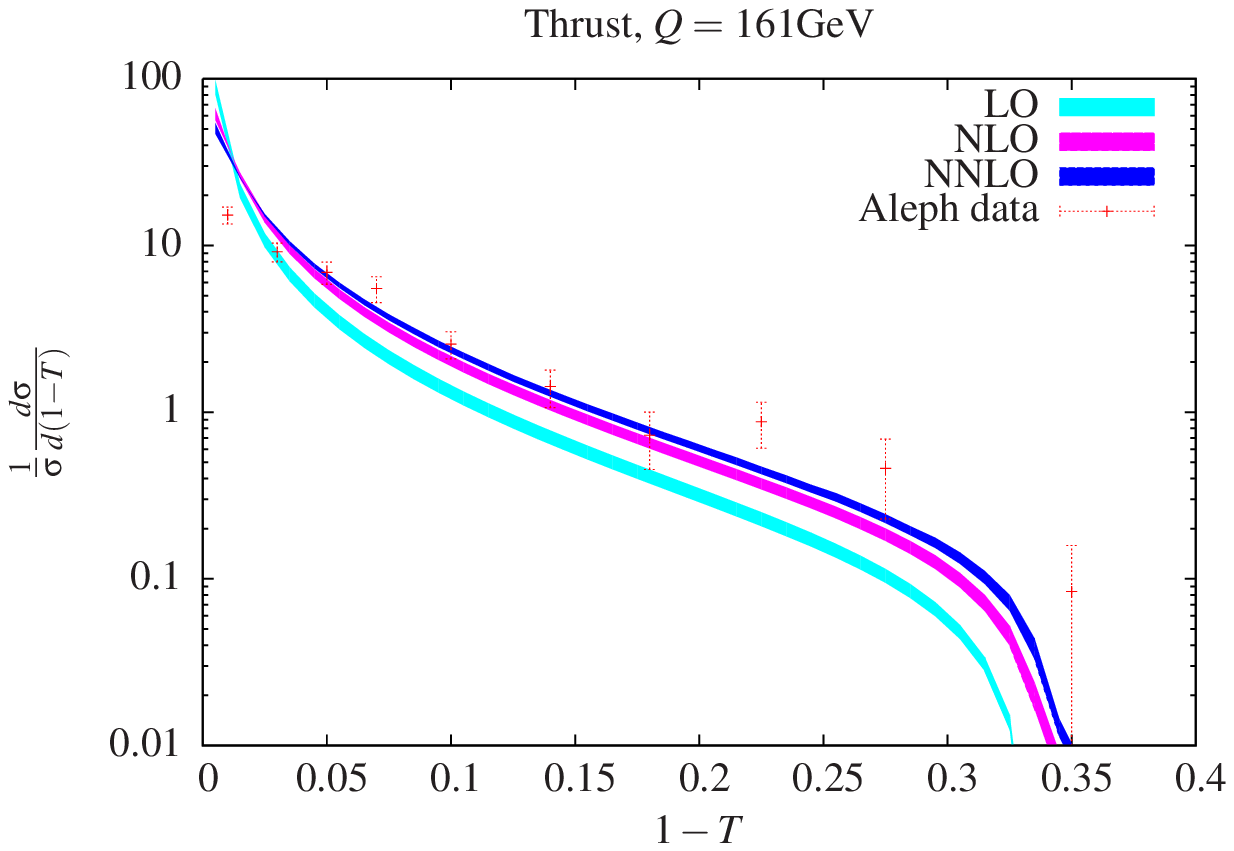}
\includegraphics[bb= 125 460 490 710,width=0.45\textwidth]{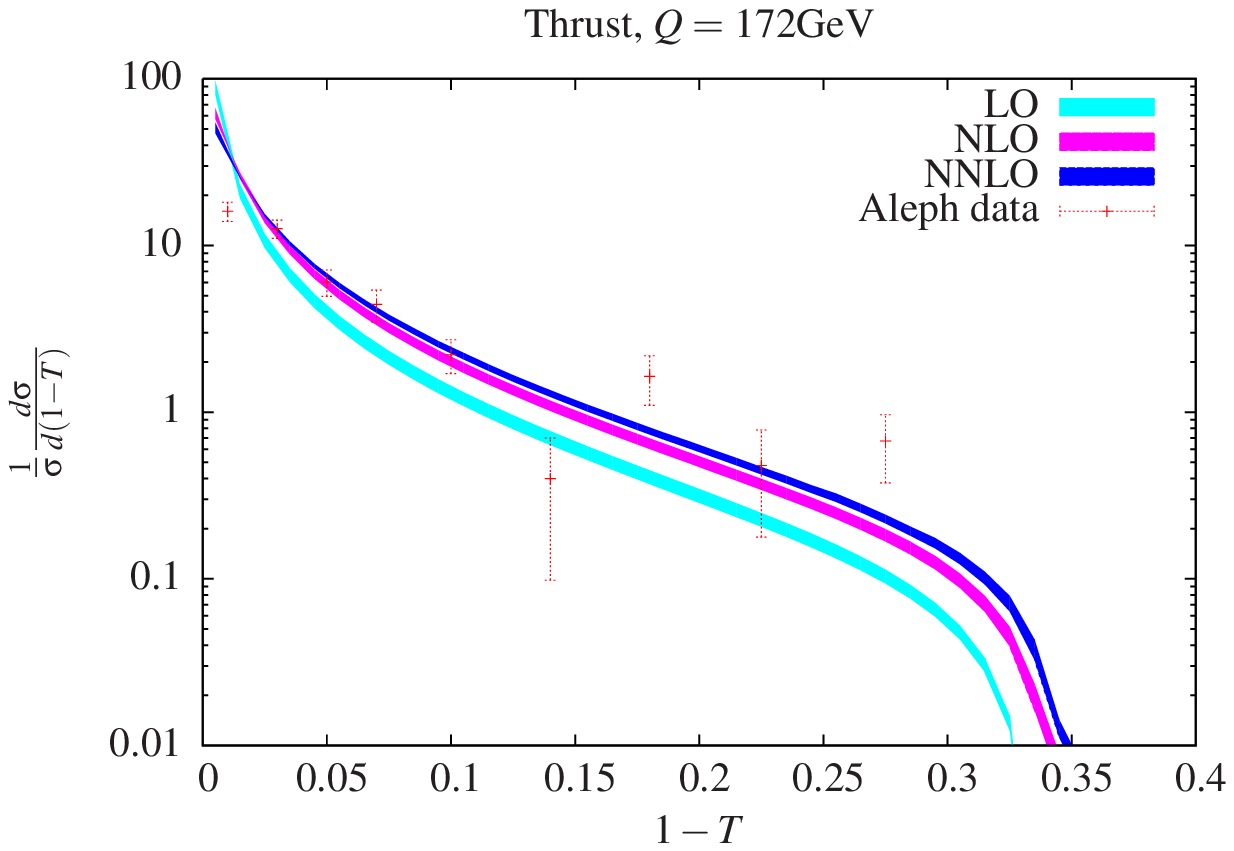}
\\
\includegraphics[bb= 125 460 490 710,width=0.45\textwidth]{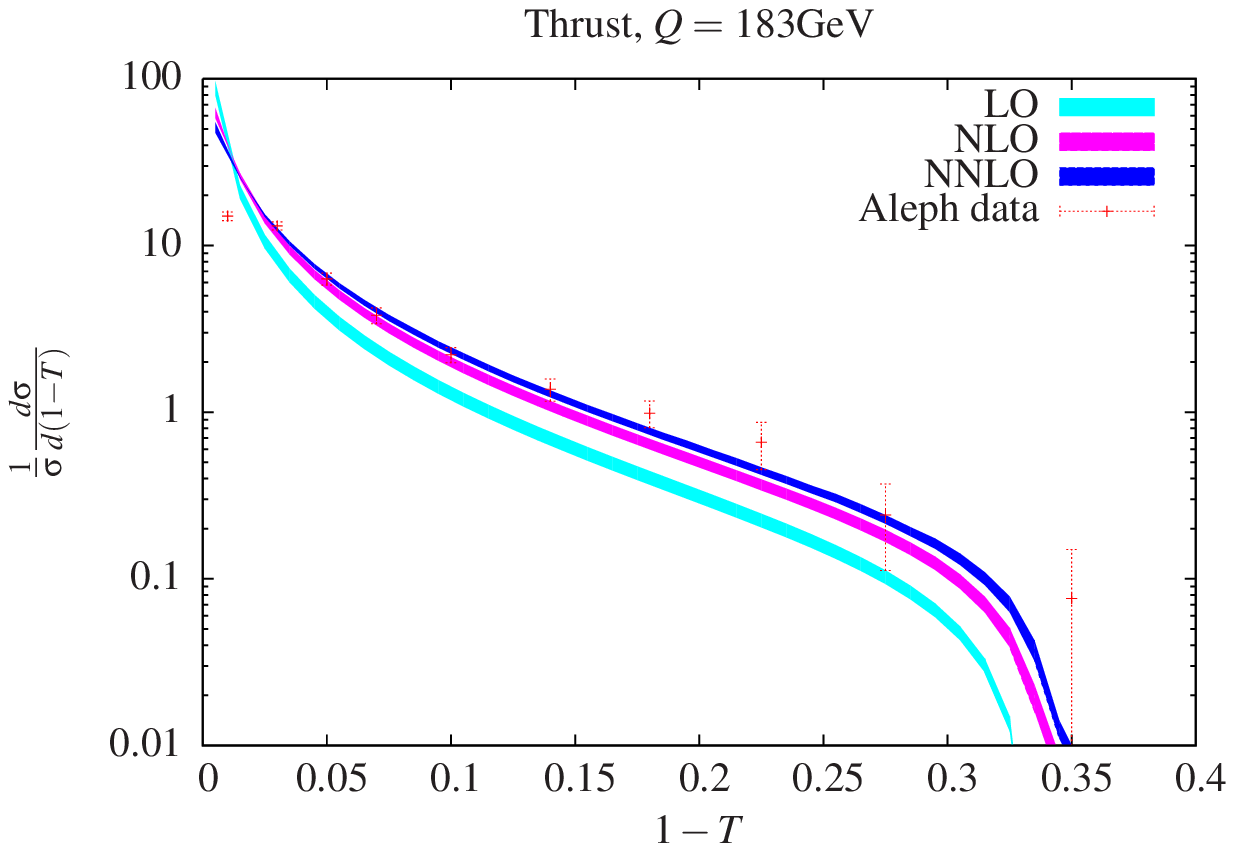}
\includegraphics[bb= 125 460 490 710,width=0.45\textwidth]{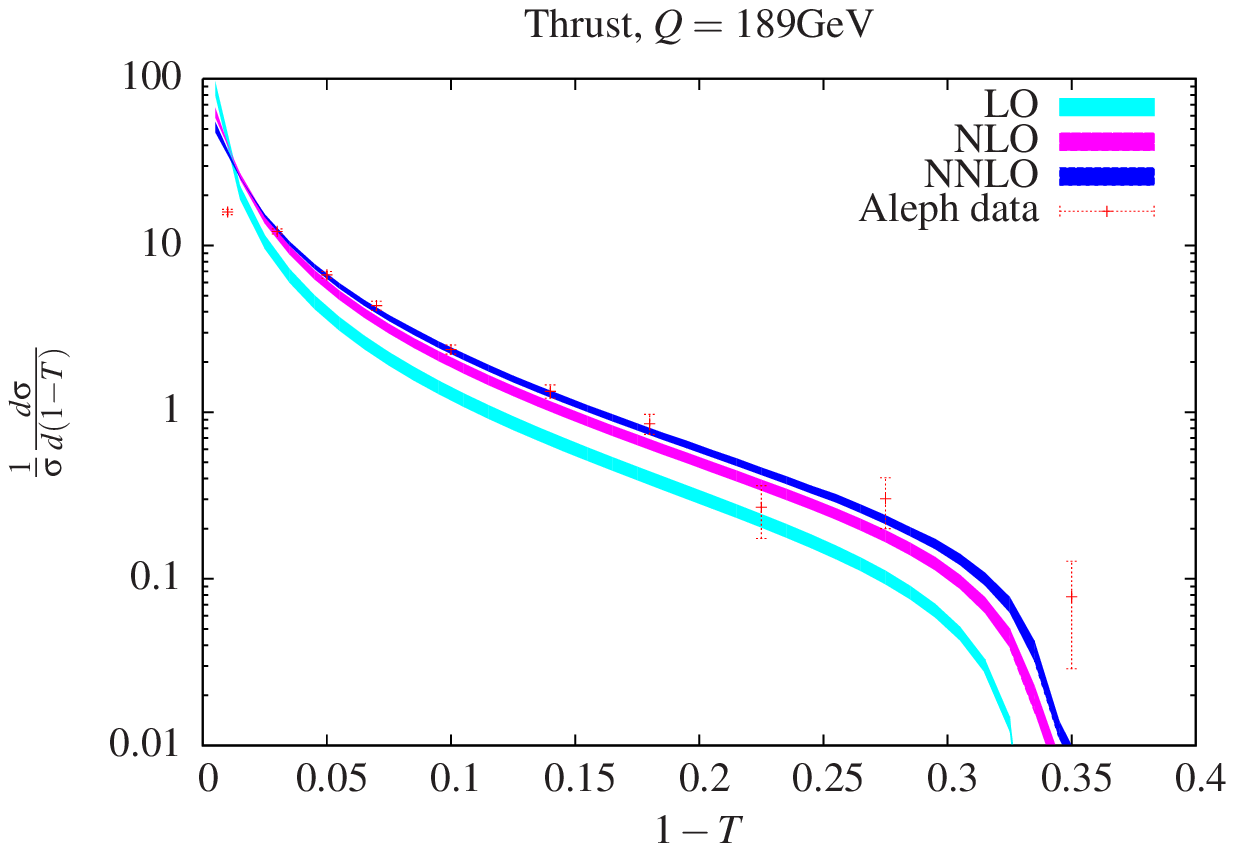}
\\
\includegraphics[bb= 125 460 490 710,width=0.45\textwidth]{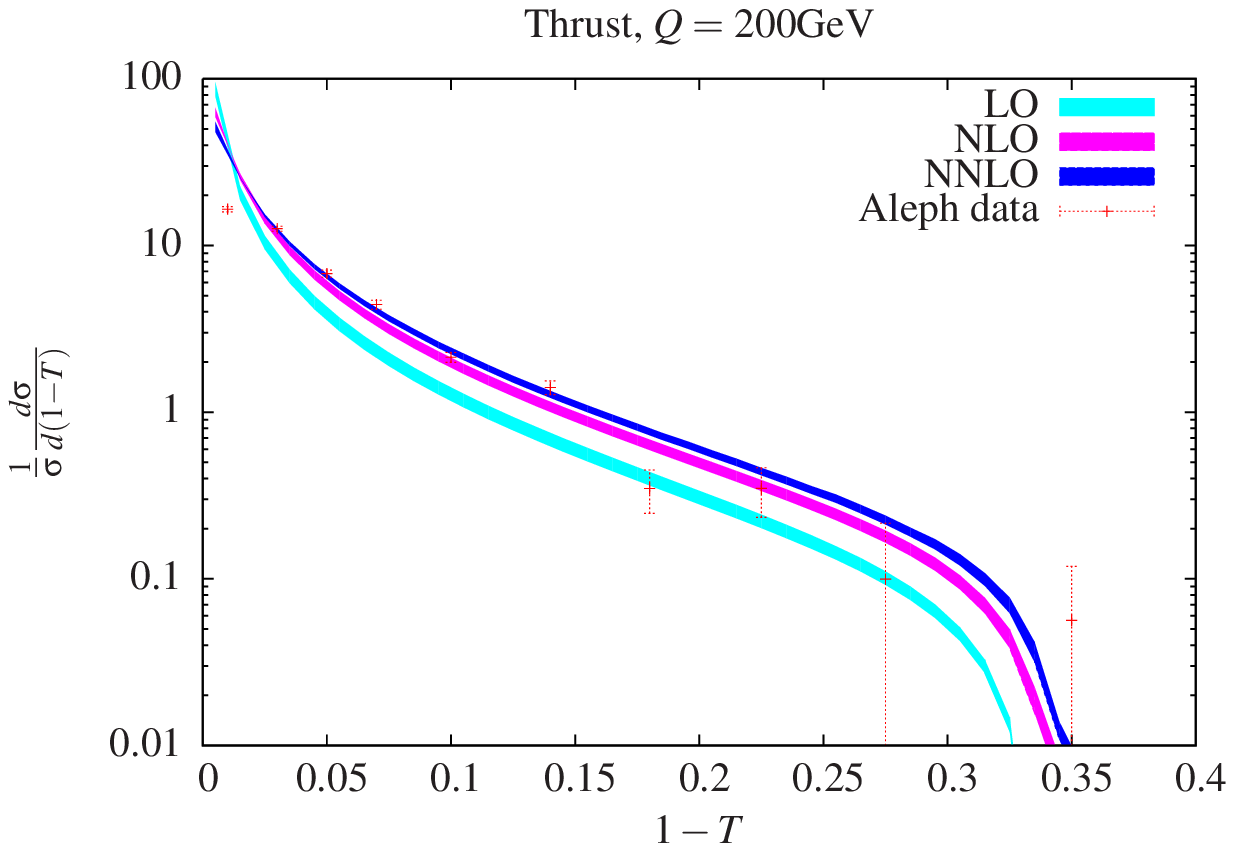}
\includegraphics[bb= 125 460 490 710,width=0.45\textwidth]{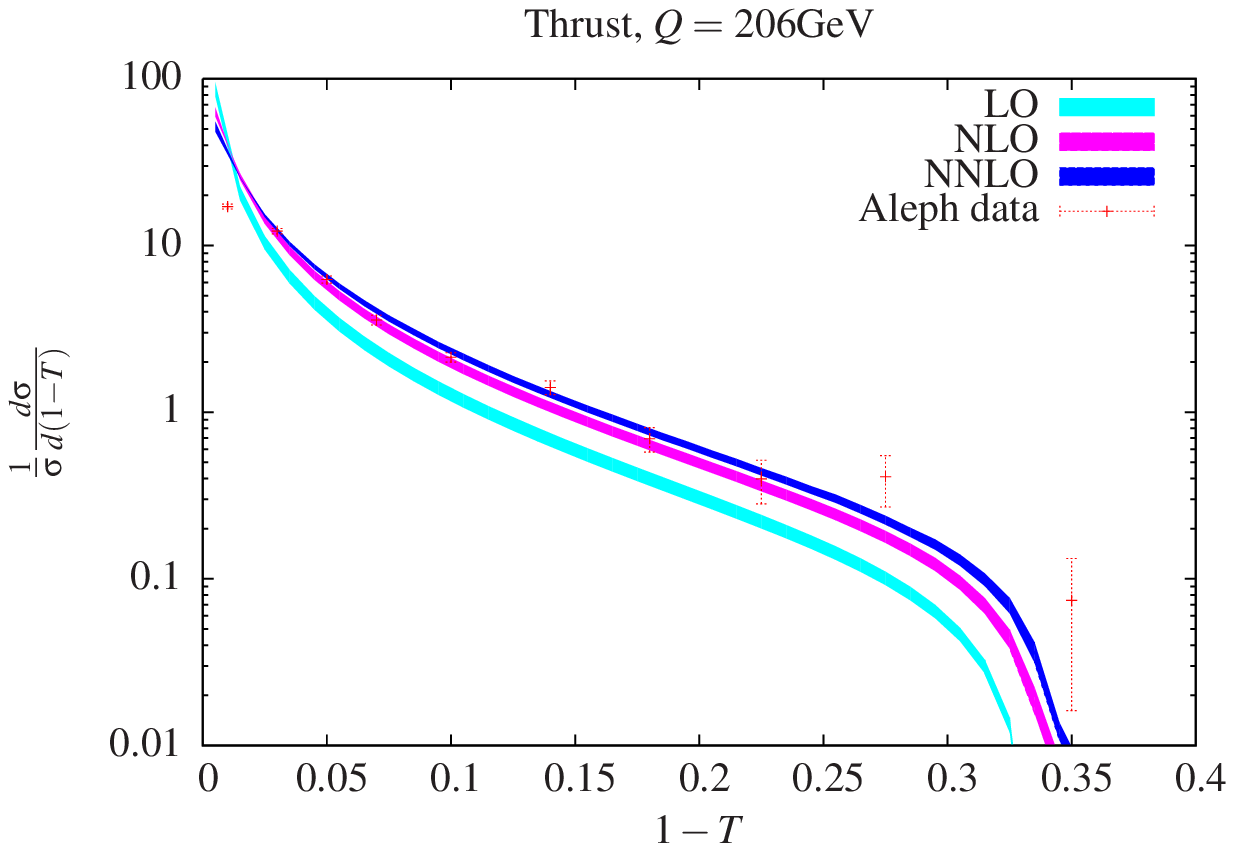}
\end{center}
\caption{The thrust distribution at LO, NLO and NNLO at various values of $\sqrt{Q^2}$.
The bands correspond to a variation of the renormalisation scale from $\mu=m_Z/2$ to $\mu=2 m_Z$.
The experimental data points are from the Aleph experiment.
}
\label{fig_thrust_lep2}
\end{figure}

Numerical results for centre-of-mass energies different from $\sqrt{Q^2}=m_Z$ are also easily obtained.
As an example I show in fig.~\ref{fig_thrust_lep2} the thrust distribution at centre-of-mass energies
of $\sqrt{Q^2}= 91.2\mbox{GeV}$, $133 \mbox{GeV}$, $161 \mbox{GeV}$, $172 \mbox{GeV}$, $183 \mbox{GeV}$, $189 \mbox{GeV}$, $200 \mbox{GeV}$, $206 \mbox{GeV}$,
again with experimental data points from the Aleph experiment.
As for the LEP II energies the bin size of the experimental data is rather large, I show in fig.~\ref{fig_thrust_lep2} the distribution
\bq
 \frac{1}{\sigma} \frac{d\sigma}{d(1-T)}
\eq
without an additional factor $(1-T)$.

The most important results of this paper are the perturbative coefficients
$A_{\cal O}$, $B_{\cal O}$ and $C_{\cal O}$.
For the six event shape variables 
(thrust, heavy jet mass, wide jet broadening, total jet broadening, $C$-parameter and the three-to-two jet transition variable)
the numerical values for the LO functions $A_{\cal O}$, 
the NLO functions $B_{\cal O}$ and the NNLO functions $C_{\cal O}$, all weighted by ${\cal O}$, are given
in tables~\ref{table_thrust} to \ref{table_y23}.
The corresponding values for the jet rates defined according to the Durham, Geneva, Jade-E0 or Cambridge algorithms
are given in tables~\ref{table_jet_rate_durham} to \ref{table_jet_rate_cambridge}.
These values are obtained by Monte-Carlo integration and the statistical Monte-Carlo integration error is indicated in these tables.
The systematic error due to the slicing parameter $\eta$ is not included in these tables, but will be discussed below.
The values for the Durham jet rate have already been given in ref.~\cite{Weinzierl:2008iv}.
Due to a typo in the numerical program which was corrected after publication of \cite{Weinzierl:2008iv}
the corrected values are repeated here.
 
\begin{figure}[p]
\begin{center}
\includegraphics[bb= 125 460 490 710,width=0.32\textwidth]{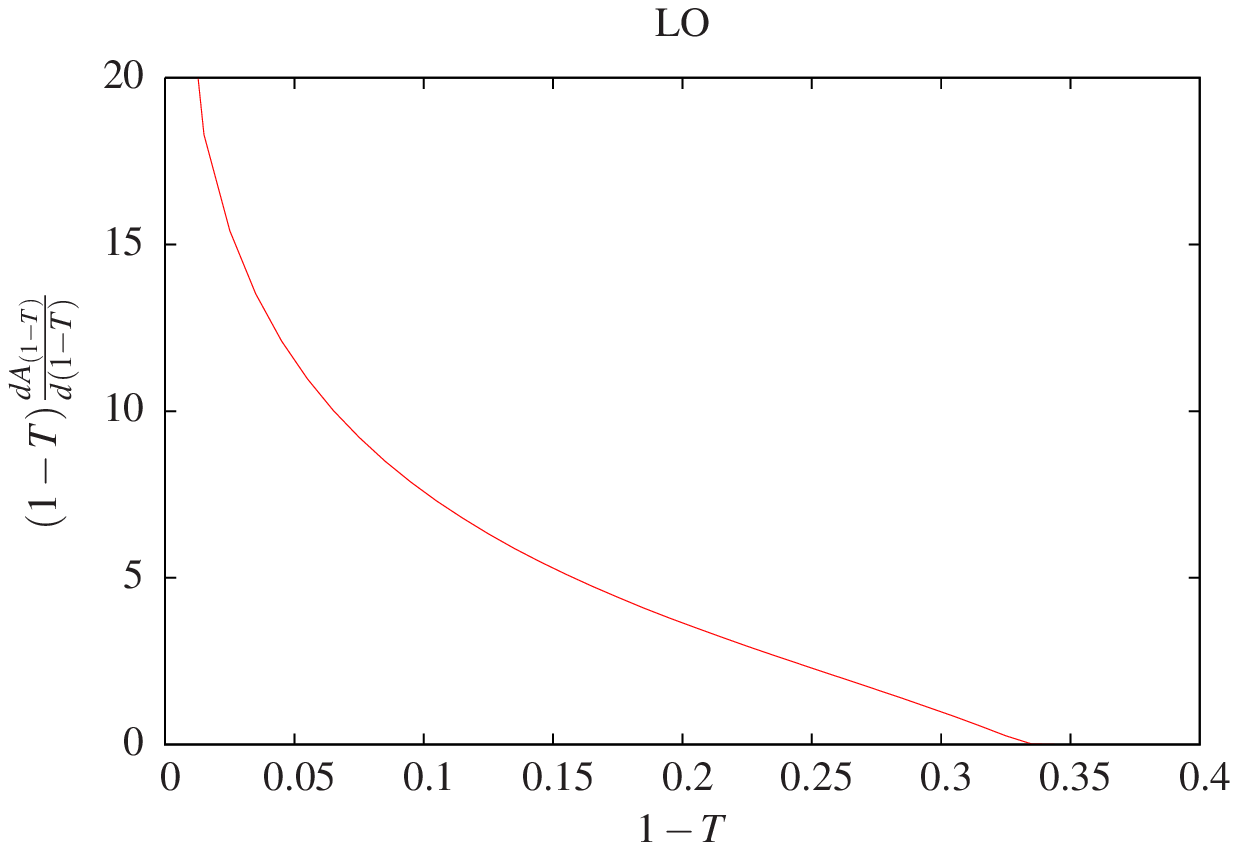}
\includegraphics[bb= 125 460 490 710,width=0.32\textwidth]{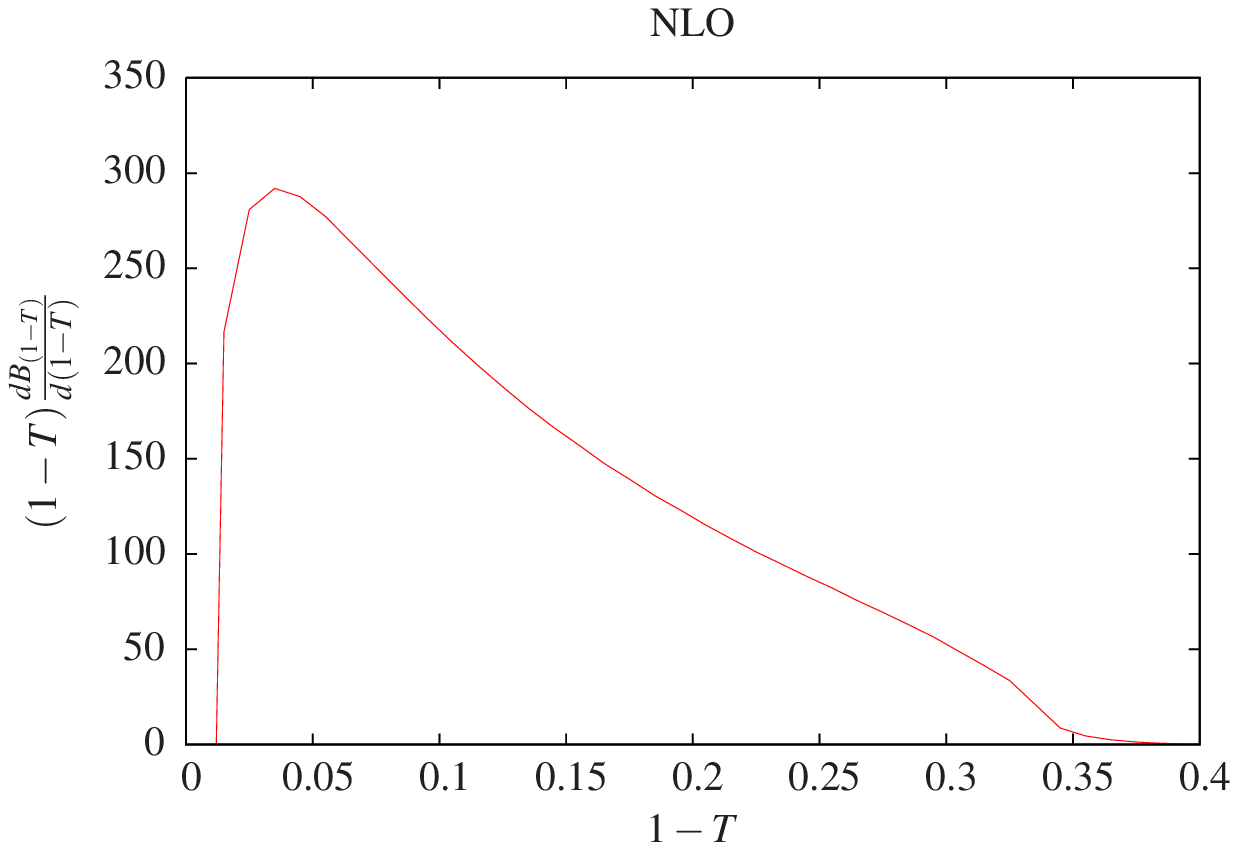}
\includegraphics[bb= 125 460 490 710,width=0.32\textwidth]{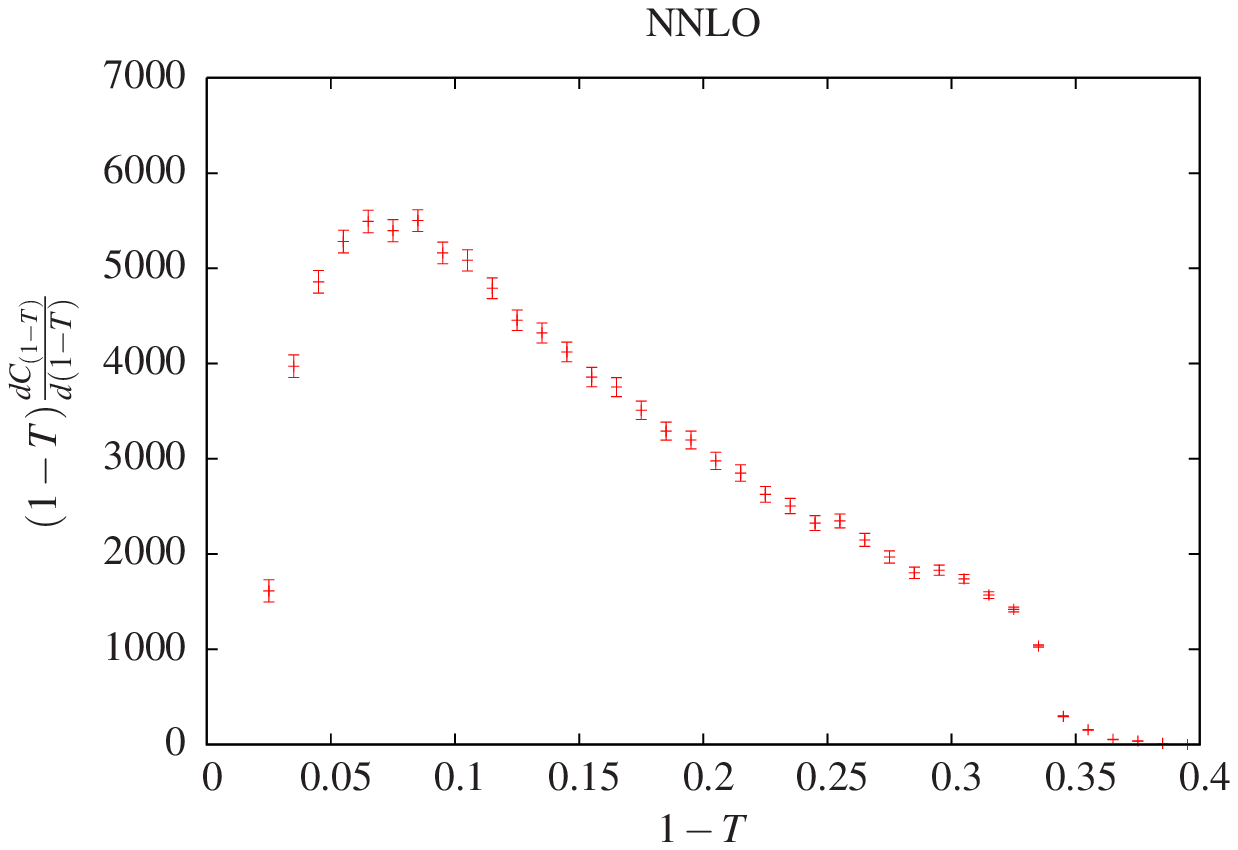}
\end{center}
\caption{
Coefficients of the leading-order ($A_{(1-T)}$, left), 
next-to-leading-order ($B_{(1-T)}$, middle)
and next-to-next-to-leading order ($C_{(1-T)}$, right)
contributions to the thrust distribution, all weighted by $(1-T)$.
For the coefficient $C_{(1-T)}$ the Monte-Carlo integration errors are also shown.
}
\label{fig_thrust_ABC}
\end{figure}

\begin{figure}[p]
\begin{center}
\includegraphics[bb= 125 460 490 710,width=0.32\textwidth]{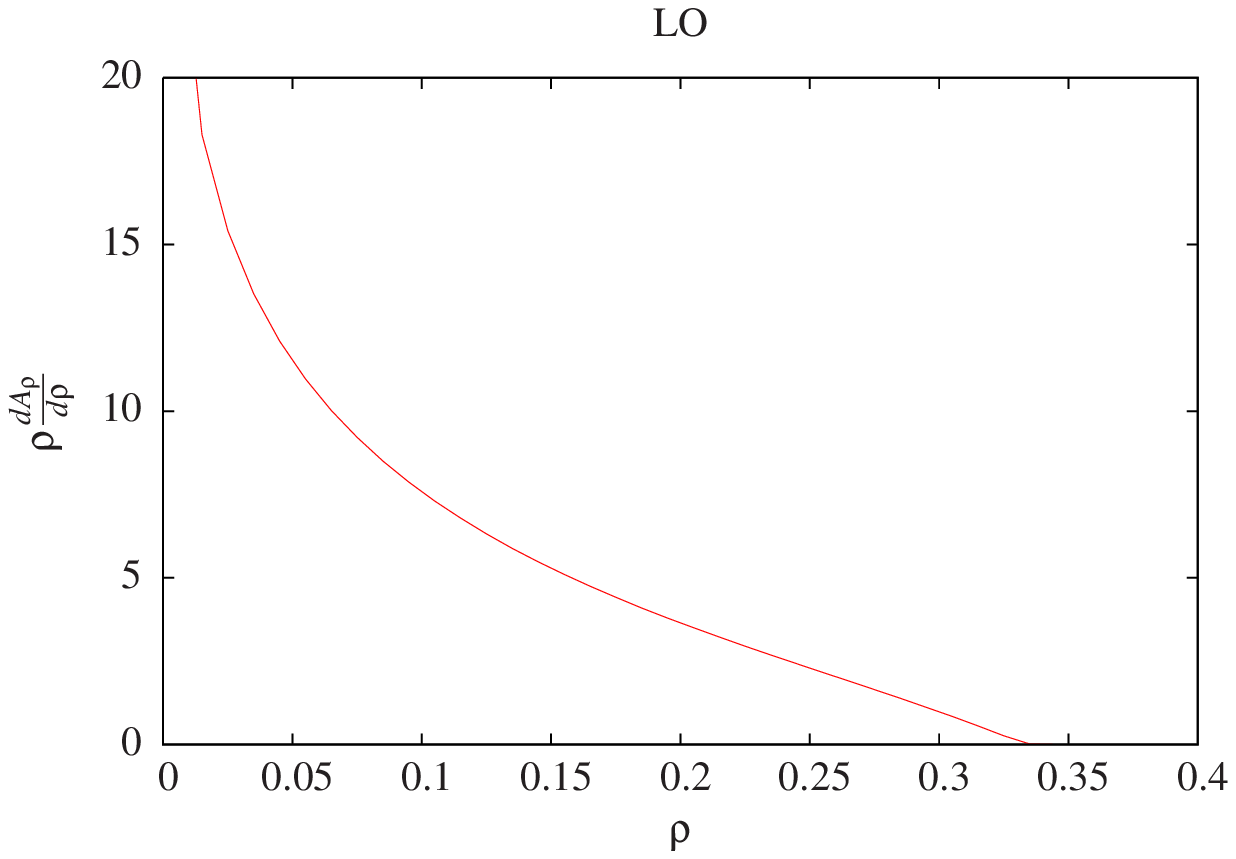}
\includegraphics[bb= 125 460 490 710,width=0.32\textwidth]{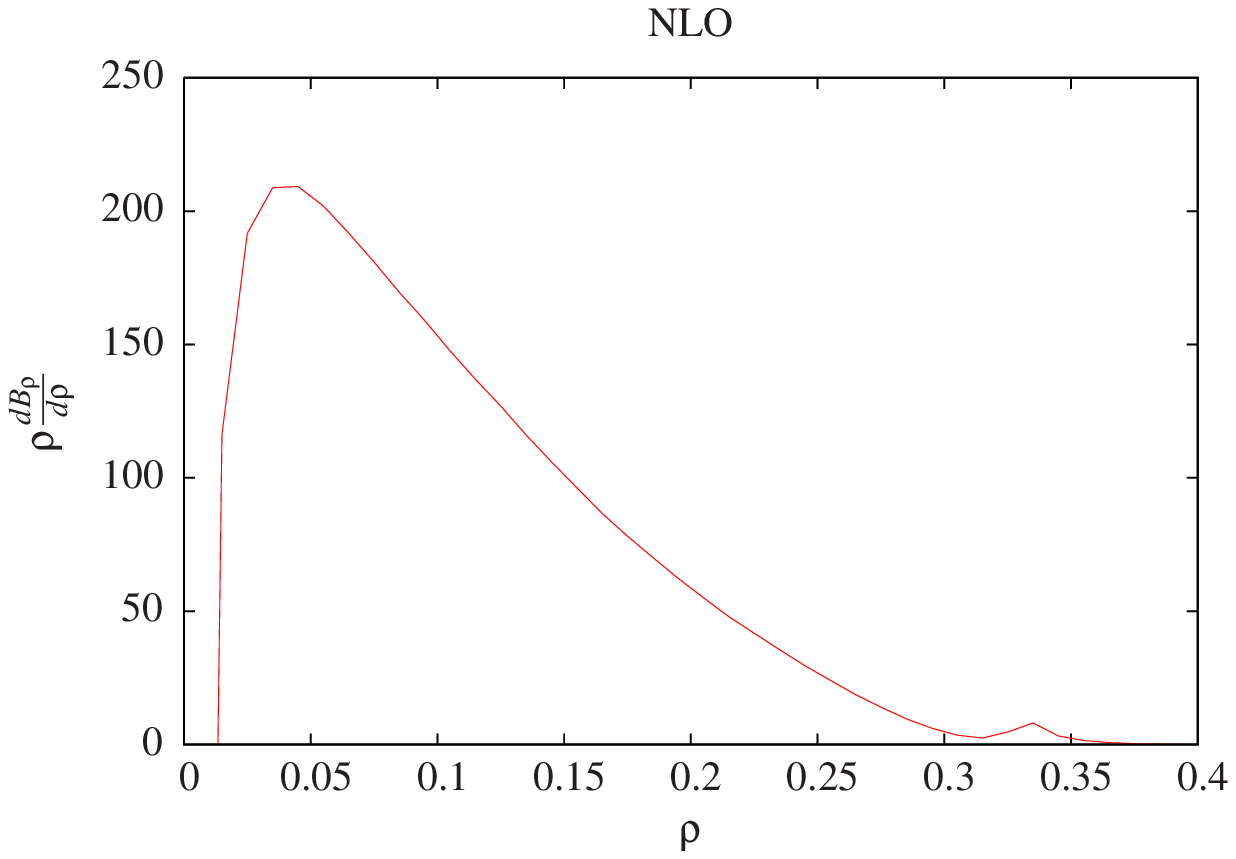}
\includegraphics[bb= 125 460 490 710,width=0.32\textwidth]{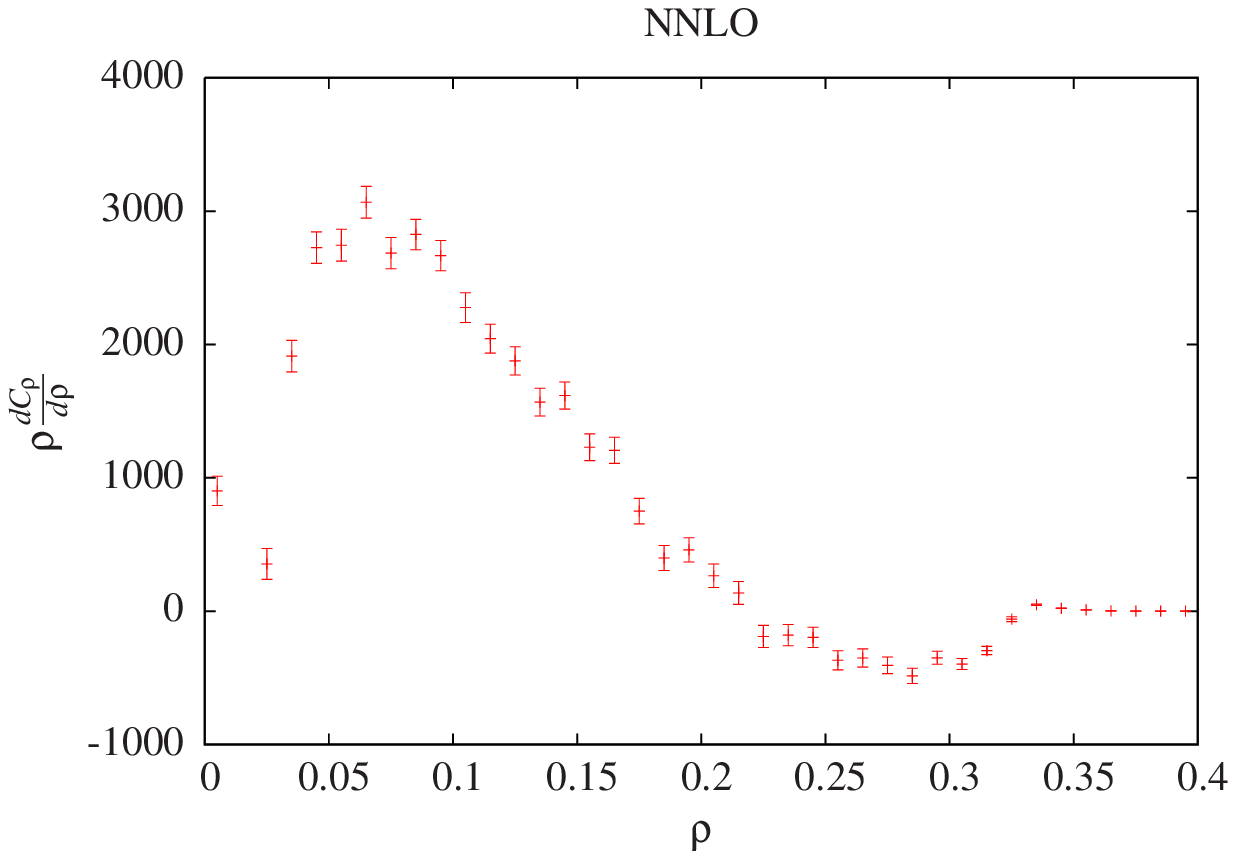}
\end{center}
\caption{
Coefficients of the leading-order ($A_{\rho}$, left), 
next-to-leading-order ($B_{\rho}$, middle)
and next-to-next-to-leading order ($C_{\rho}$, right)
contributions to the heavy jet mass distribution, all weighted by $\rho$.
For the coefficient $C_{\rho}$ the Monte-Carlo integration errors are also shown.
}
\label{fig_heavyjetmass_ABC}
\end{figure}

\begin{figure}[p]
\begin{center}
\includegraphics[bb= 125 460 490 710,width=0.32\textwidth]{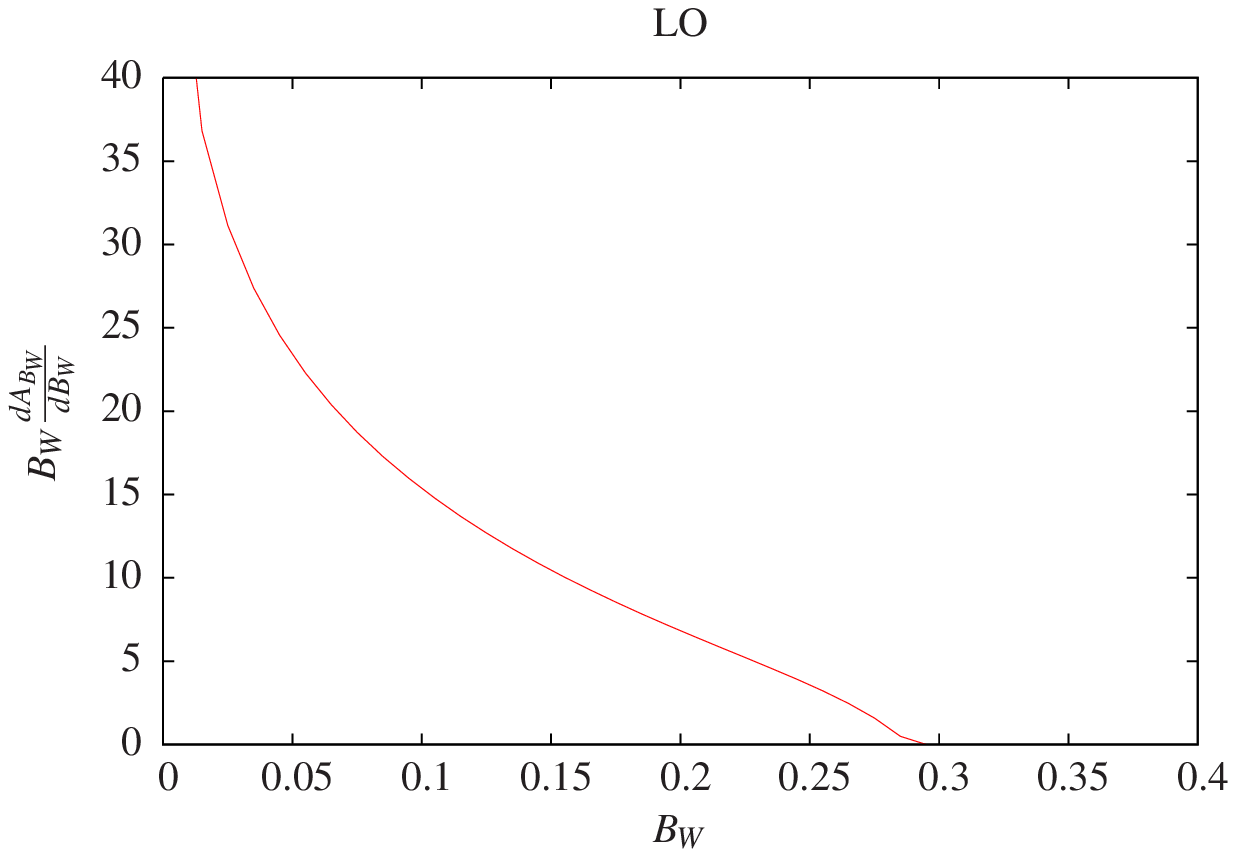}
\includegraphics[bb= 125 460 490 710,width=0.32\textwidth]{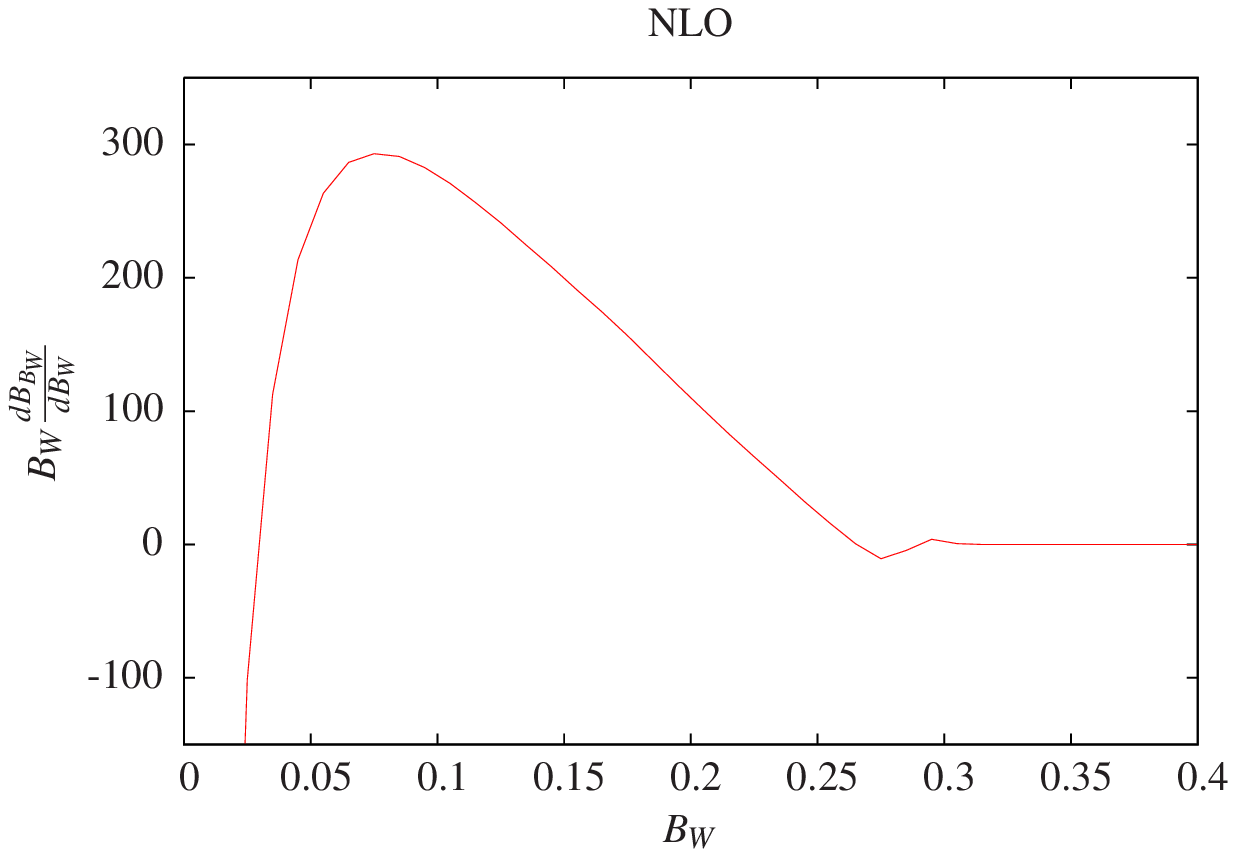}
\includegraphics[bb= 125 460 490 710,width=0.32\textwidth]{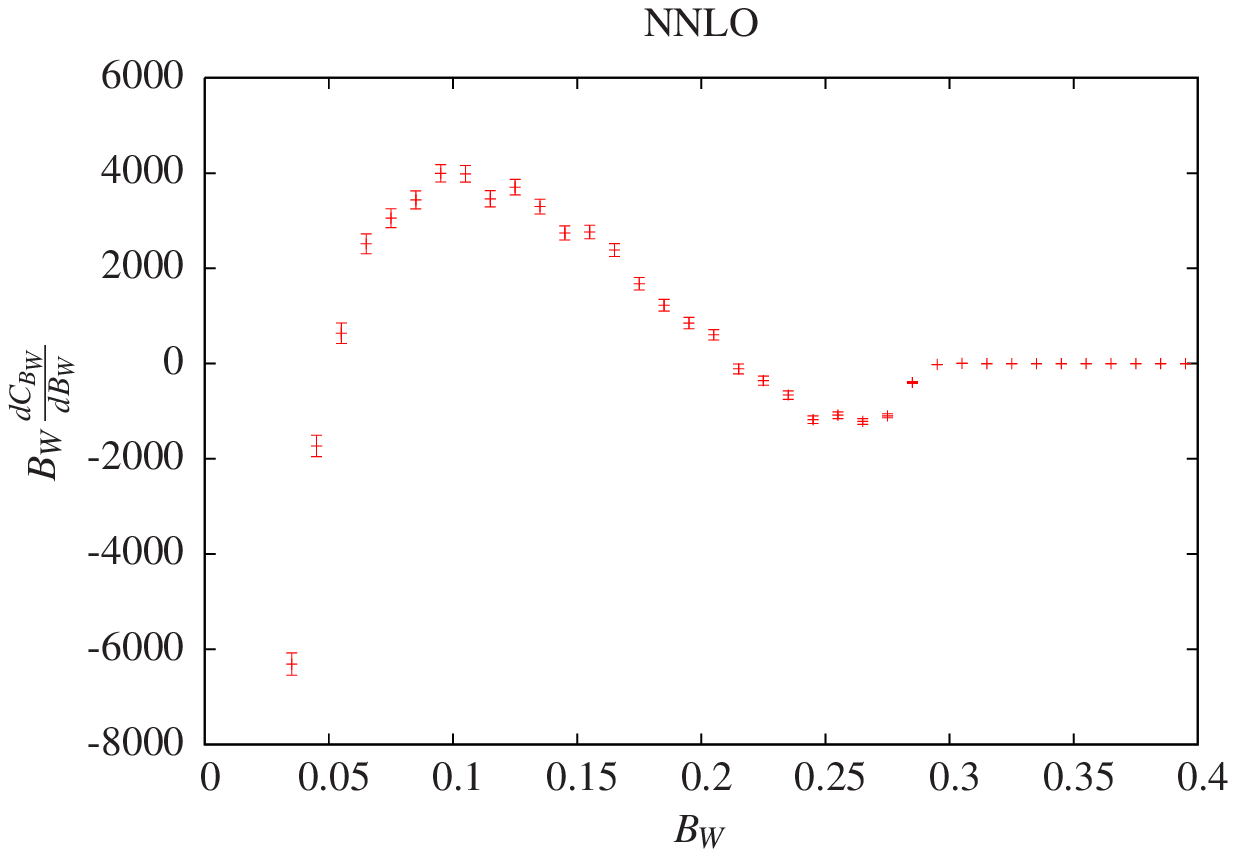}
\end{center}
\caption{
Coefficients of the leading-order ($A_{B_W}$, left), 
next-to-leading-order ($B_{B_W}$, middle)
and next-to-next-to-leading order ($C_{B_W}$, right)
contributions to the wide jet broadening distribution, all weighted by $B_W$.
For the coefficient $C_{B_W}$ the Monte-Carlo integration errors are also shown.
}
\label{fig_widejetbroadening_ABC}
\end{figure}

\begin{figure}[p]
\begin{center}
\includegraphics[bb= 125 460 490 710,width=0.32\textwidth]{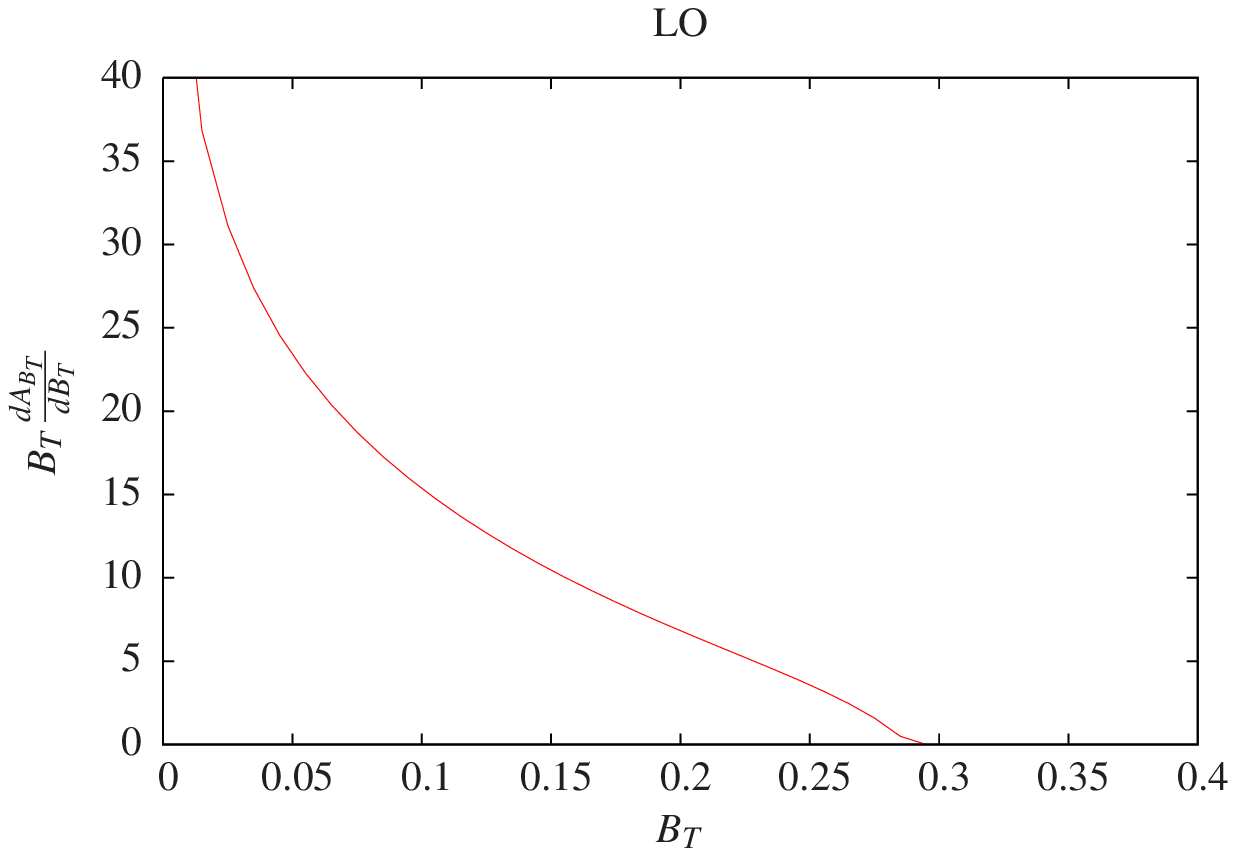}
\includegraphics[bb= 125 460 490 710,width=0.32\textwidth]{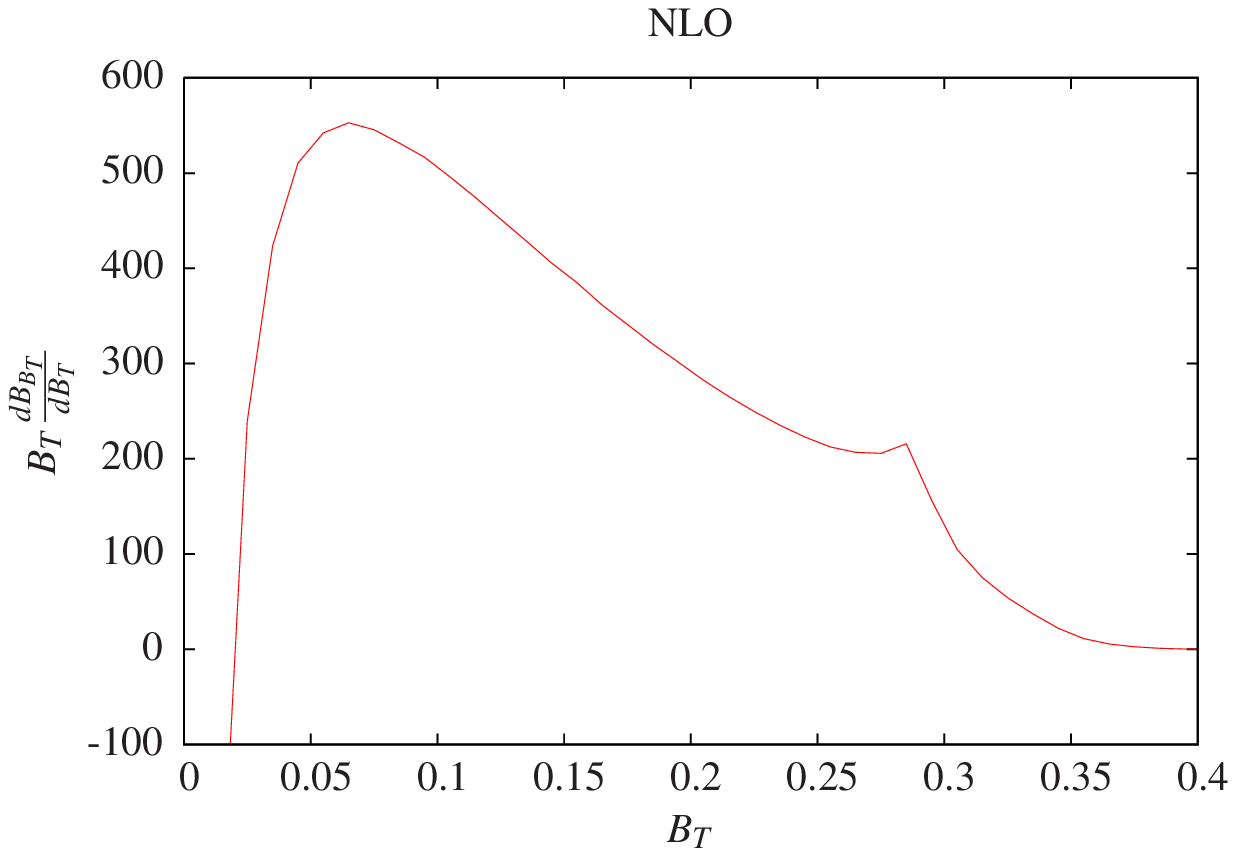}
\includegraphics[bb= 125 460 490 710,width=0.32\textwidth]{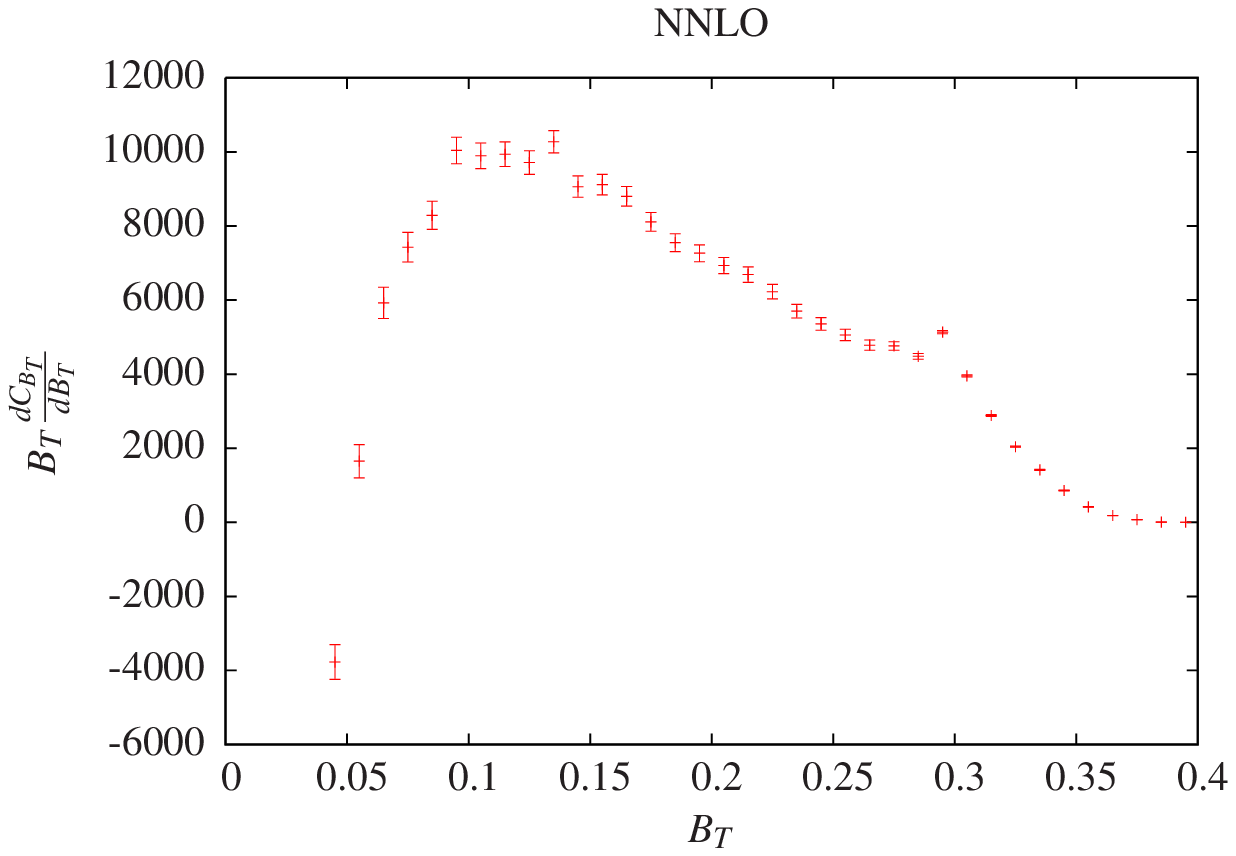}
\end{center}
\caption{
Coefficients of the leading-order ($A_{B_T}$, left), 
next-to-leading-order ($B_{B_T}$, middle)
and next-to-next-to-leading order ($C_{B_T}$, right)
contributions to the total jet broadening distribution, all weighted by $B_T$.
For the coefficient $C_{B_T}$ the Monte-Carlo integration errors are also shown.
}
\label{fig_totaljetbroadening_ABC}
\end{figure}

\begin{figure}[p]
\begin{center}
\includegraphics[bb= 125 460 490 710,width=0.32\textwidth]{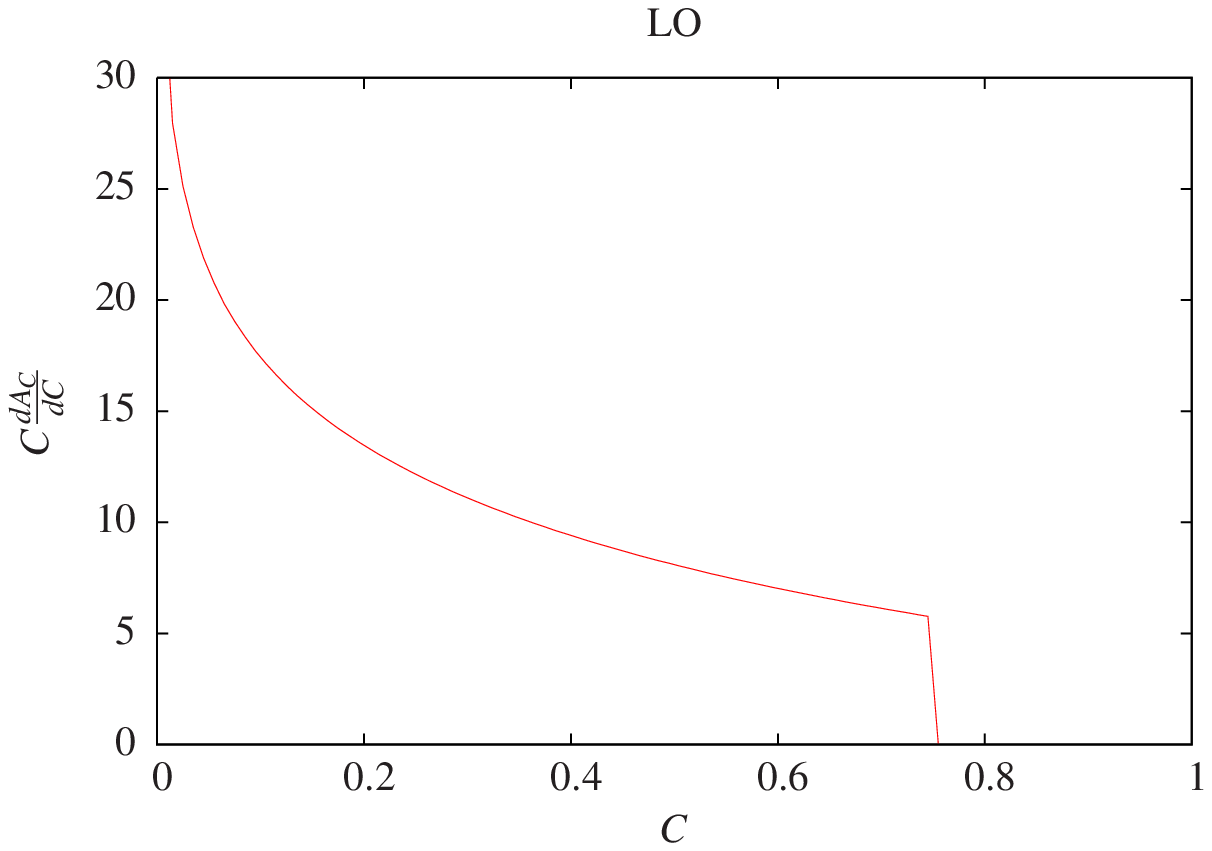}
\includegraphics[bb= 125 460 490 710,width=0.32\textwidth]{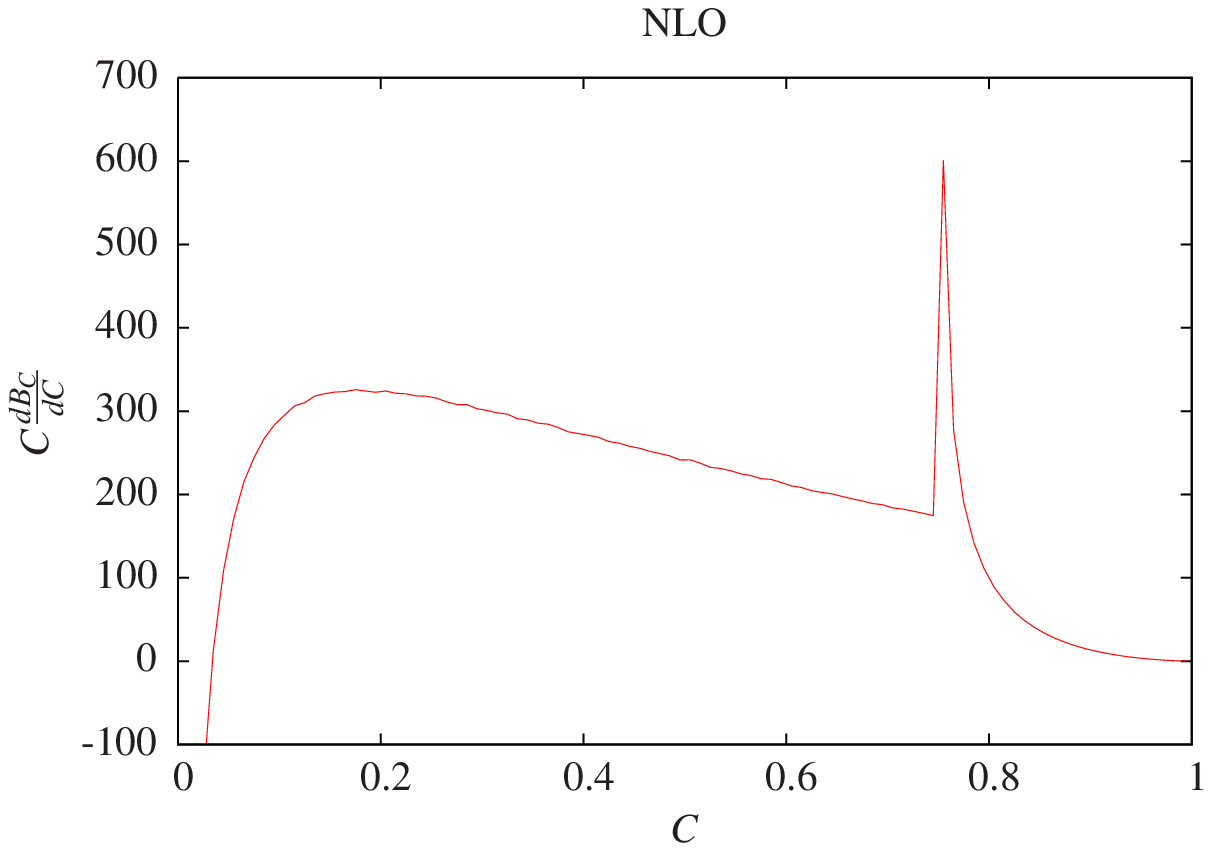}
\includegraphics[bb= 125 460 490 710,width=0.32\textwidth]{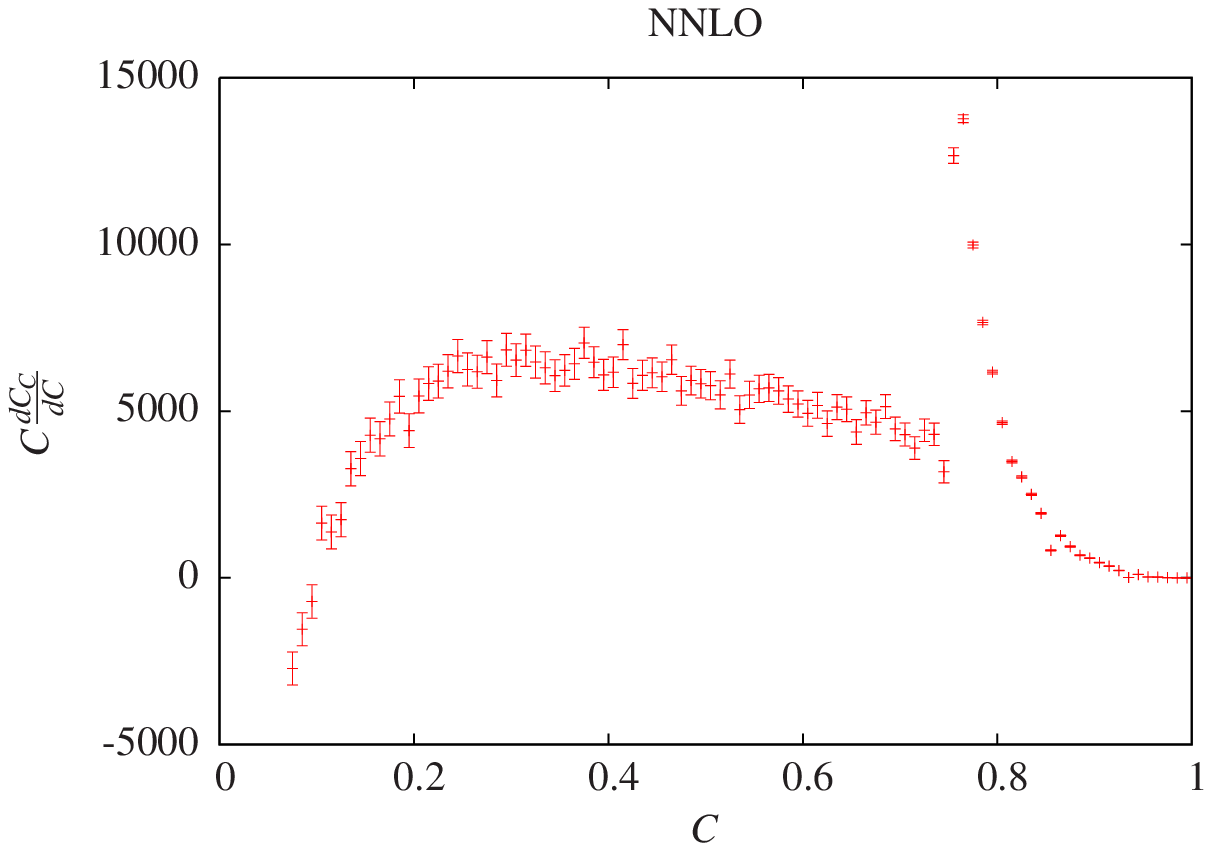}
\end{center}
\caption{
Coefficients of the leading-order ($A_{C}$, left), 
next-to-leading-order ($B_{C}$, middle)
and next-to-next-to-leading order ($C_{C}$, right)
contributions to the $C$-parameter distribution, all weighted by $C$.
For the coefficient $C_{C}$ the Monte-Carlo integration errors are also shown.
}
\label{fig_Cparameter_ABC}
\end{figure}

\begin{figure}[p]
\begin{center}
\includegraphics[bb= 125 460 490 710,width=0.32\textwidth]{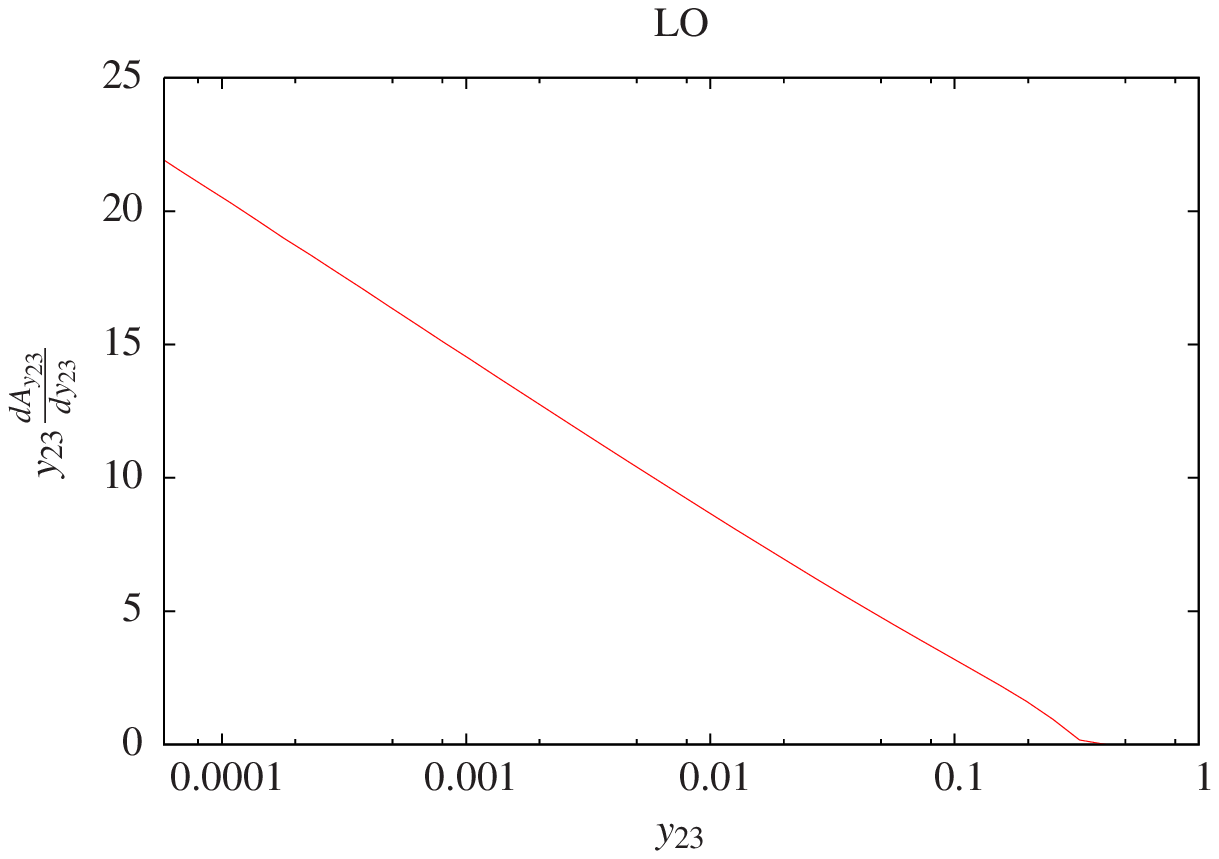}
\includegraphics[bb= 125 460 490 710,width=0.32\textwidth]{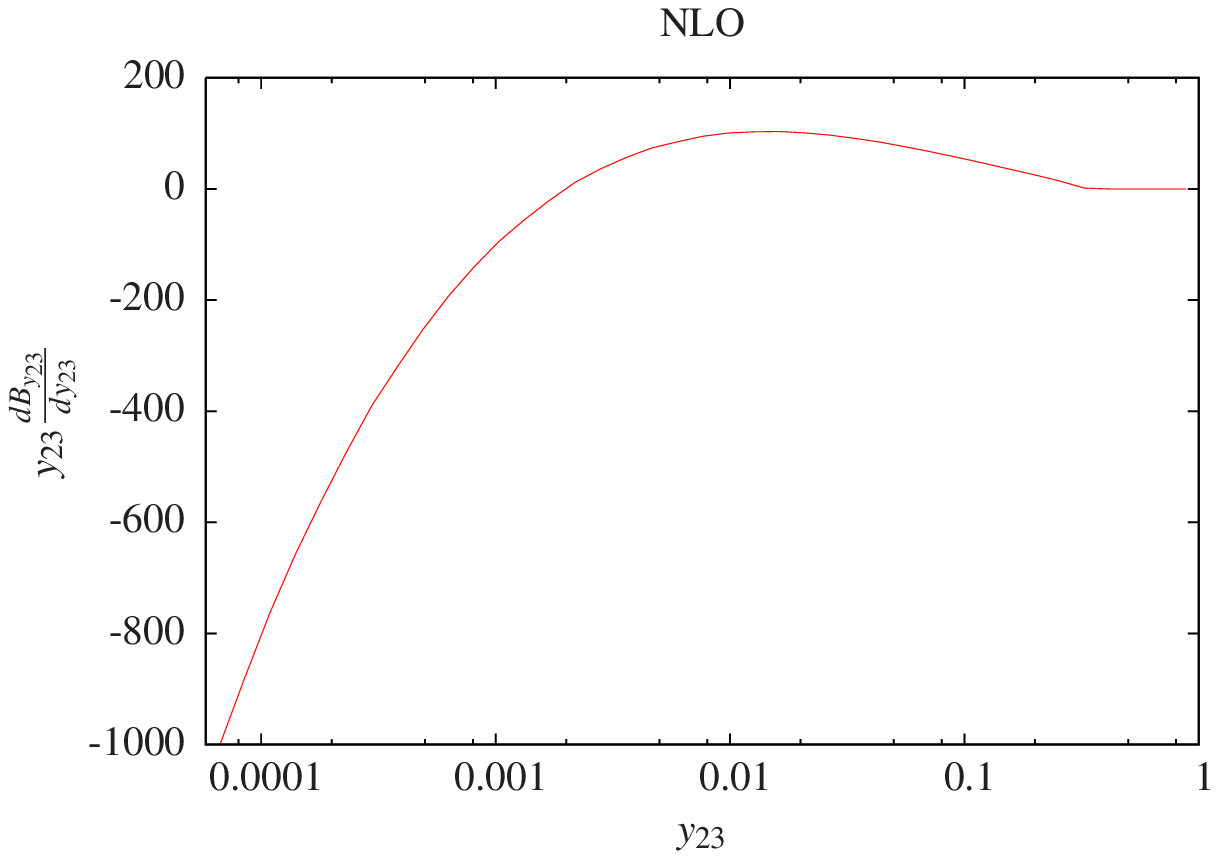}
\includegraphics[bb= 125 460 490 710,width=0.32\textwidth]{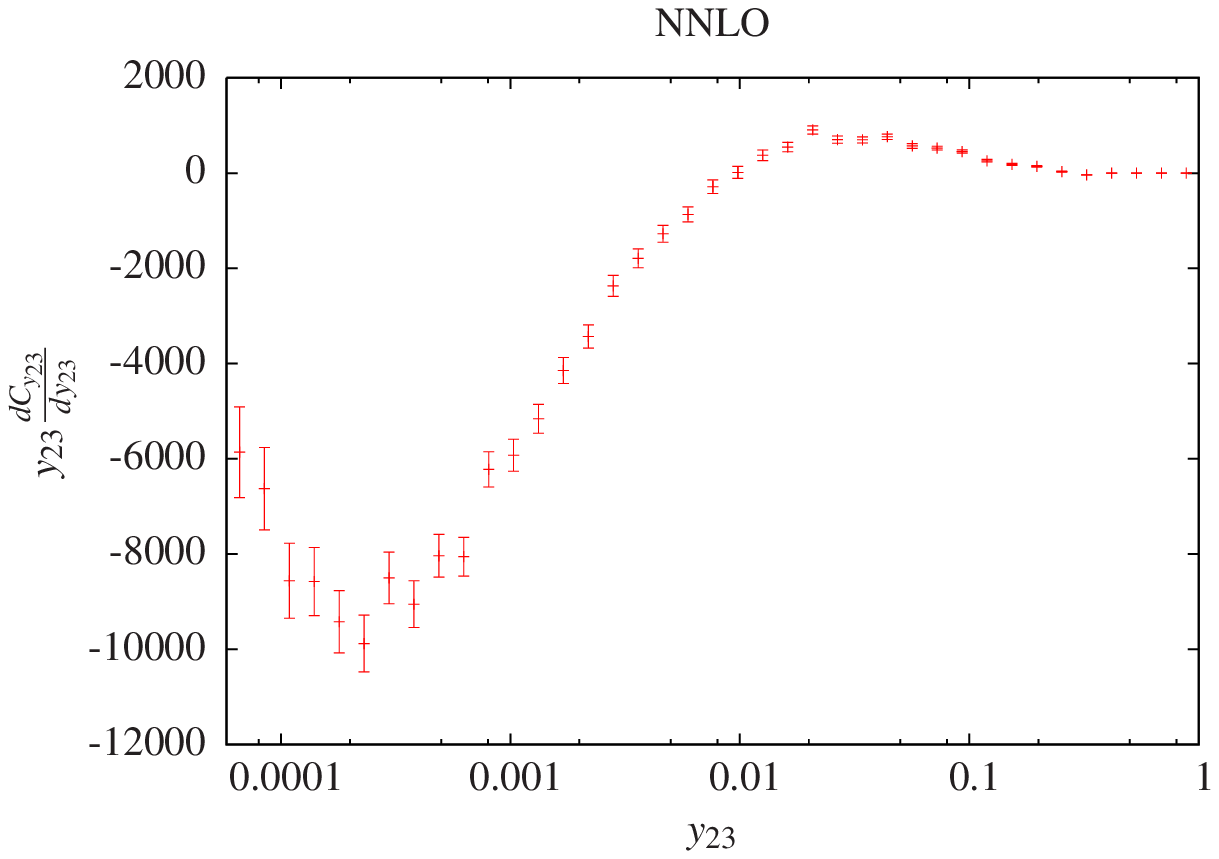}
\end{center}
\caption{
Coefficients of the leading-order ($A_{y_{23}}$, left), 
next-to-leading-order ($B_{y_{23}}$, middle)
and next-to-next-to-leading order ($C_{y_{23}}$, right)
contributions to the three-to-two jet transition distribution, all weighted by $y_{23}$.
For the coefficient $C_{y_{23}}$ the Monte-Carlo integration errors are also shown.
}
\label{fig_y23_ABC}
\end{figure}

For the six event shape variables the LO functions $A_{\cal O}$, 
the NLO functions $B_{\cal O}$ and the NNLO functions $C_{\cal O}$ are also shown in 
in figs.~\ref{fig_thrust_ABC} to \ref{fig_y23_ABC}.
The leading-order function $A_{\cal O}$ and the next-to-leading order function $B_{\cal O}$
can be computed with high precision and the graphs are shown simply with solid lines 
in figs.~\ref{fig_thrust_ABC}-\ref{fig_y23_ABC}.
For the next-to-next-to-leading order function $C_{\cal O}$ the Monte-Carlo integration errors
are typically at the per cent level, and this function is shown with errorbars corresponding to the Monte-Carlo
integration errors in figs.~\ref{fig_thrust_ABC}-\ref{fig_y23_ABC}.

\begin{figure}[p]
\begin{center}
\includegraphics[bb= 125 460 490 710,width=0.32\textwidth]{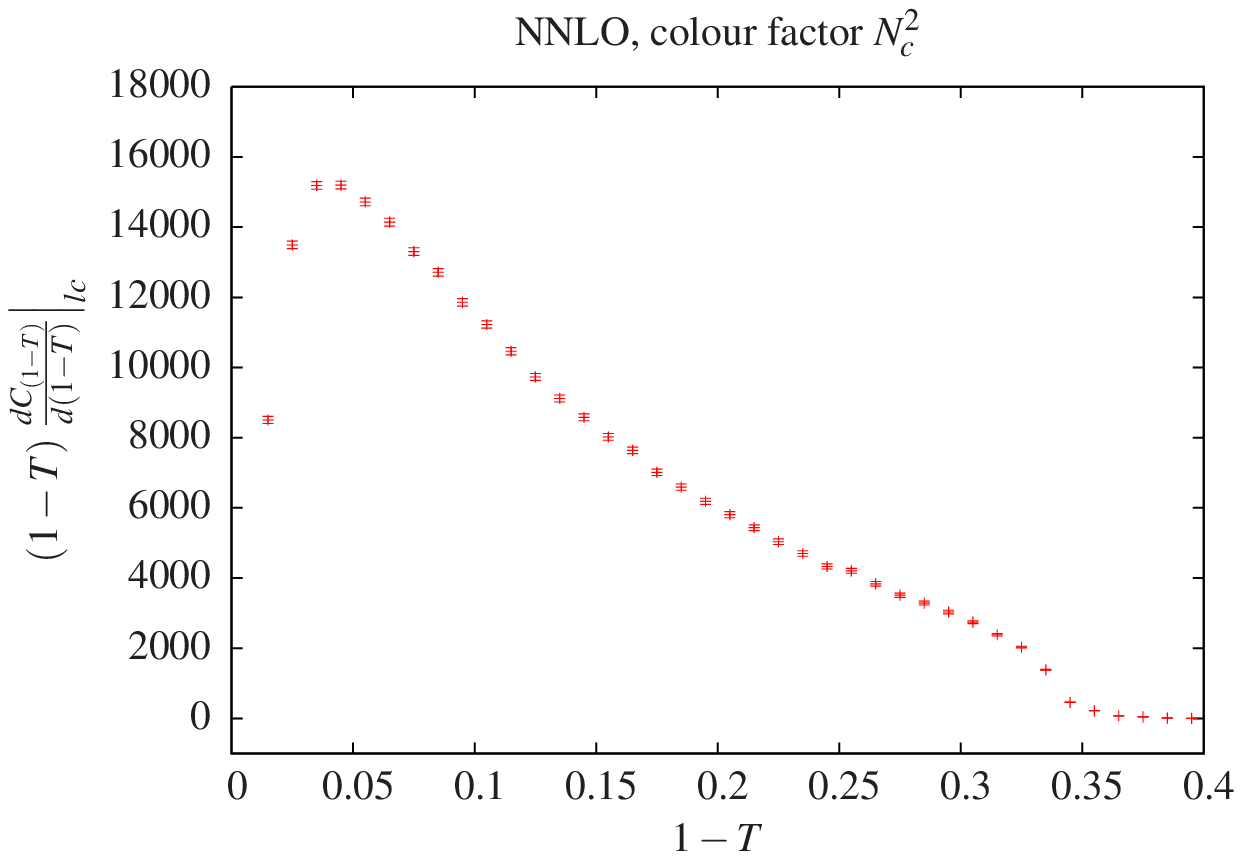}
\includegraphics[bb= 125 460 490 710,width=0.32\textwidth]{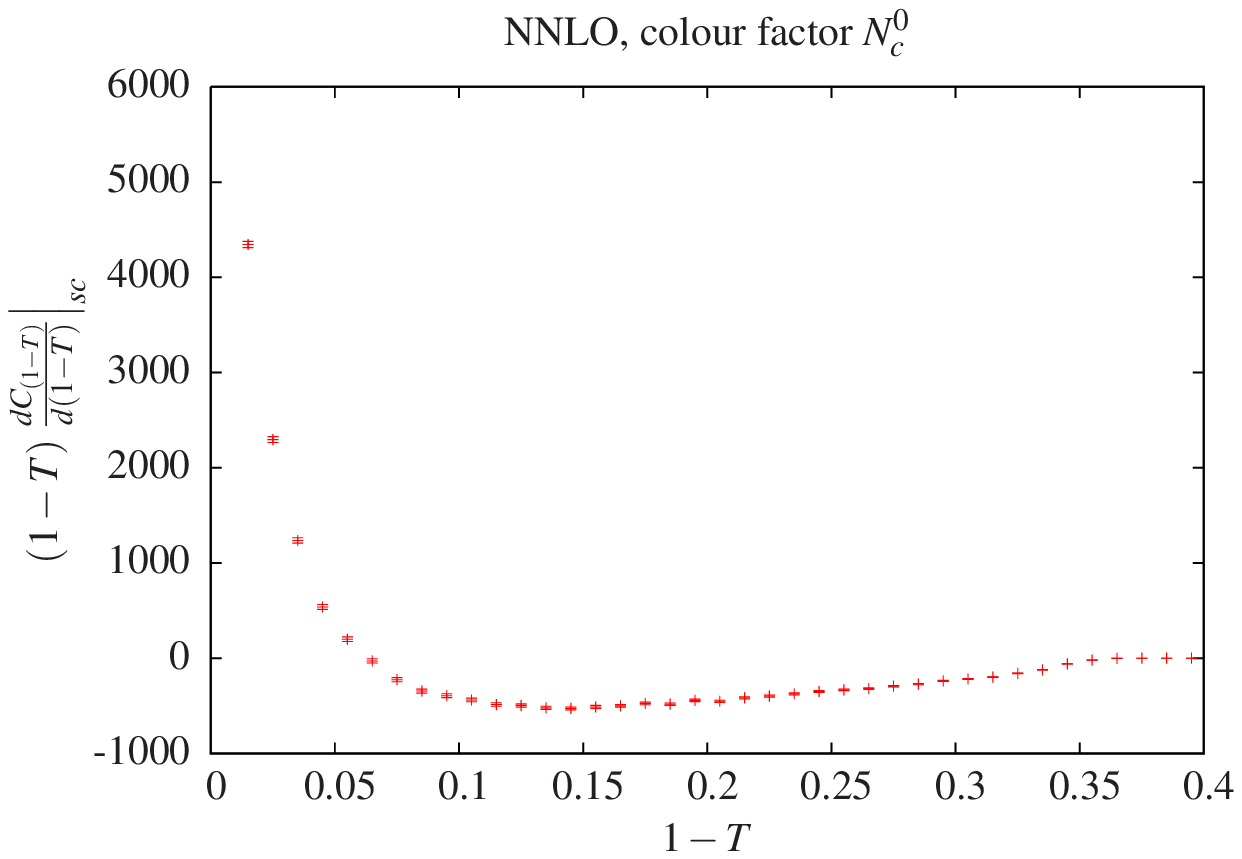}
\includegraphics[bb= 125 460 490 710,width=0.32\textwidth]{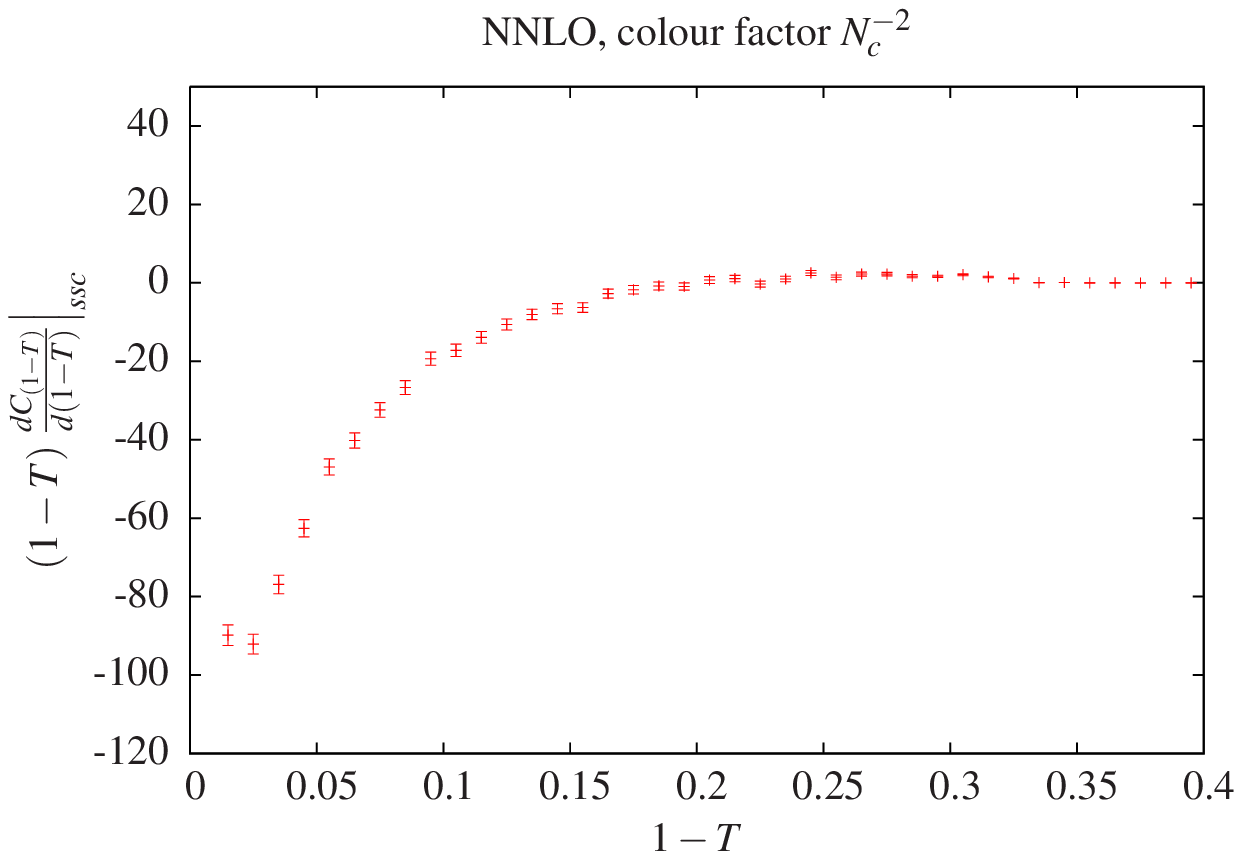}
\\
\includegraphics[bb= 125 460 490 710,width=0.32\textwidth]{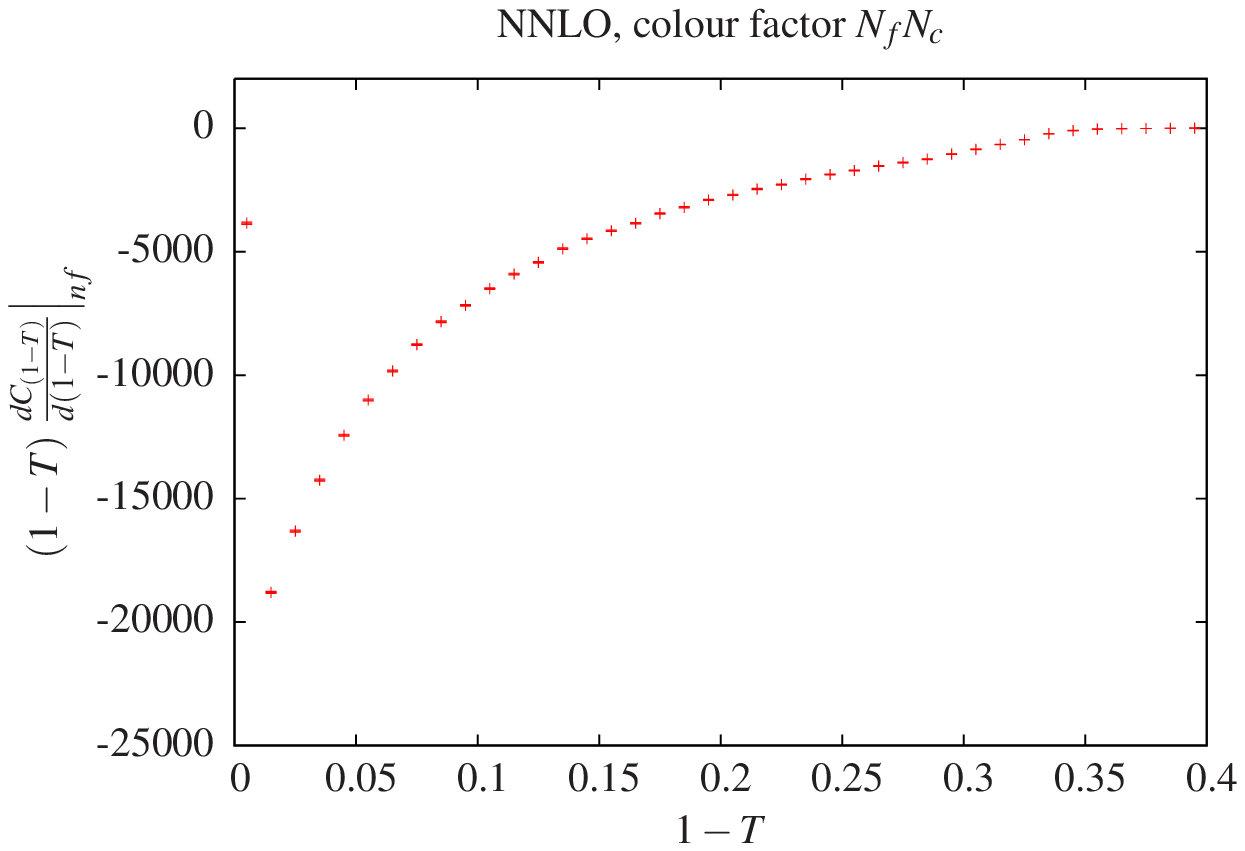}
\includegraphics[bb= 125 460 490 710,width=0.32\textwidth]{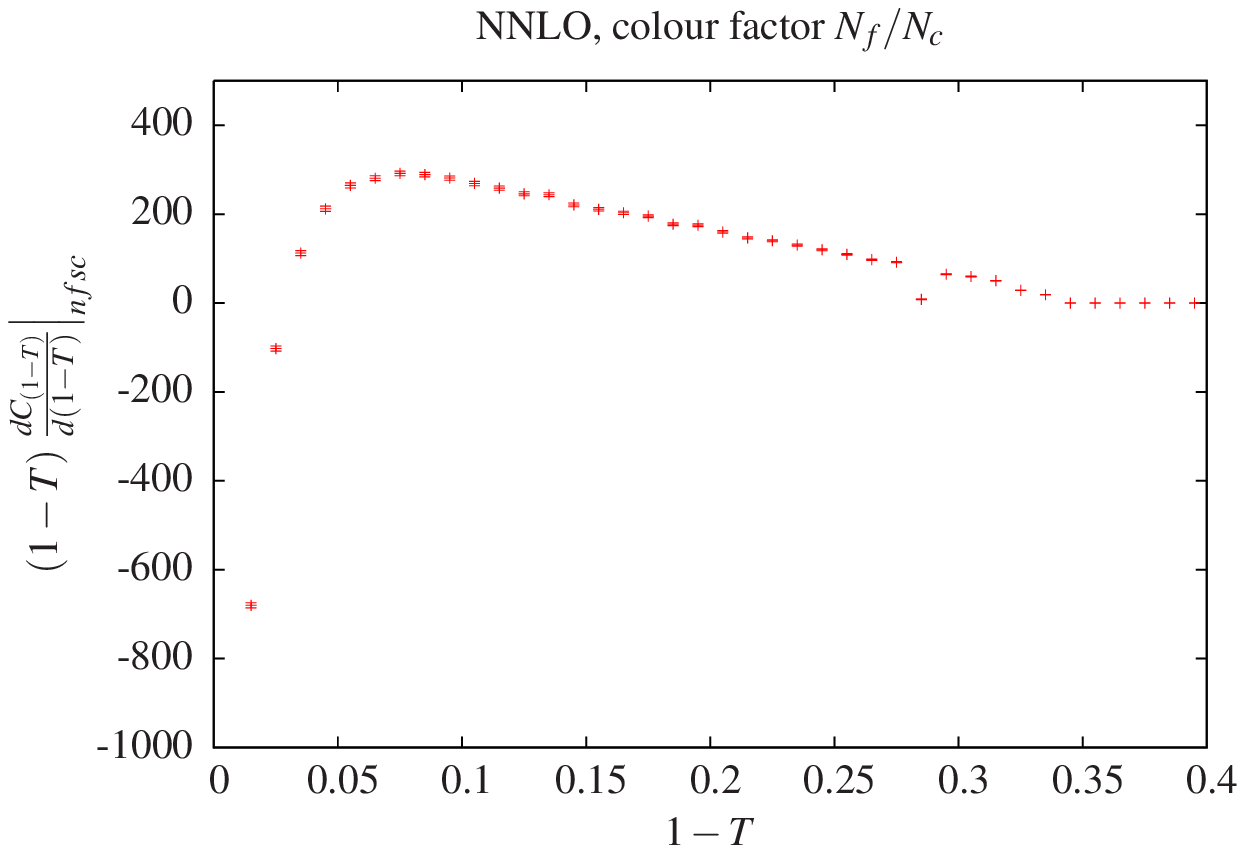}
\includegraphics[bb= 125 460 490 710,width=0.32\textwidth]{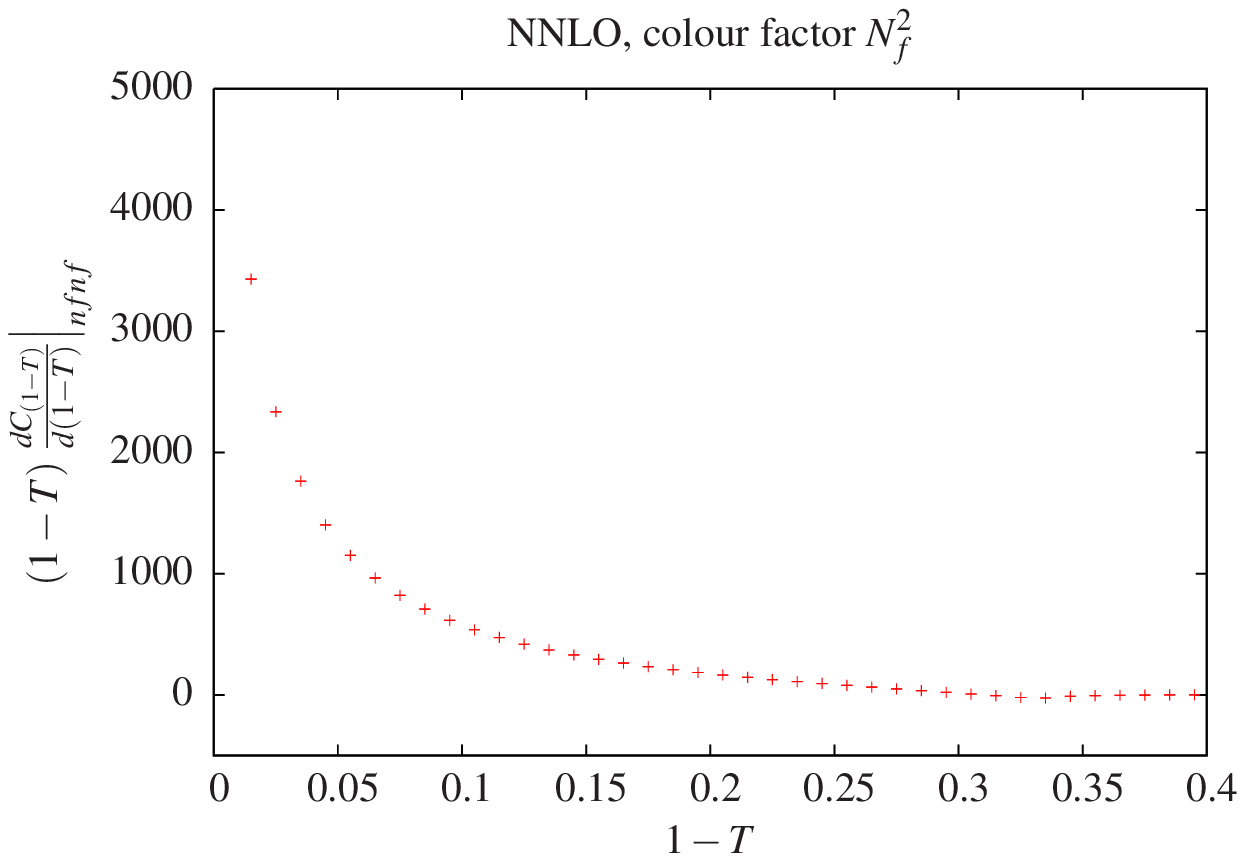}
\end{center}
\caption{
The NNLO coefficient $C_{(1-T)}$ for the thrust distribution split up into
individual colour factors.
}
\label{fig_thrust_C_col}
\end{figure}

\begin{figure}[p]
\begin{center}
\includegraphics[bb= 125 460 490 710,width=0.32\textwidth]{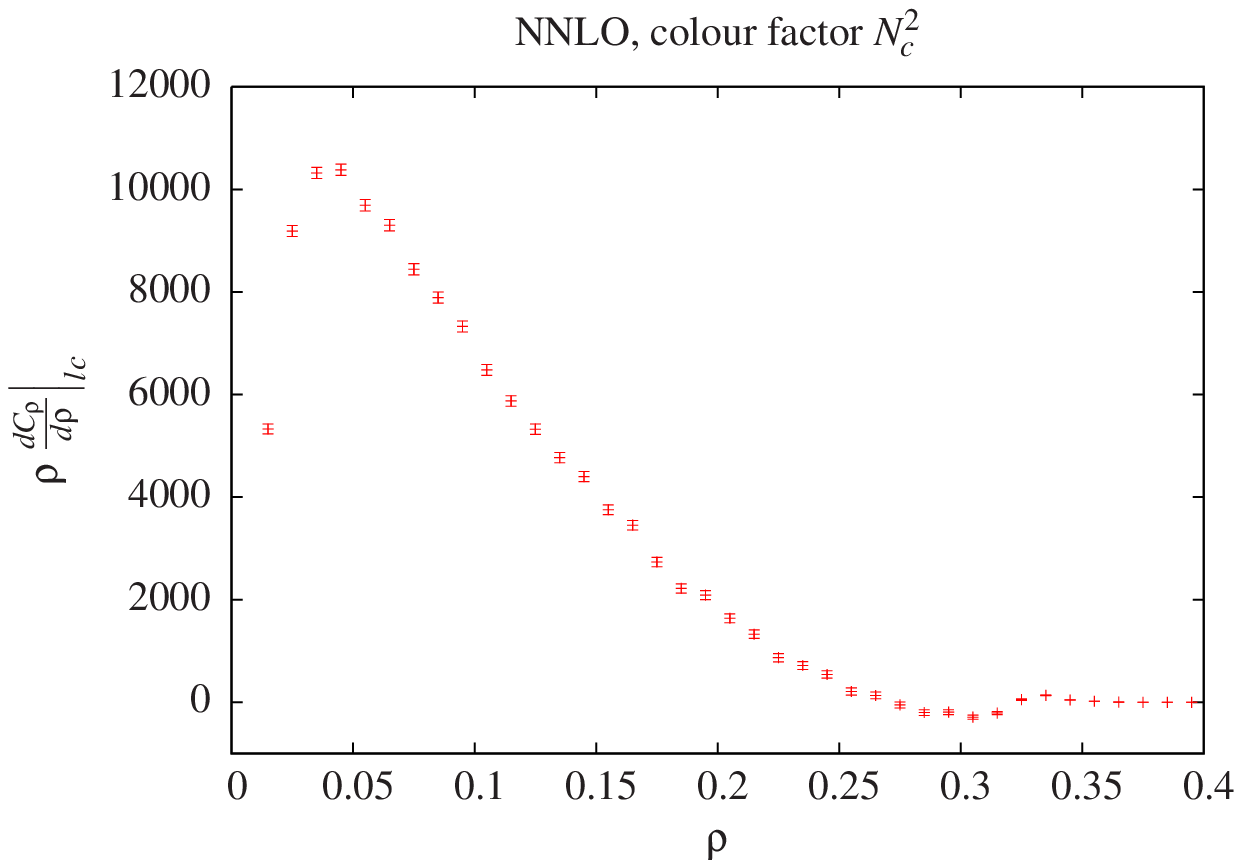}
\includegraphics[bb= 125 460 490 710,width=0.32\textwidth]{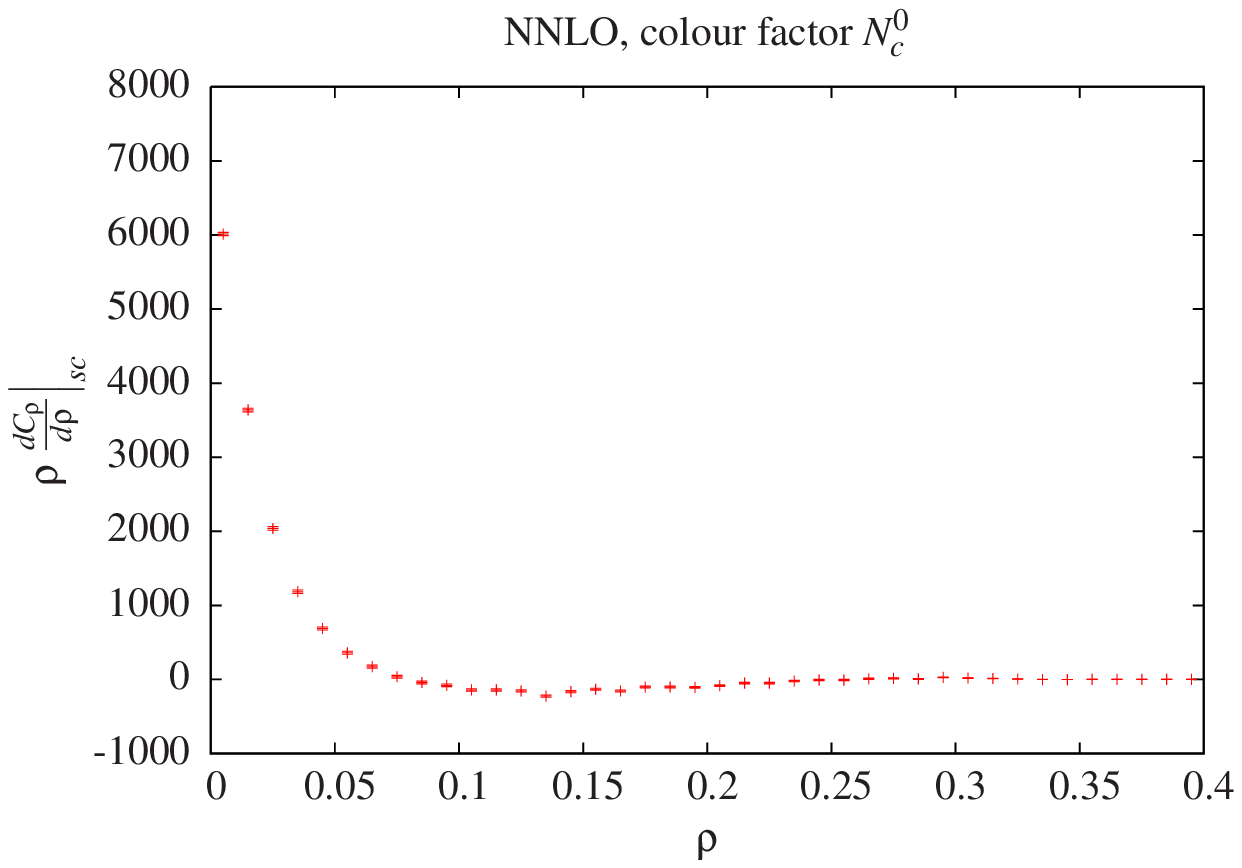}
\includegraphics[bb= 125 460 490 710,width=0.32\textwidth]{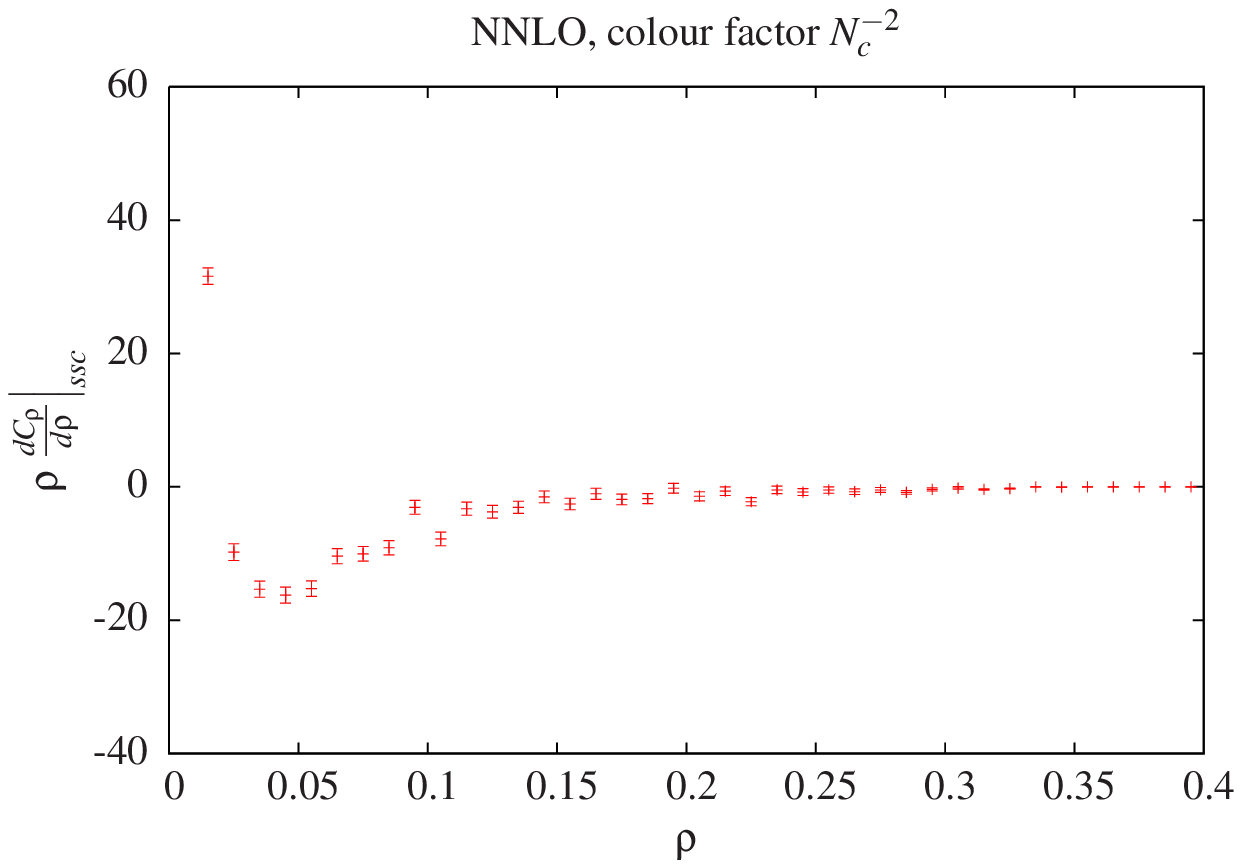}
\\
\includegraphics[bb= 125 460 490 710,width=0.32\textwidth]{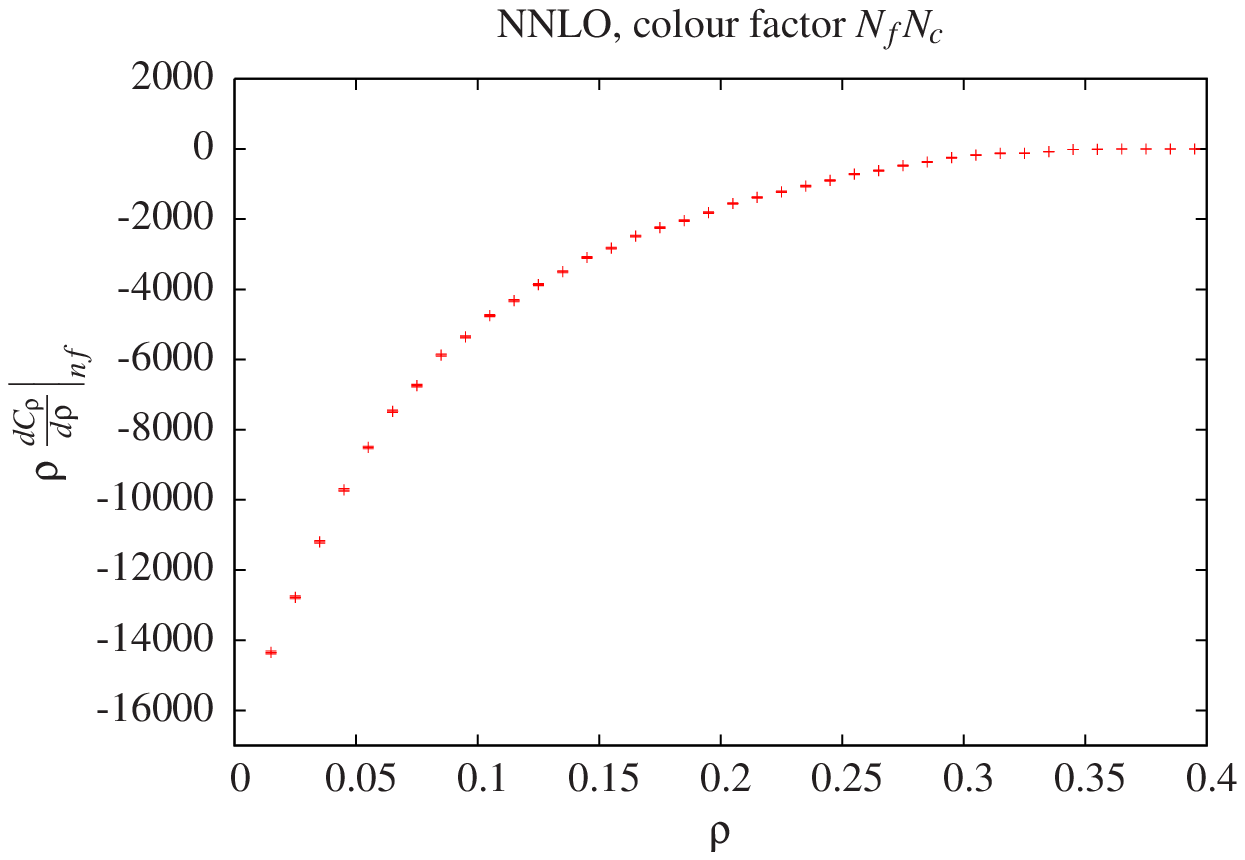}
\includegraphics[bb= 125 460 490 710,width=0.32\textwidth]{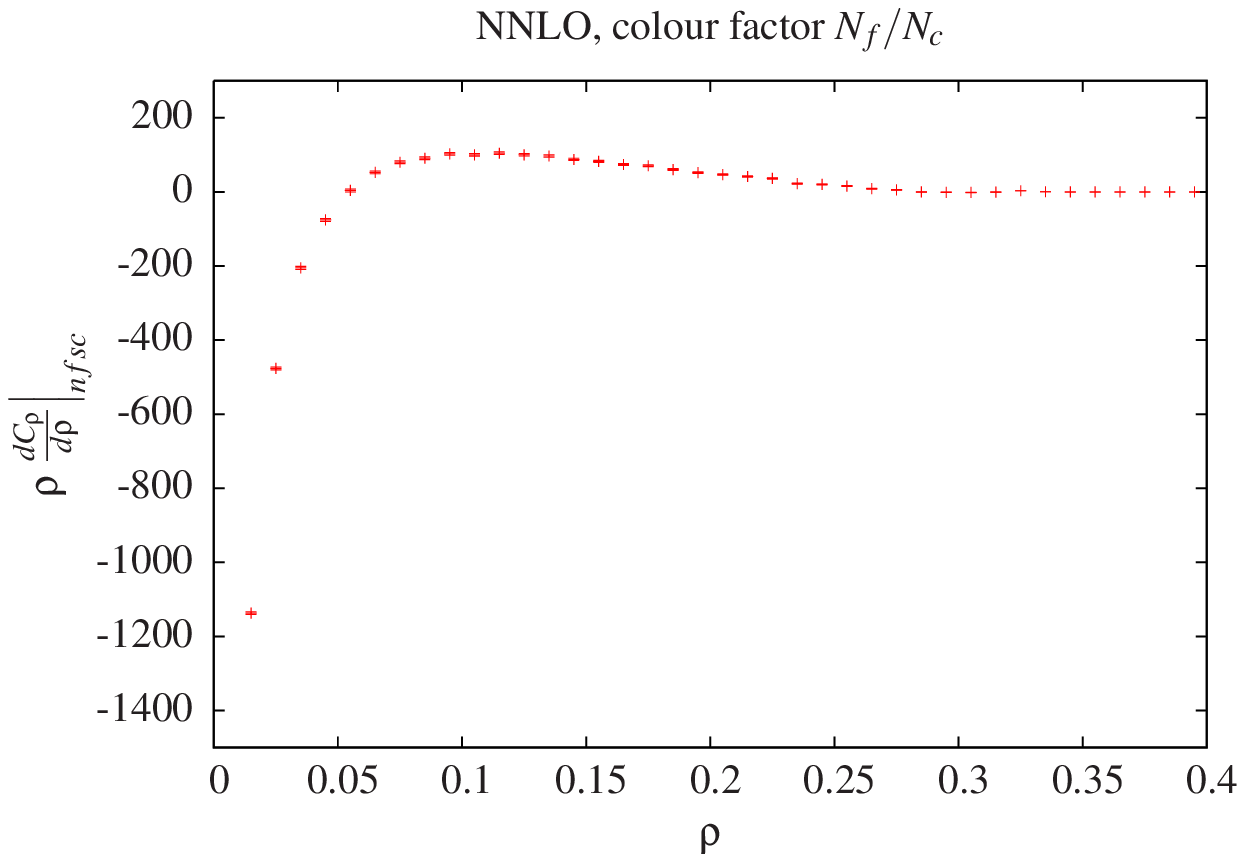}
\includegraphics[bb= 125 460 490 710,width=0.32\textwidth]{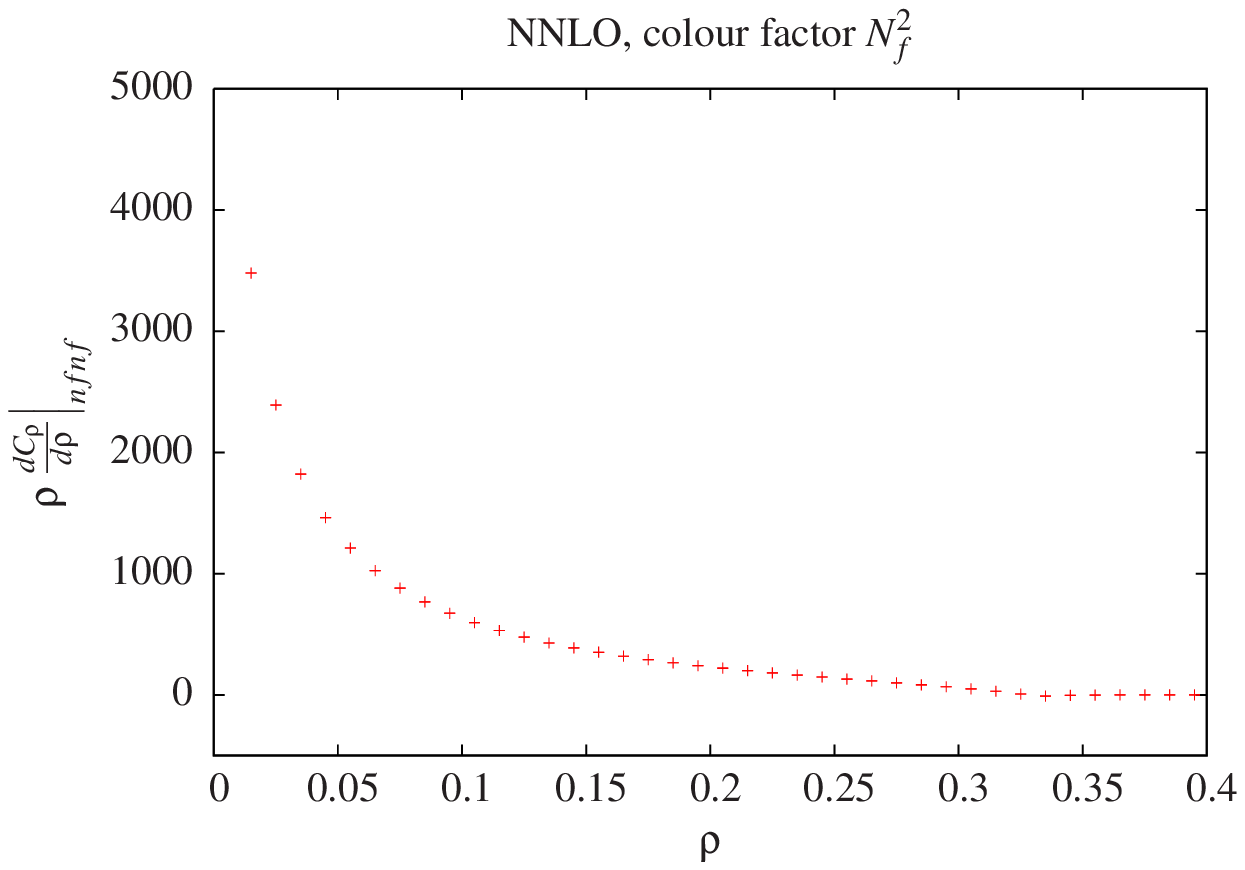}
\end{center}
\caption{
The NNLO coefficient $C_{\rho}$ for the heavy jet mass distribution split up into
individual colour factors.
}
\label{fig_heavyjetmass_C_col}
\end{figure}

\begin{figure}[p]
\begin{center}
\includegraphics[bb= 125 460 490 710,width=0.32\textwidth]{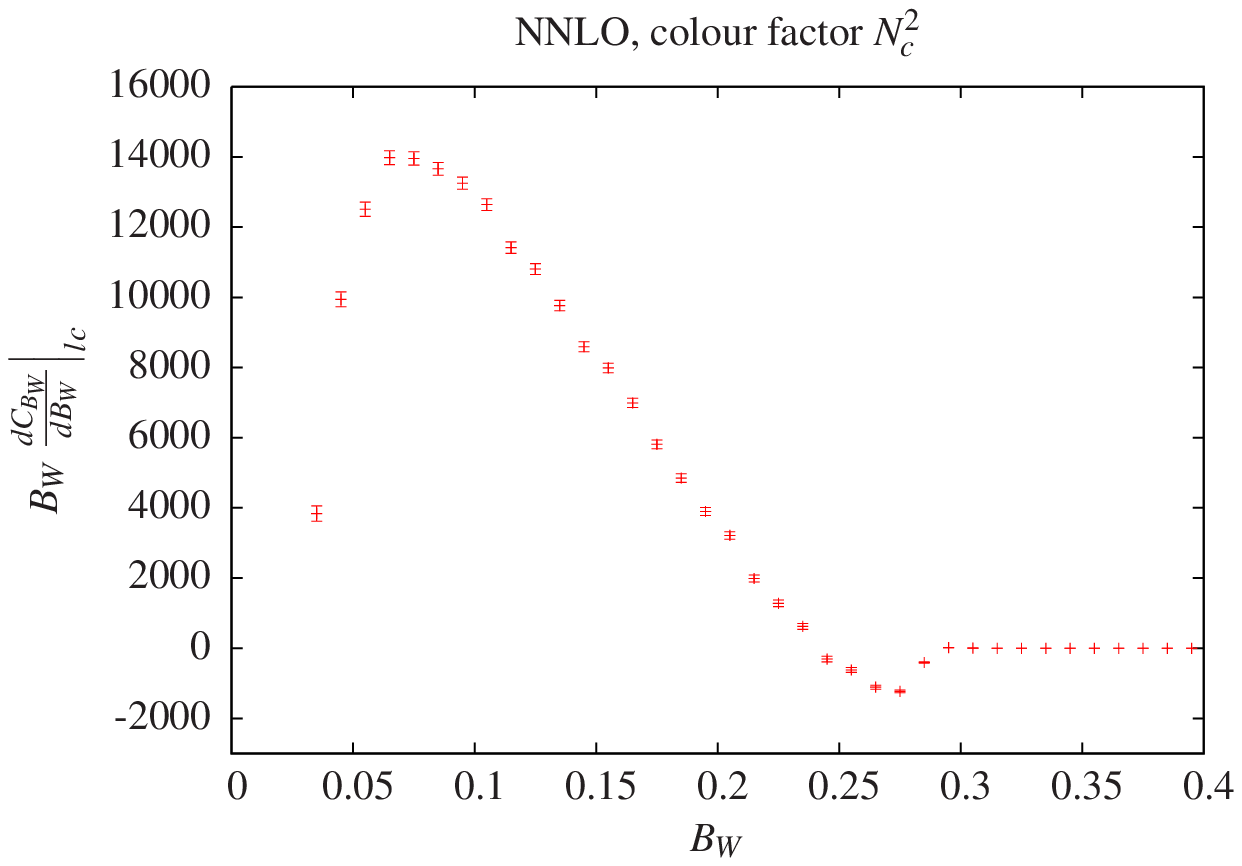}
\includegraphics[bb= 125 460 490 710,width=0.32\textwidth]{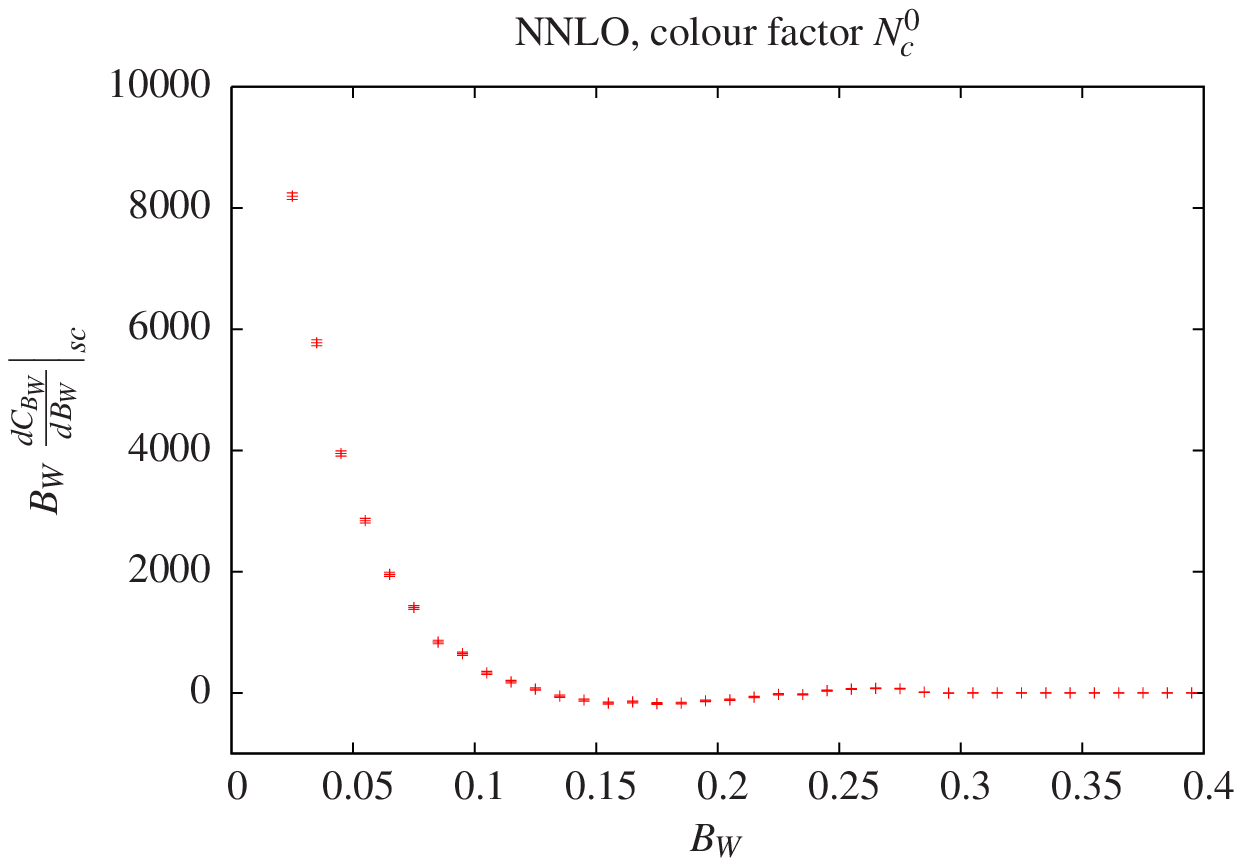}
\includegraphics[bb= 125 460 490 710,width=0.32\textwidth]{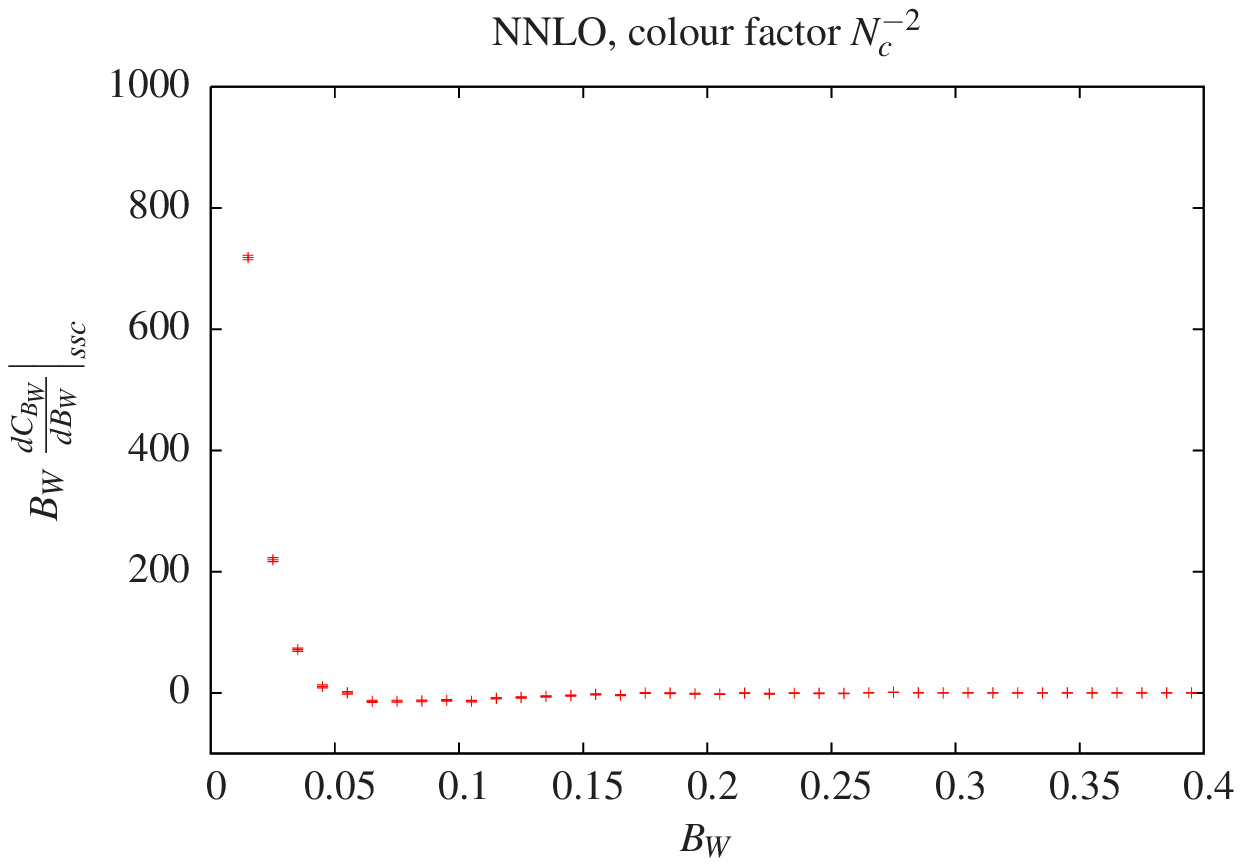}
\\
\includegraphics[bb= 125 460 490 710,width=0.32\textwidth]{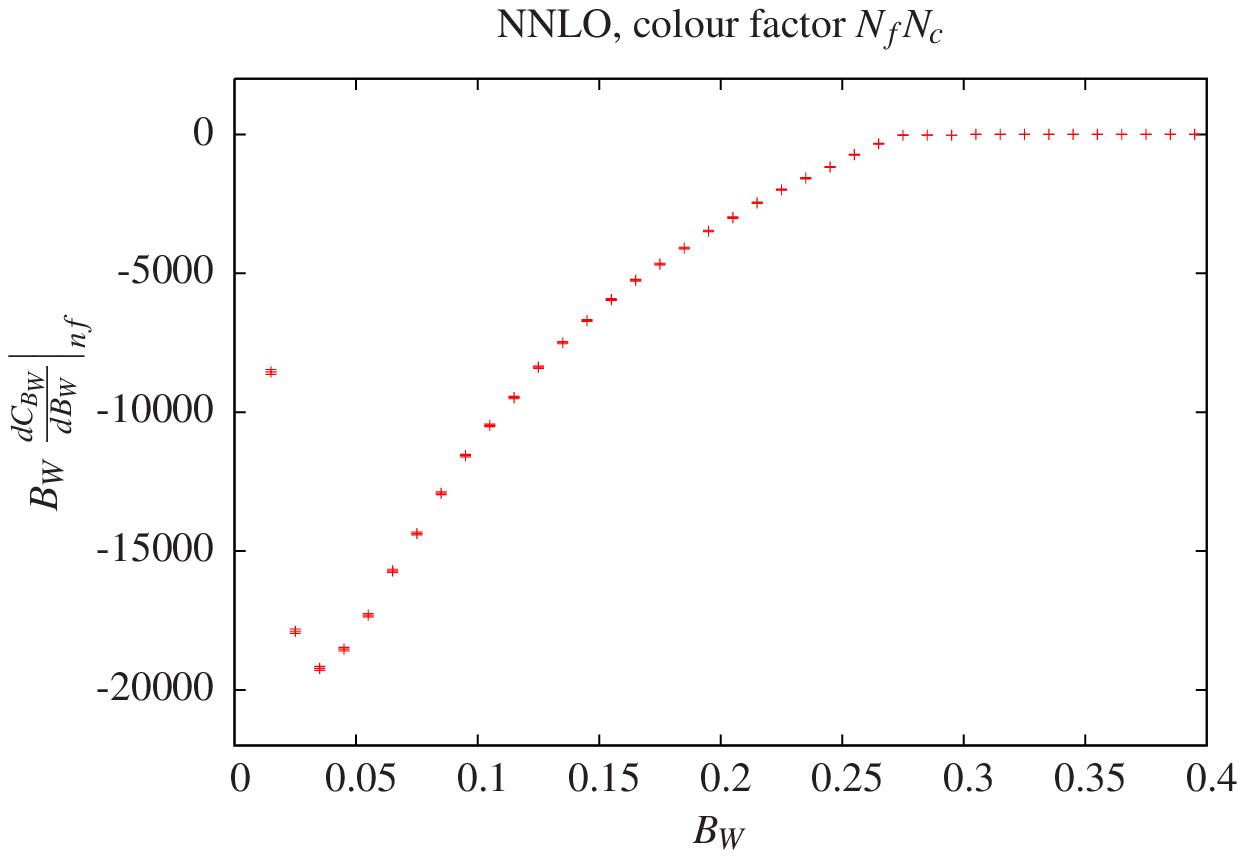}
\includegraphics[bb= 125 460 490 710,width=0.32\textwidth]{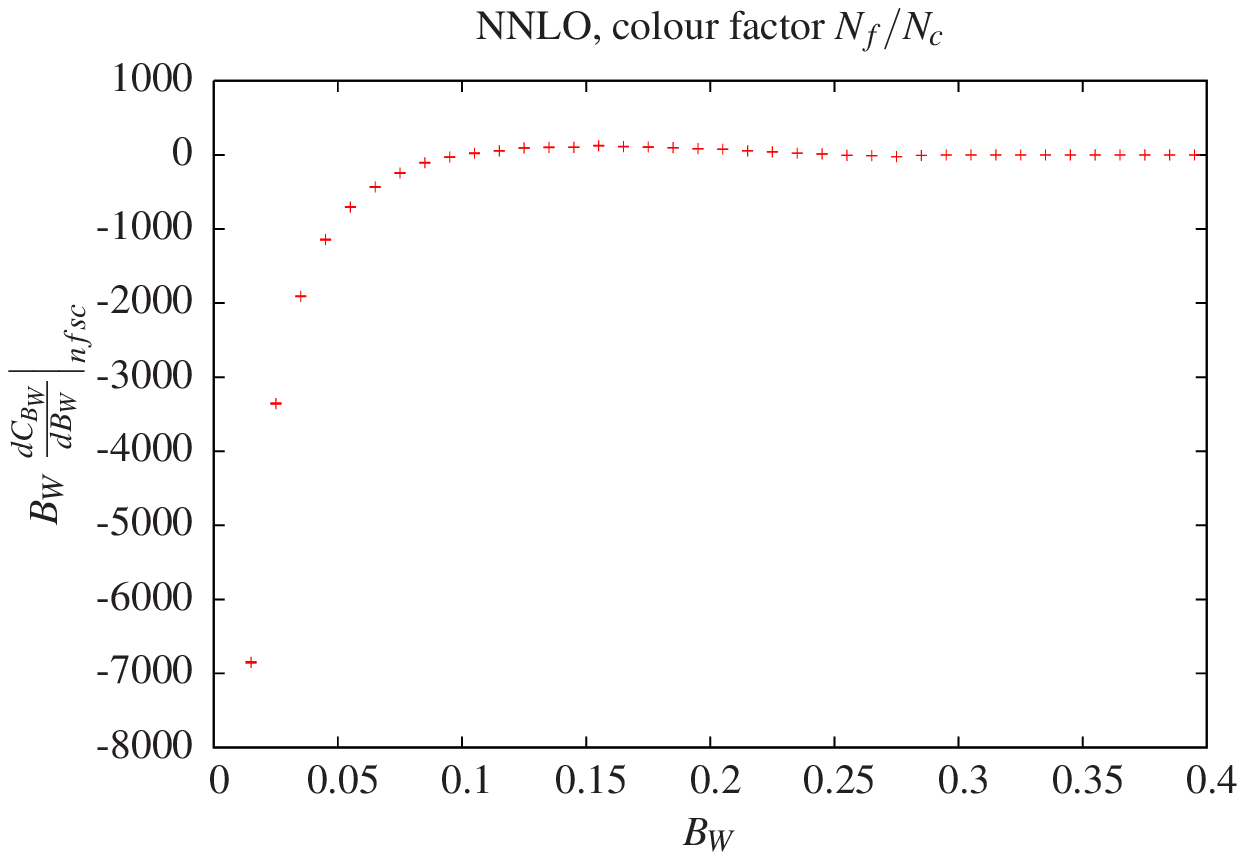}
\includegraphics[bb= 125 460 490 710,width=0.32\textwidth]{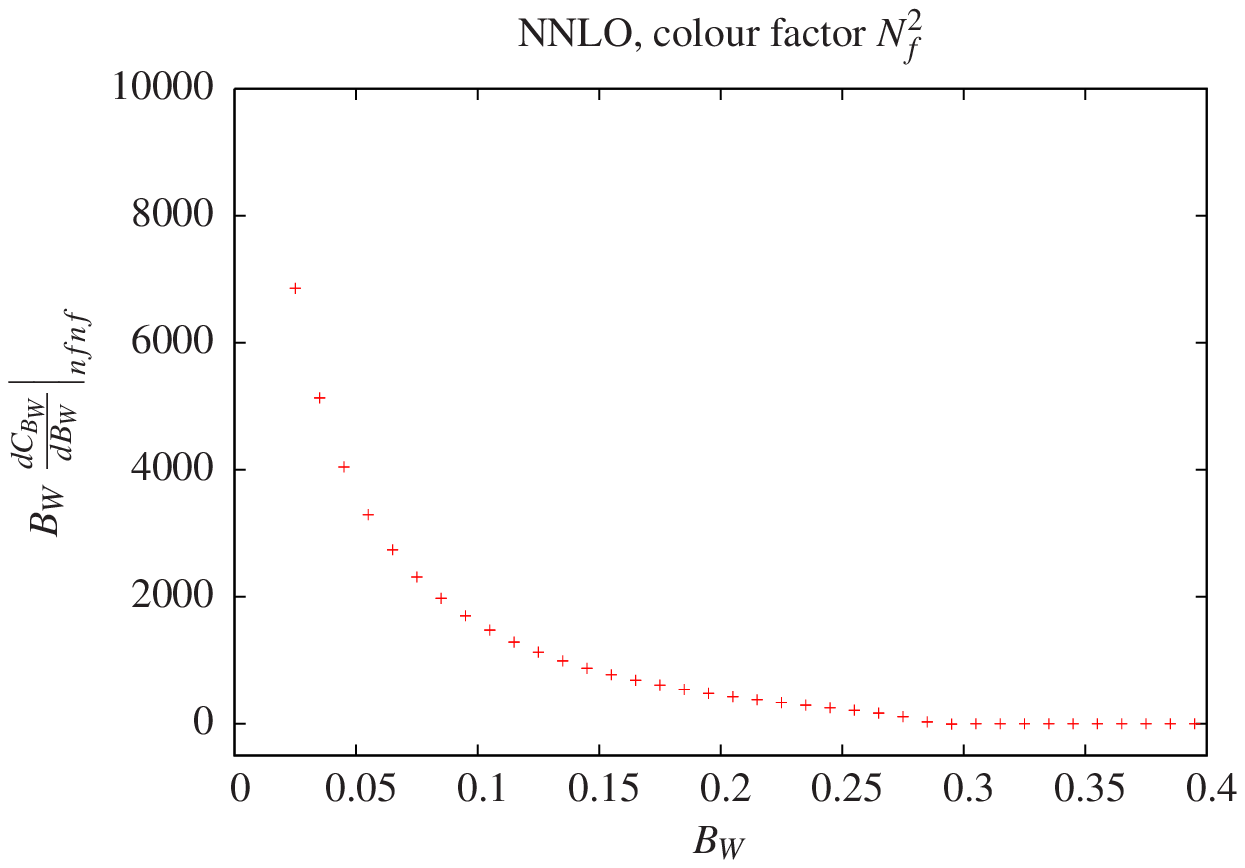}
\end{center}
\caption{
The NNLO coefficient $C_{B_W}$ for the wide jet broadening distribution split up into
individual colour factors.
}
\label{fig_widejetbroadening_C_col}
\end{figure}

\begin{figure}[p]
\begin{center}
\includegraphics[bb= 125 460 490 710,width=0.32\textwidth]{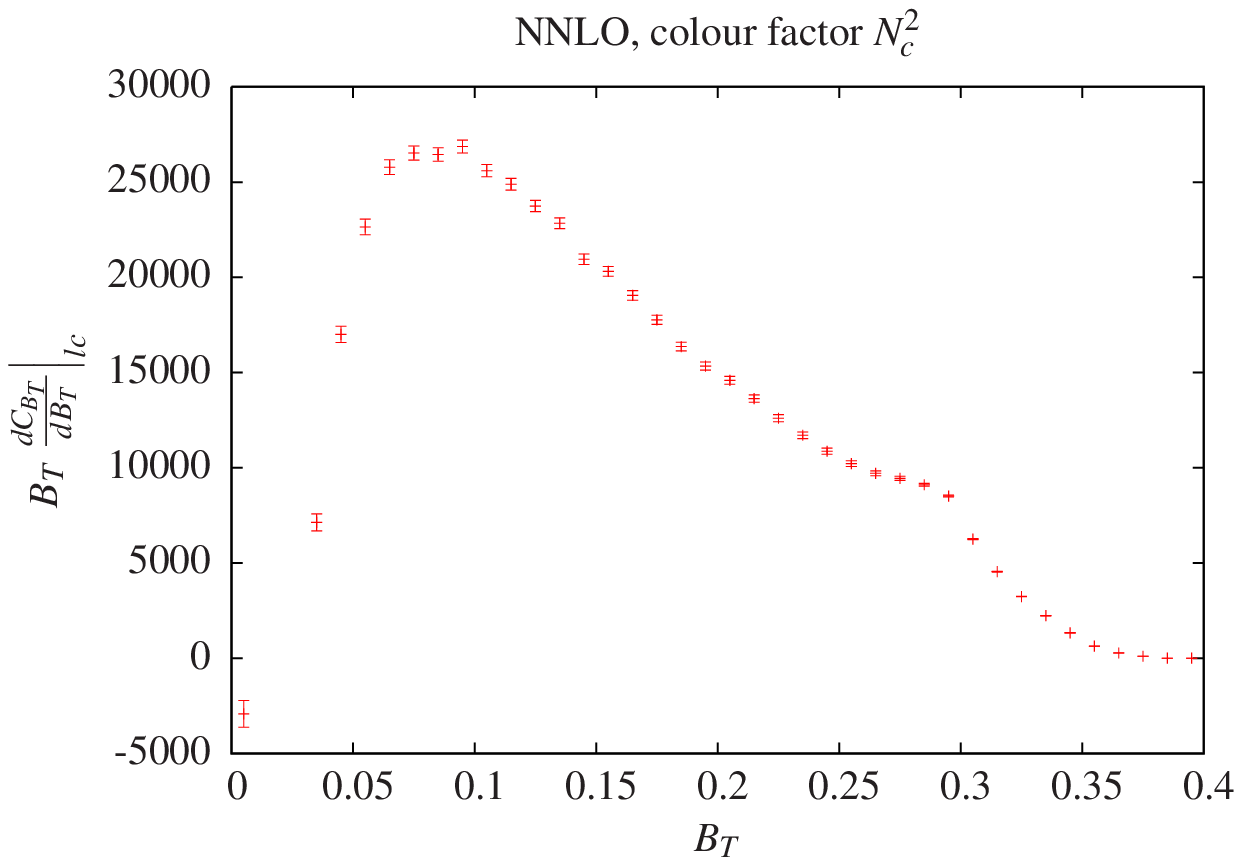}
\includegraphics[bb= 125 460 490 710,width=0.32\textwidth]{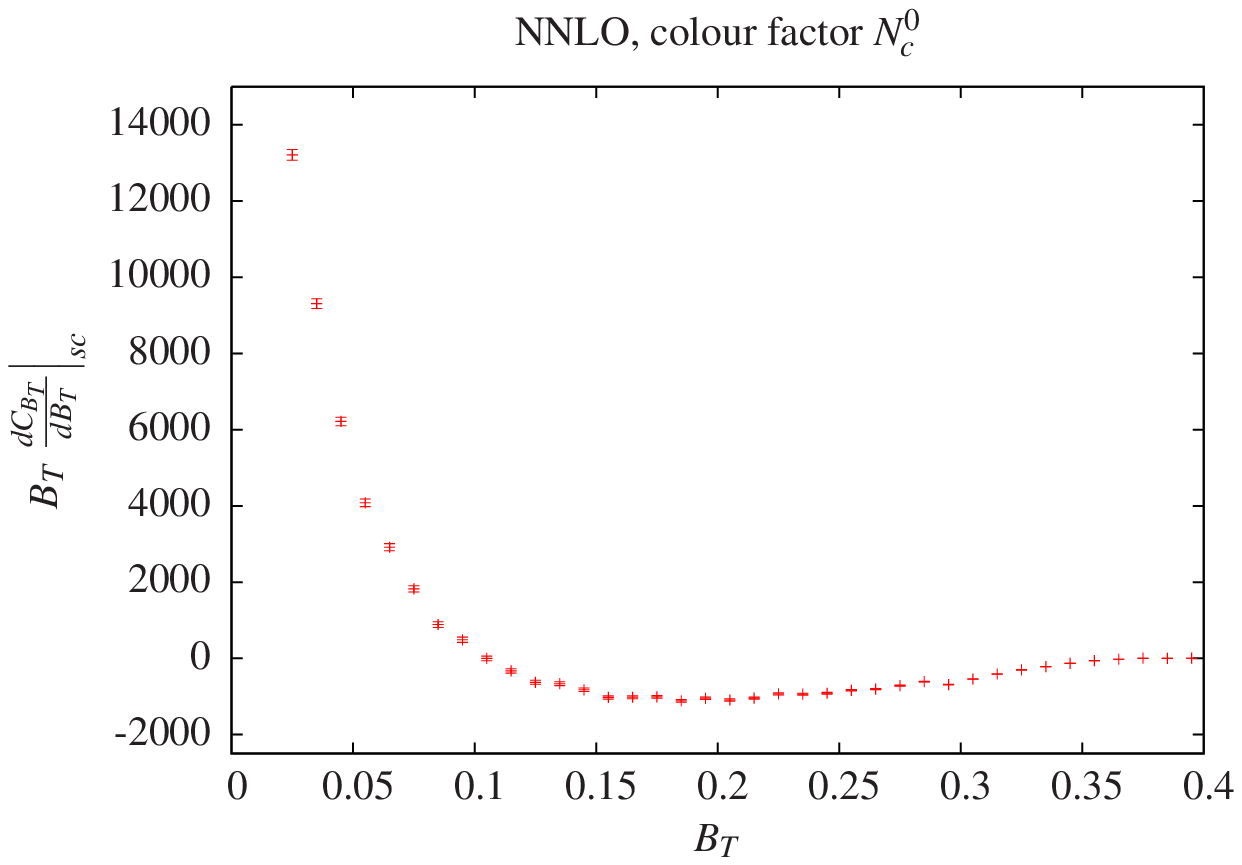}
\includegraphics[bb= 125 460 490 710,width=0.32\textwidth]{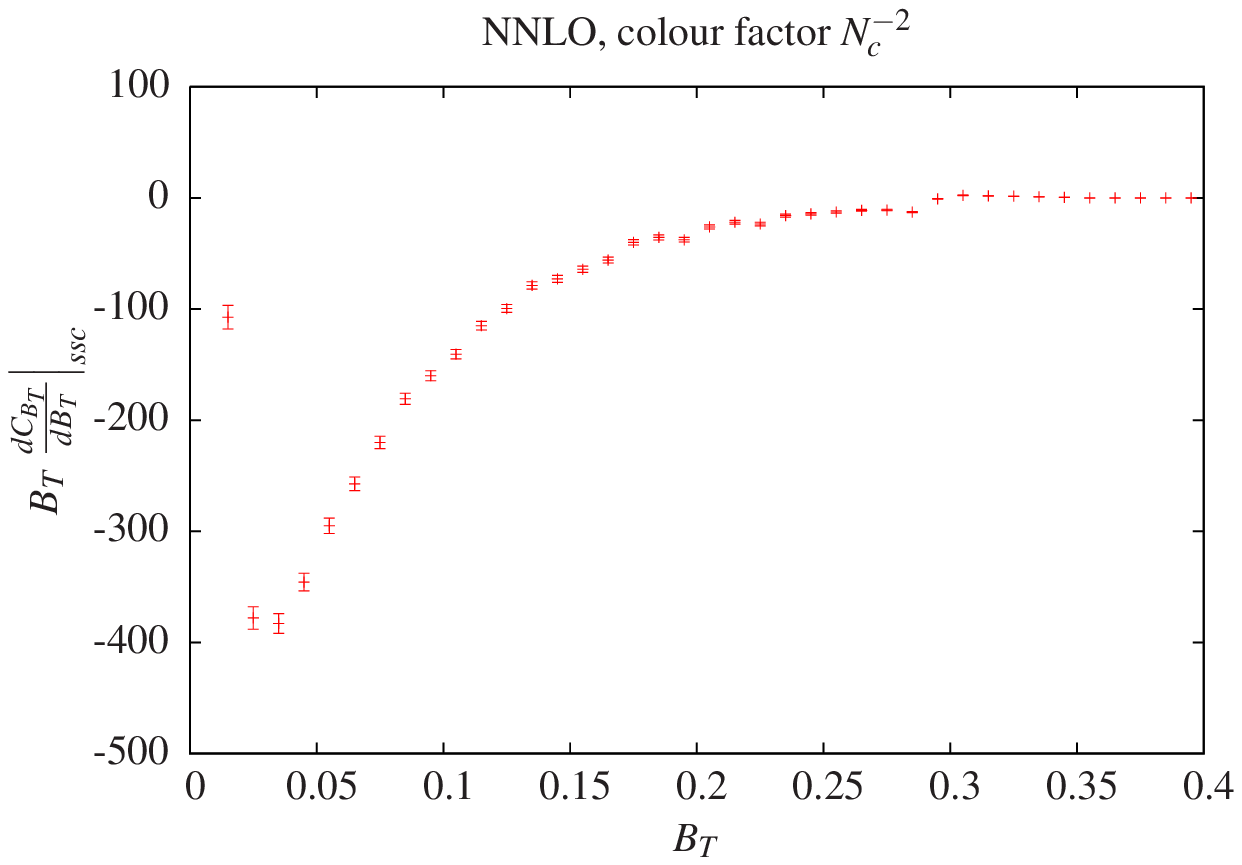}
\\
\includegraphics[bb= 125 460 490 710,width=0.32\textwidth]{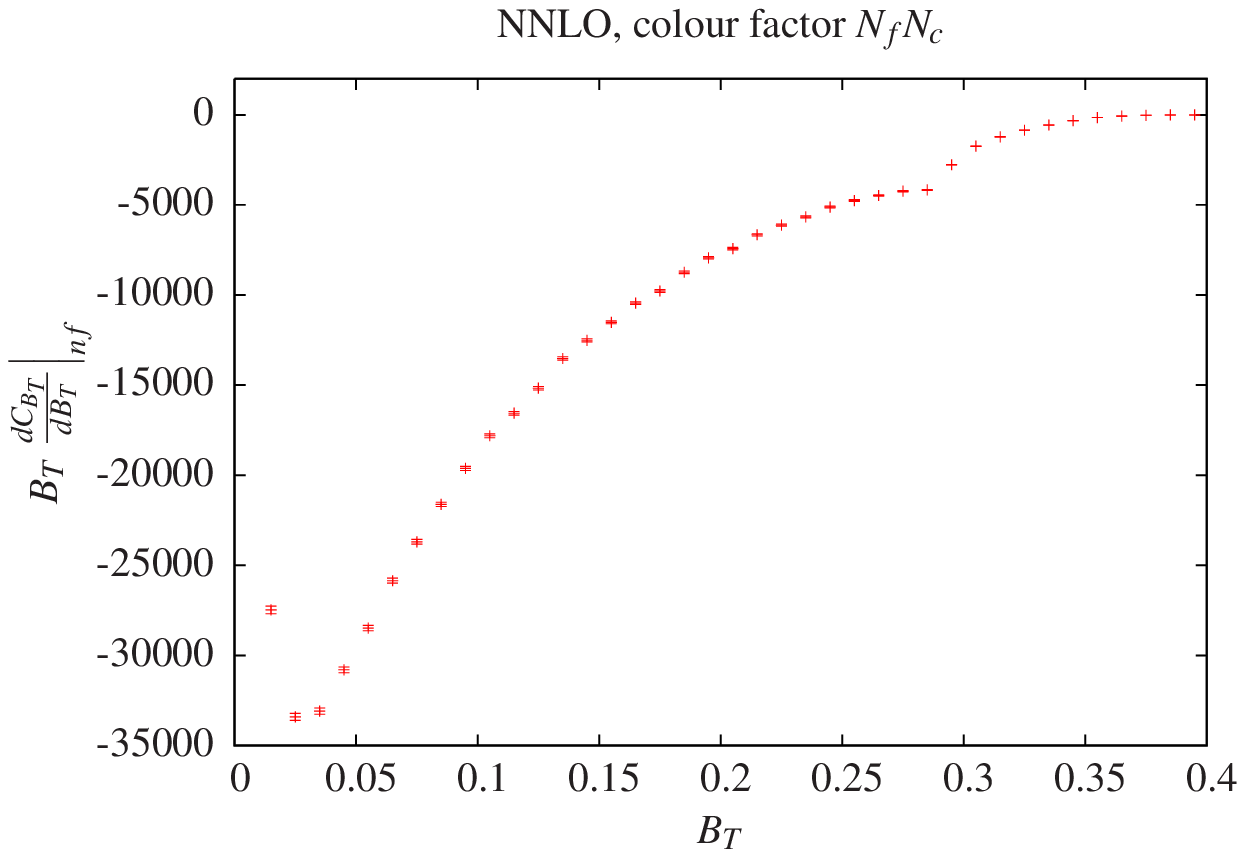}
\includegraphics[bb= 125 460 490 710,width=0.32\textwidth]{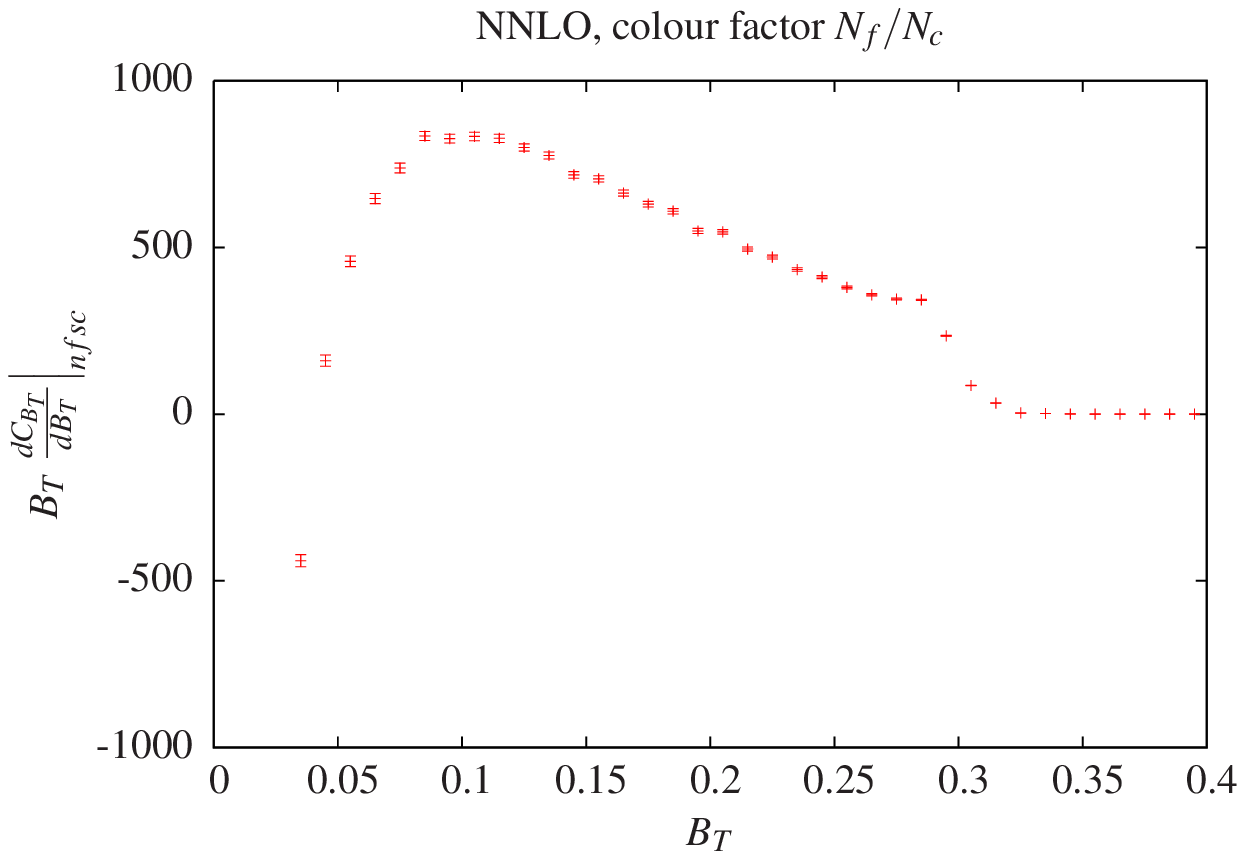}
\includegraphics[bb= 125 460 490 710,width=0.32\textwidth]{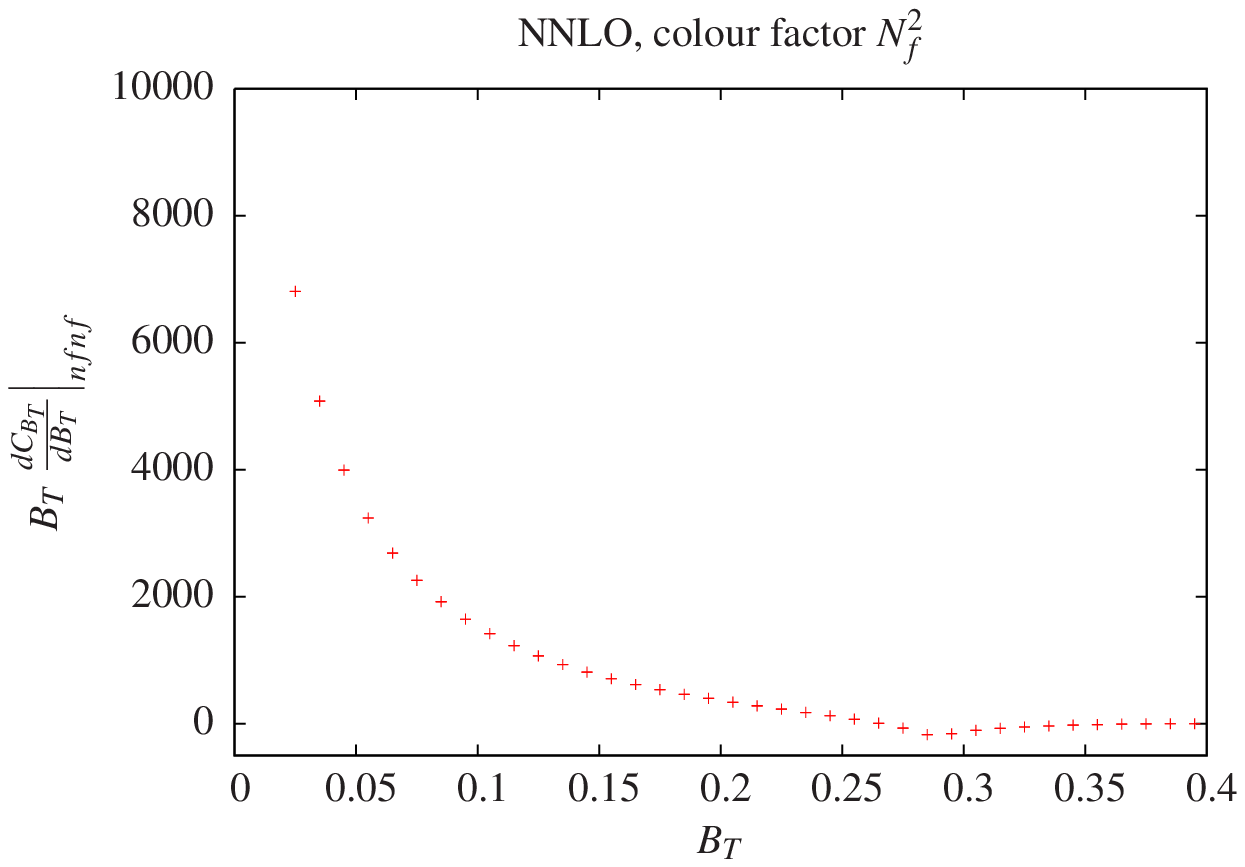}
\end{center}
\caption{
The NNLO coefficient $C_{B_T}$ for the total jet broadening distribution split up into
individual colour factors.
}
\label{fig_totaljetbroadening_C_col}
\end{figure}

\begin{figure}[p]
\begin{center}
\includegraphics[bb= 125 460 490 710,width=0.32\textwidth]{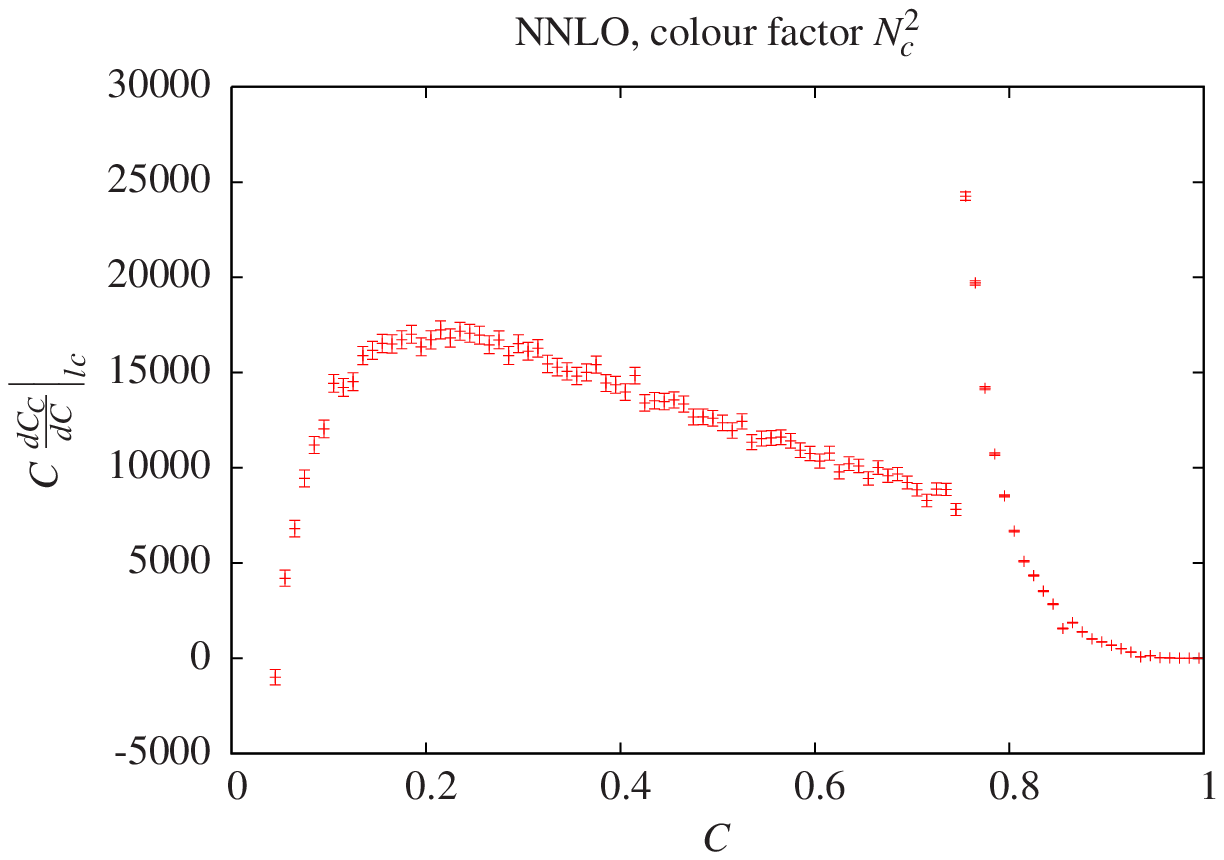}
\includegraphics[bb= 125 460 490 710,width=0.32\textwidth]{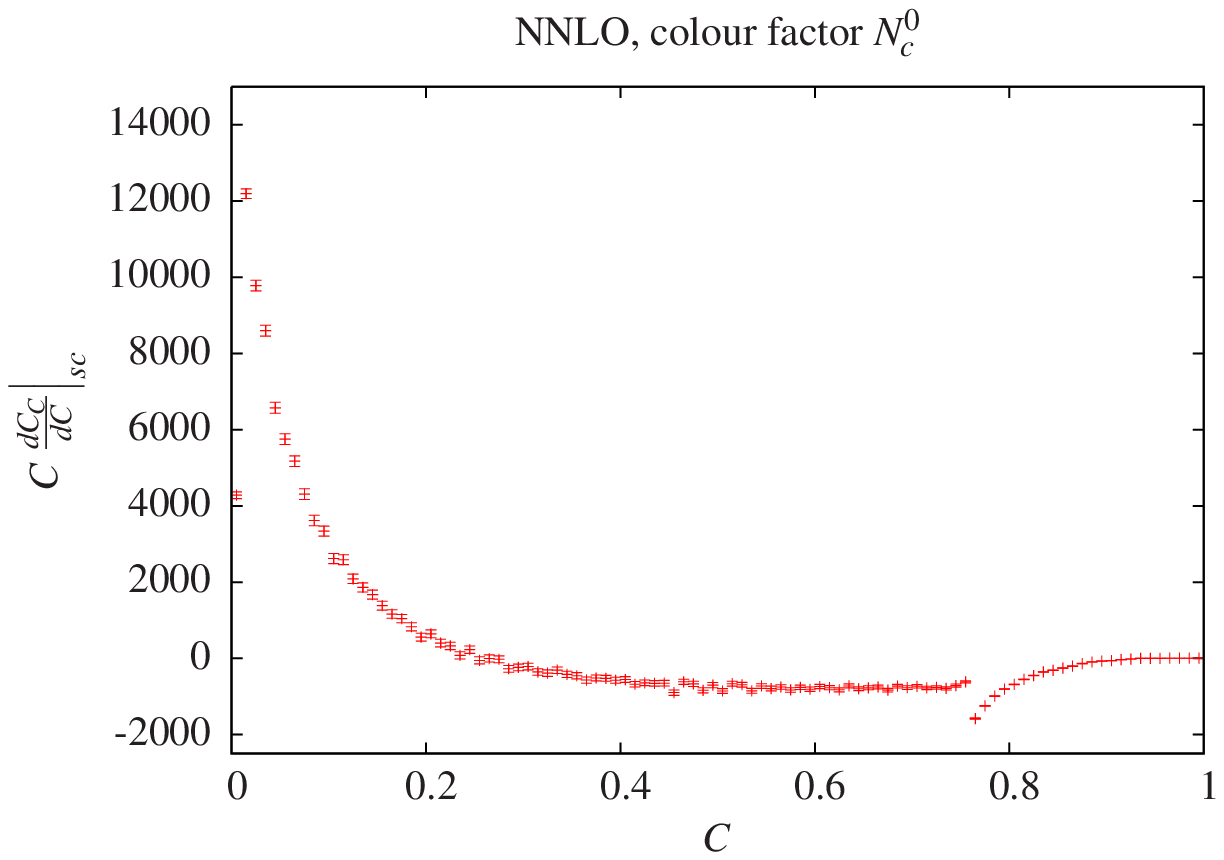}
\includegraphics[bb= 125 460 490 710,width=0.32\textwidth]{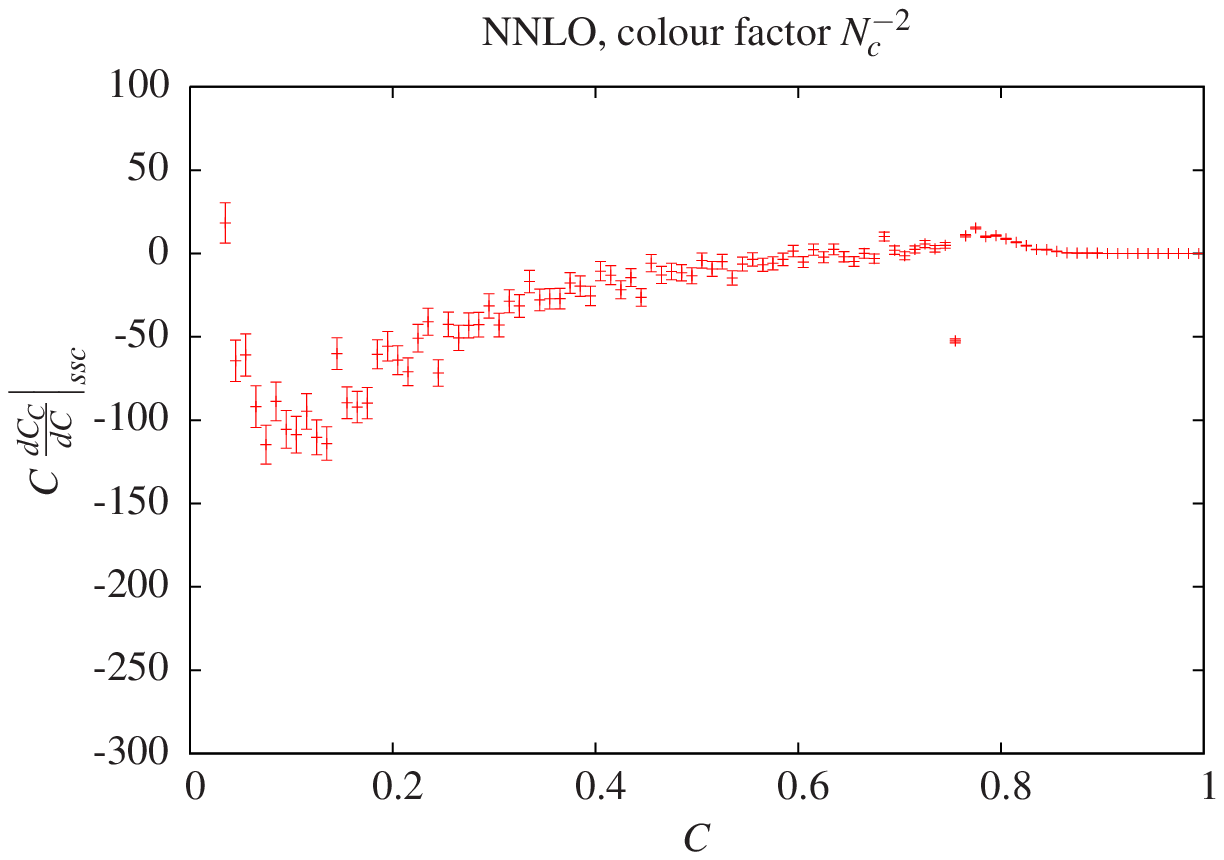}
\\
\includegraphics[bb= 125 460 490 710,width=0.32\textwidth]{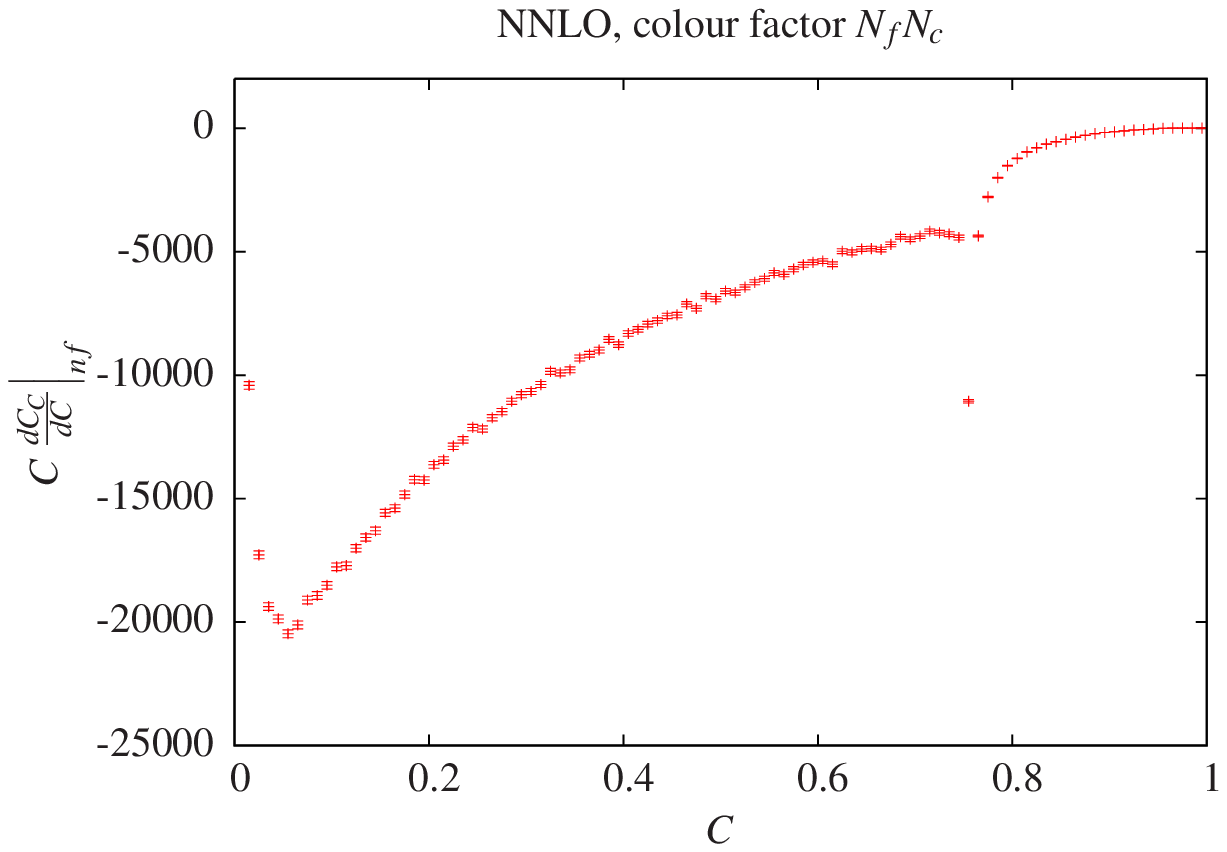}
\includegraphics[bb= 125 460 490 710,width=0.32\textwidth]{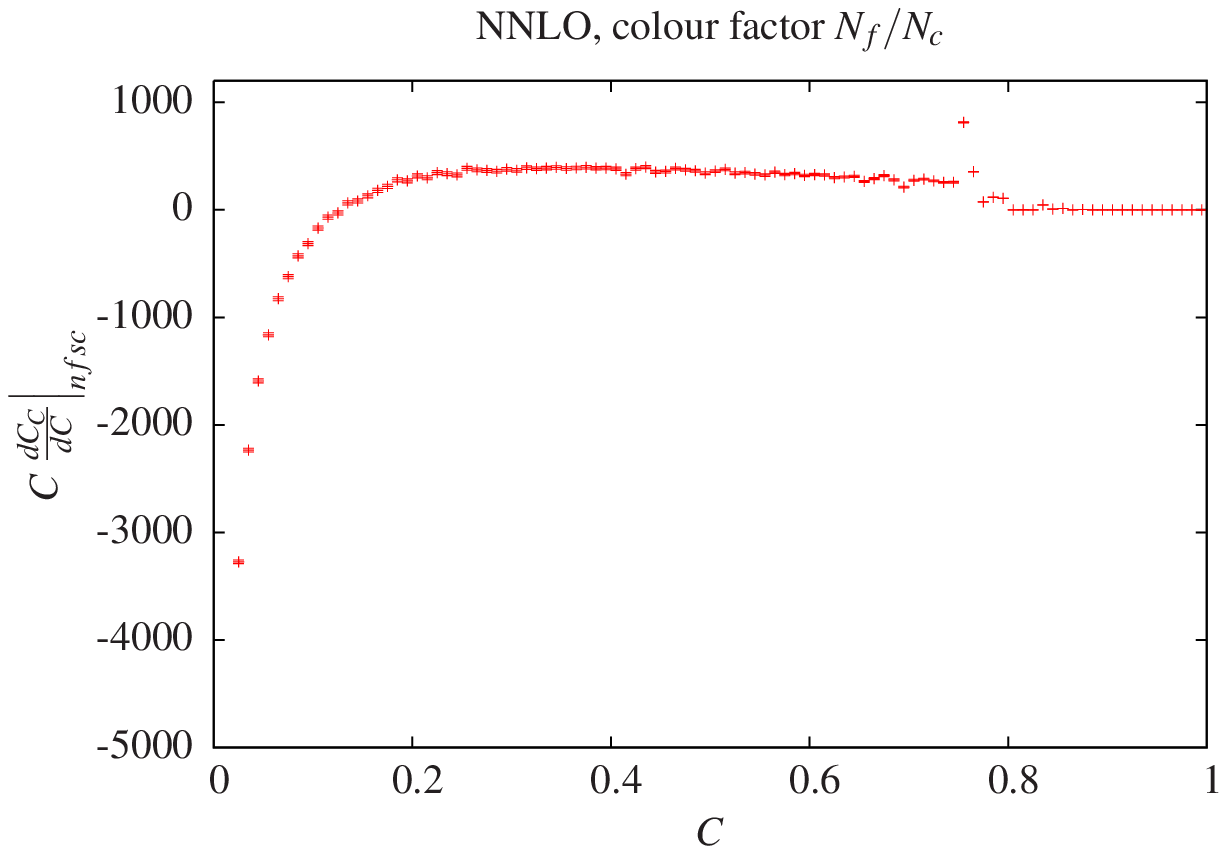}
\includegraphics[bb= 125 460 490 710,width=0.32\textwidth]{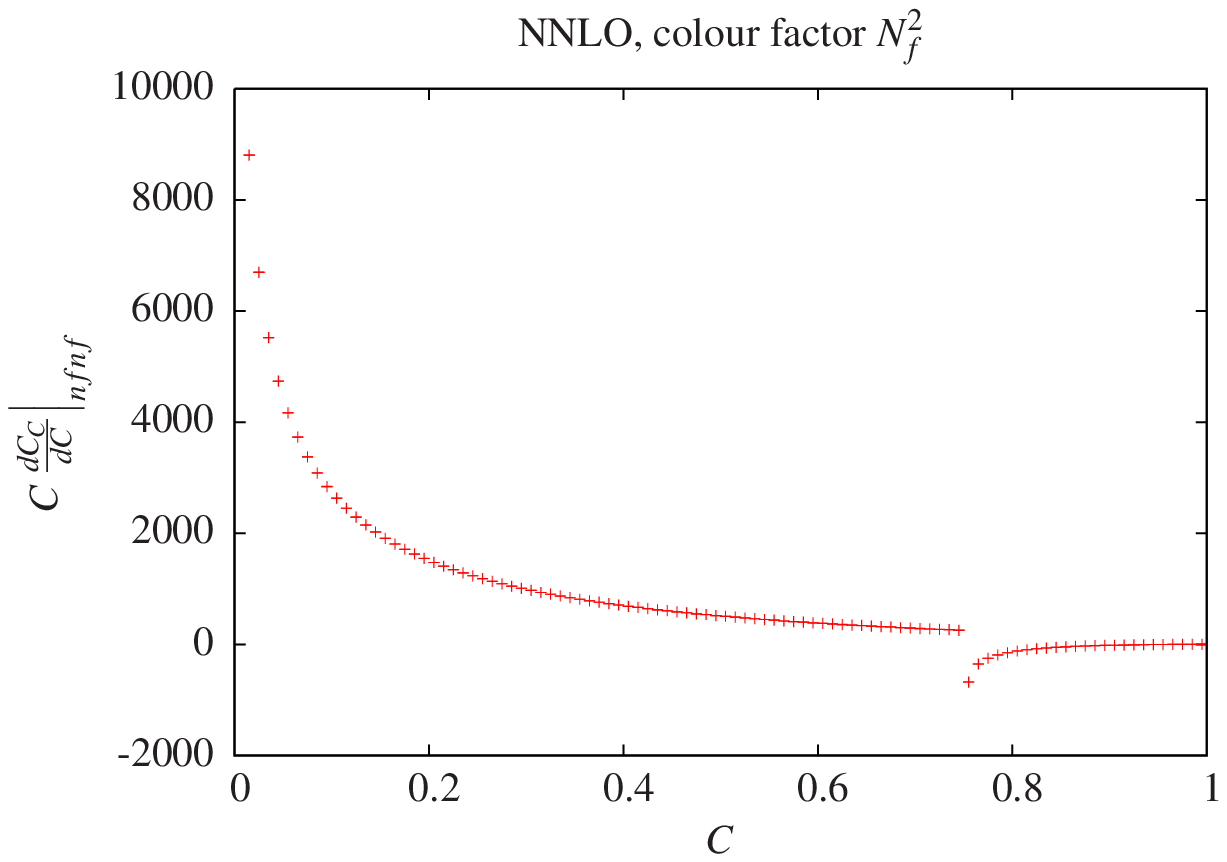}
\end{center}
\caption{
The NNLO coefficient $C_{C}$ for the $C$-parameter distribution split up into
individual colour factors.
}
\label{fig_Cparameter_C_col}
\end{figure}

\begin{figure}[p]
\begin{center}
\includegraphics[bb= 125 460 490 710,width=0.32\textwidth]{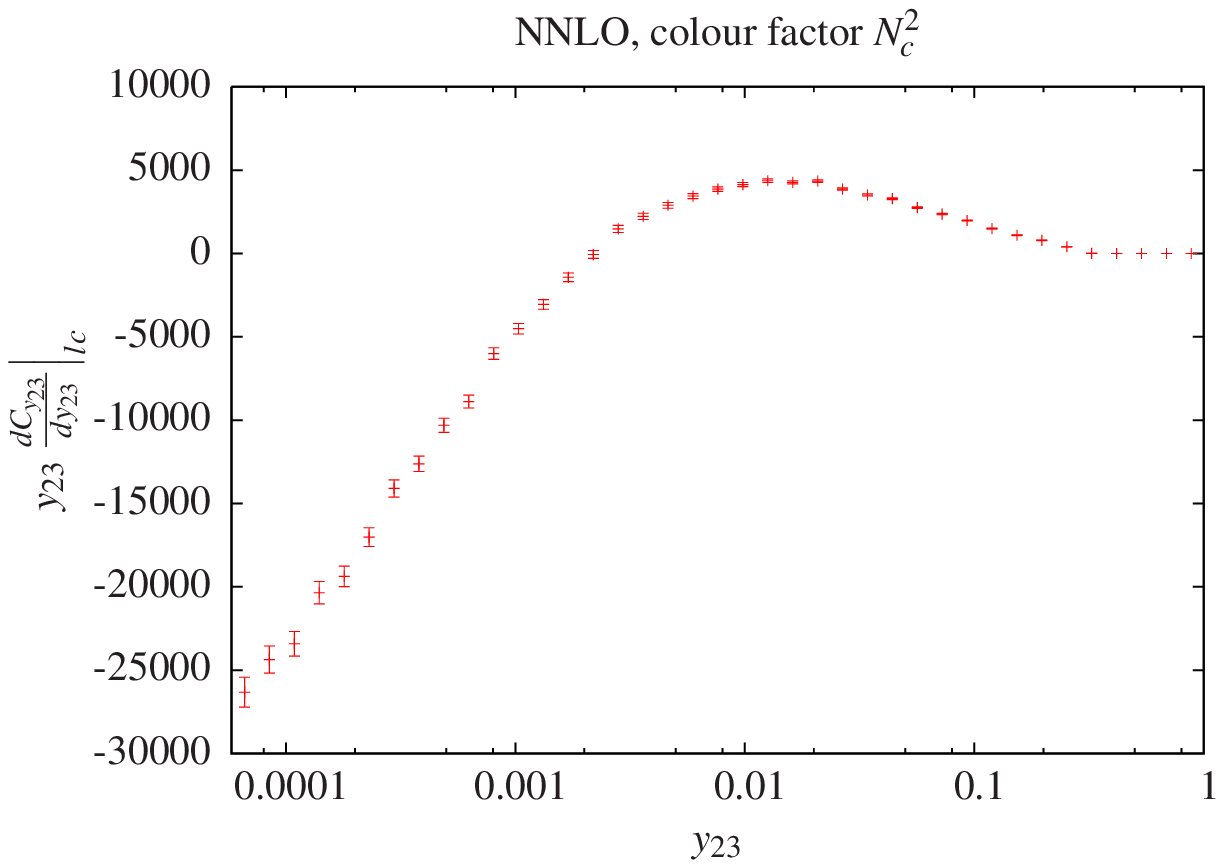}
\includegraphics[bb= 125 460 490 710,width=0.32\textwidth]{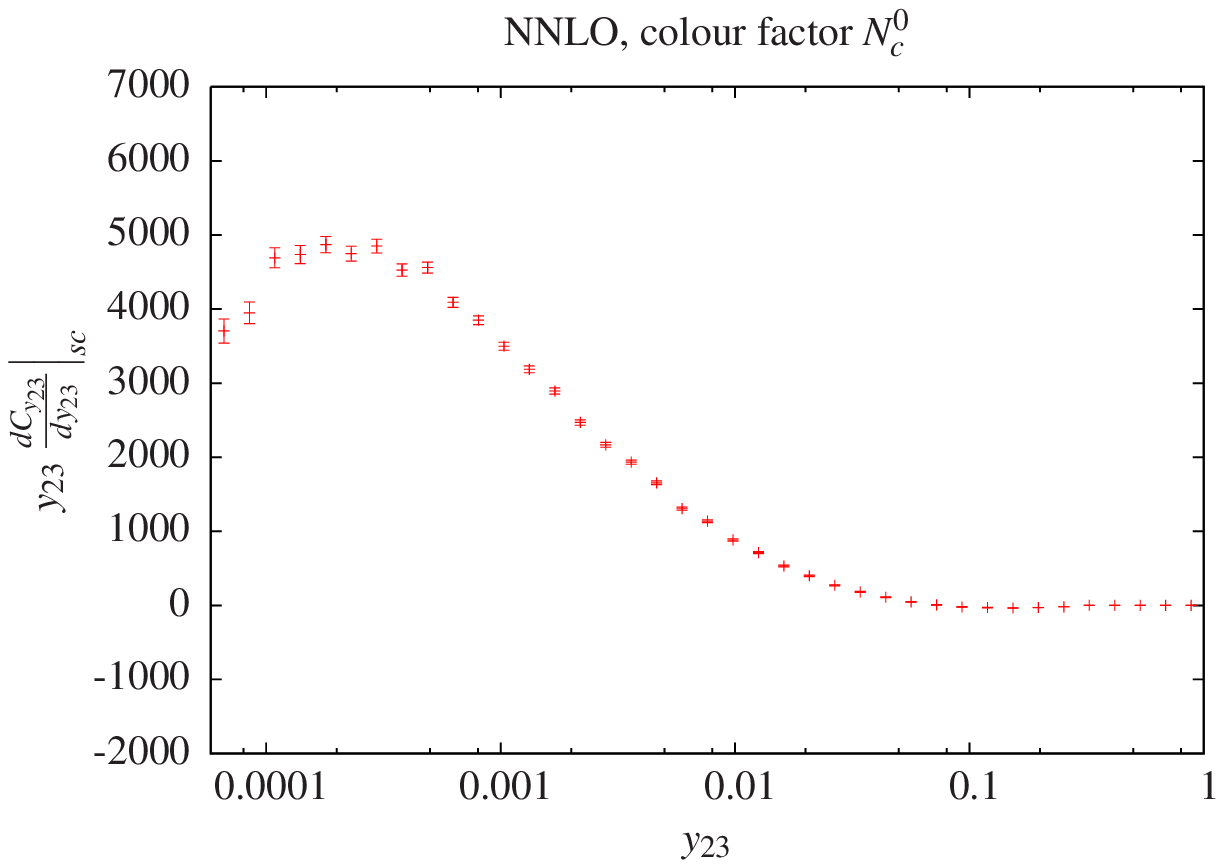}
\includegraphics[bb= 125 460 490 710,width=0.32\textwidth]{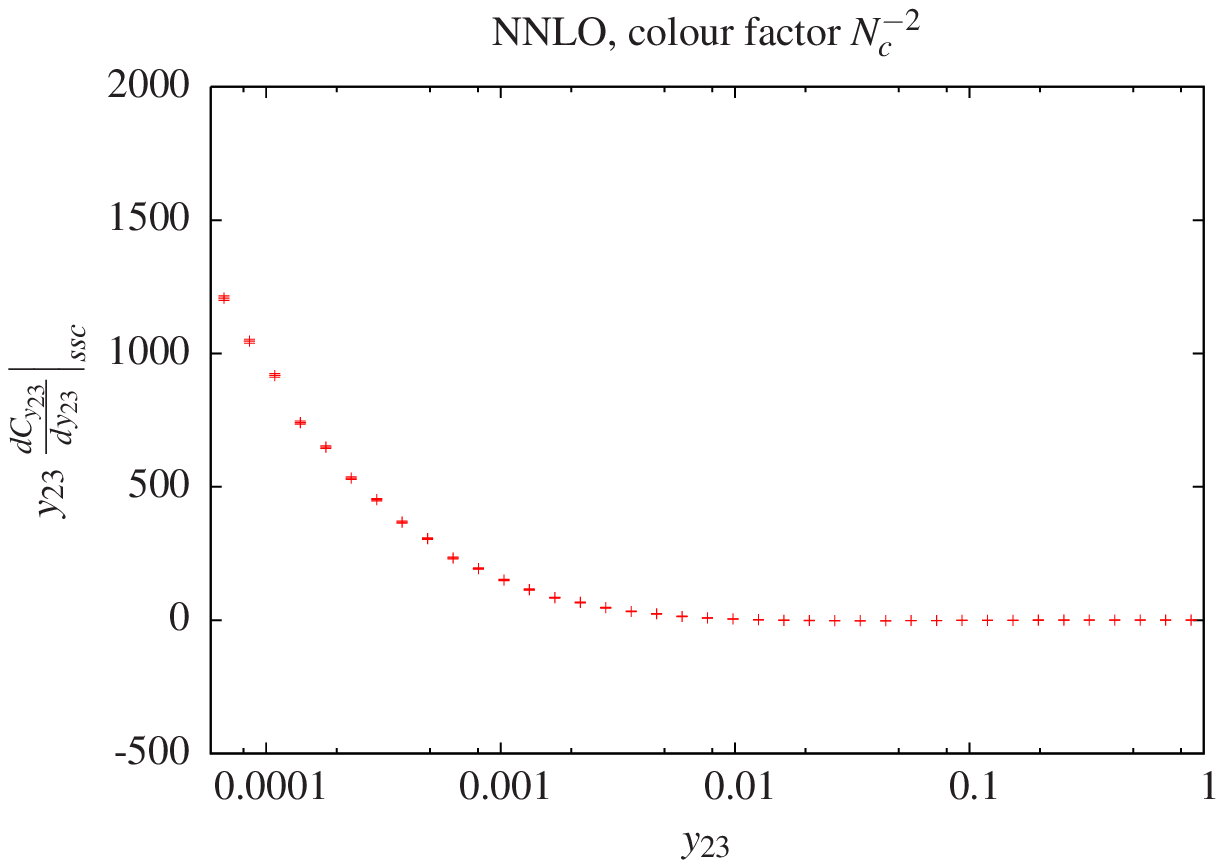}
\\
\includegraphics[bb= 125 460 490 710,width=0.32\textwidth]{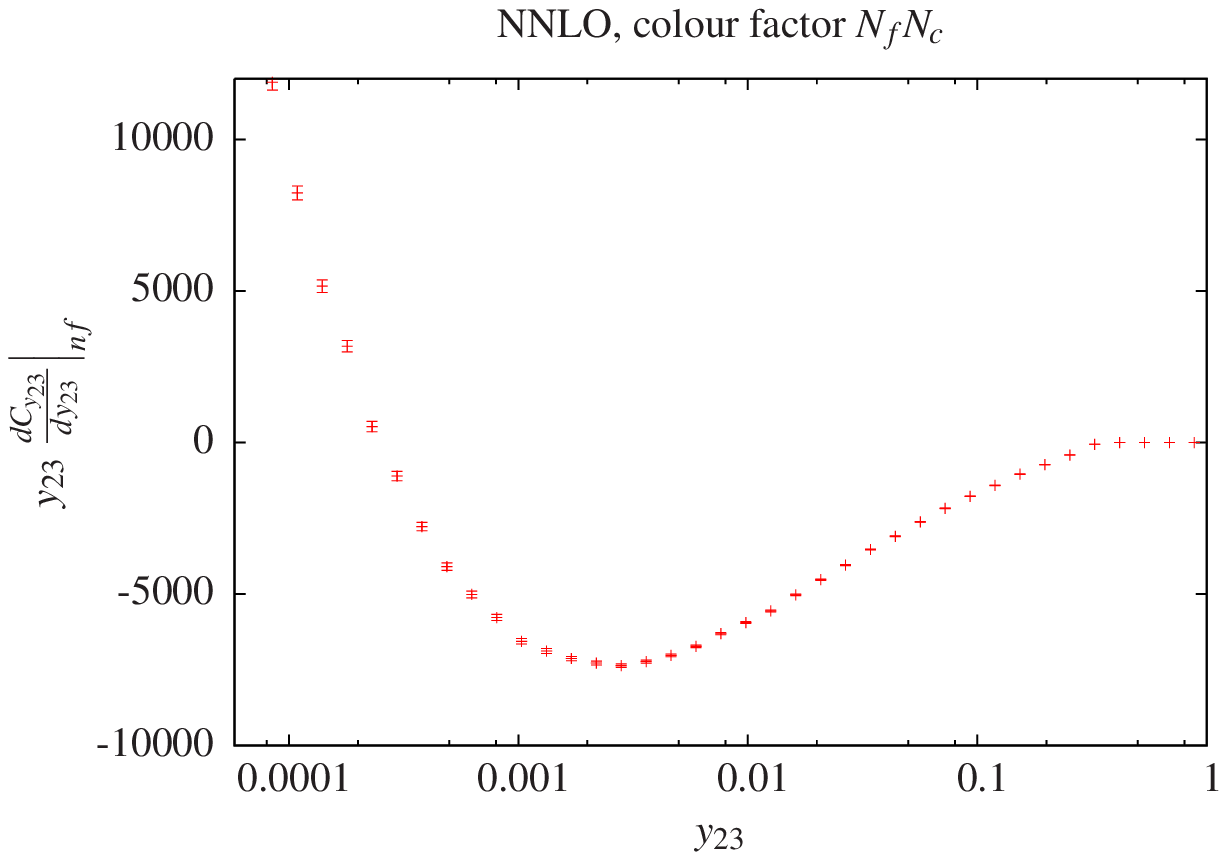}
\includegraphics[bb= 125 460 490 710,width=0.32\textwidth]{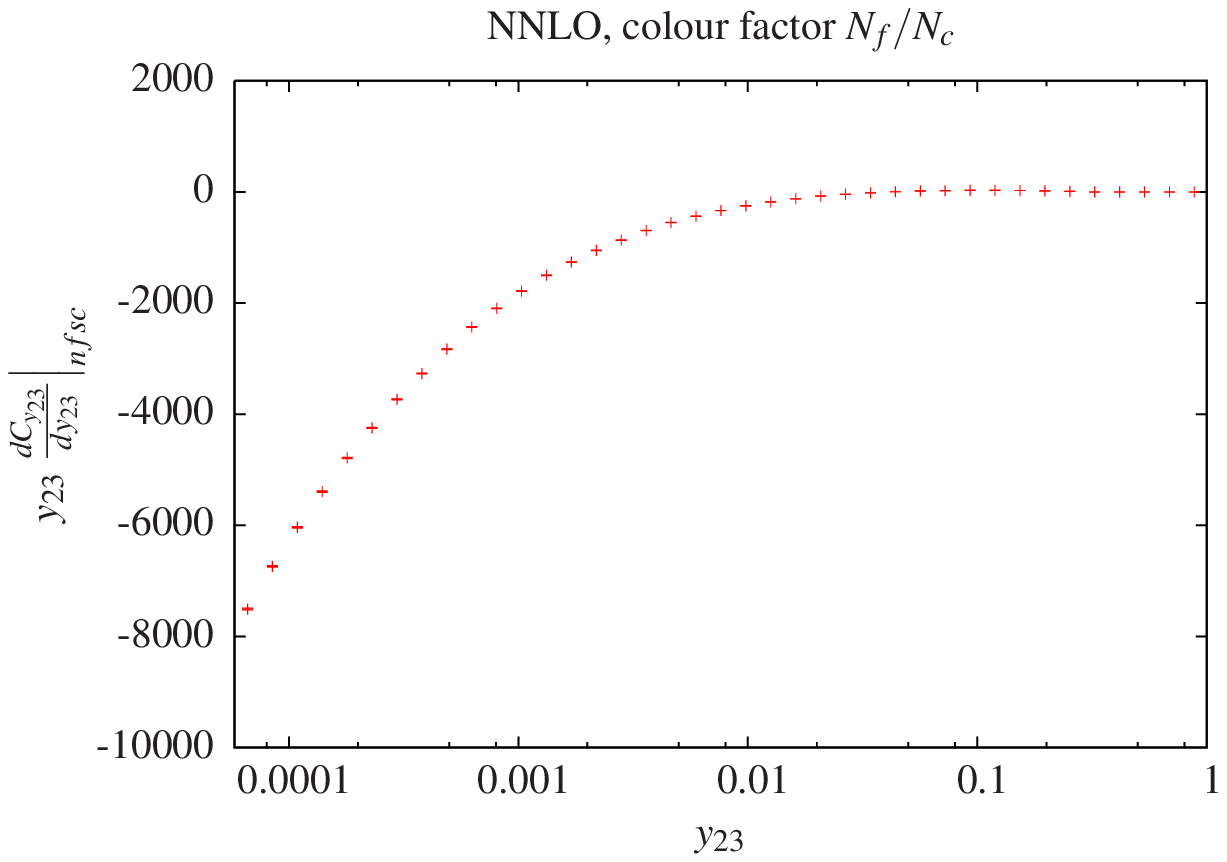}
\includegraphics[bb= 125 460 490 710,width=0.32\textwidth]{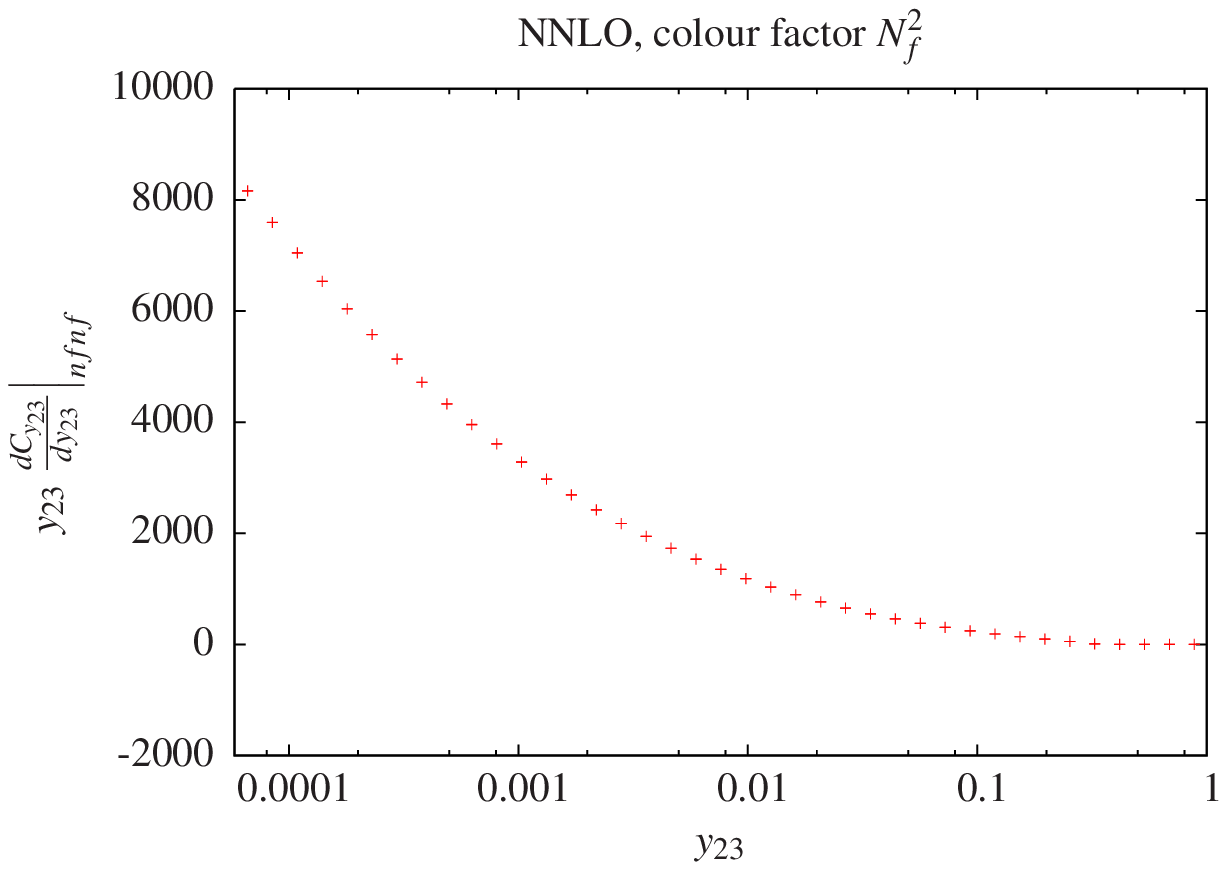}
\end{center}
\caption{
The NNLO coefficient $C_{y_{23}}$ for the three-to-two jet transition distribution split up into
individual colour factors.
}
\label{fig_y23_C_col}
\end{figure}

The NNLO function $C_{\cal O}$ can be split up into 
the contributions from the individual colour factors
\bq
 \left. C_{\cal O} \right|_{lc},
 \;\;\;
 \left. C_{\cal O} \right|_{sc},
 \;\;\;
 \left. C_{\cal O} \right|_{ssc},
 \;\;\;
 \left. C_{\cal O} \right|_{nf},
 \;\;\;
 \left. C_{\cal O} \right|_{nfsc},
 \;\;\;
 \left. C_{\cal O} \right|_{nfnf}
\eq
defined in eq.~(\ref{def_colour_factors}).
For the event shape variables the contributions from the individual colour factors to $C_{\cal O}$ are given
in tables~\ref{table_thrust_colour} to \ref{table_y23_colour}.
For the jet rates the contributions from the individual colour factors to $C_{\cal O}$ are given
in tables~\ref{table_jet_rate_durham_colour} to \ref{table_jet_rate_geneva_colour}.
In addition the individual colour contributions are  plotted for the six event shape variables in
figs.~\ref{fig_thrust_C_col}-\ref{fig_y23_C_col}.

\begin{figure}[p]
\begin{center}
\includegraphics[bb= 125 460 490 710,width=0.32\textwidth]{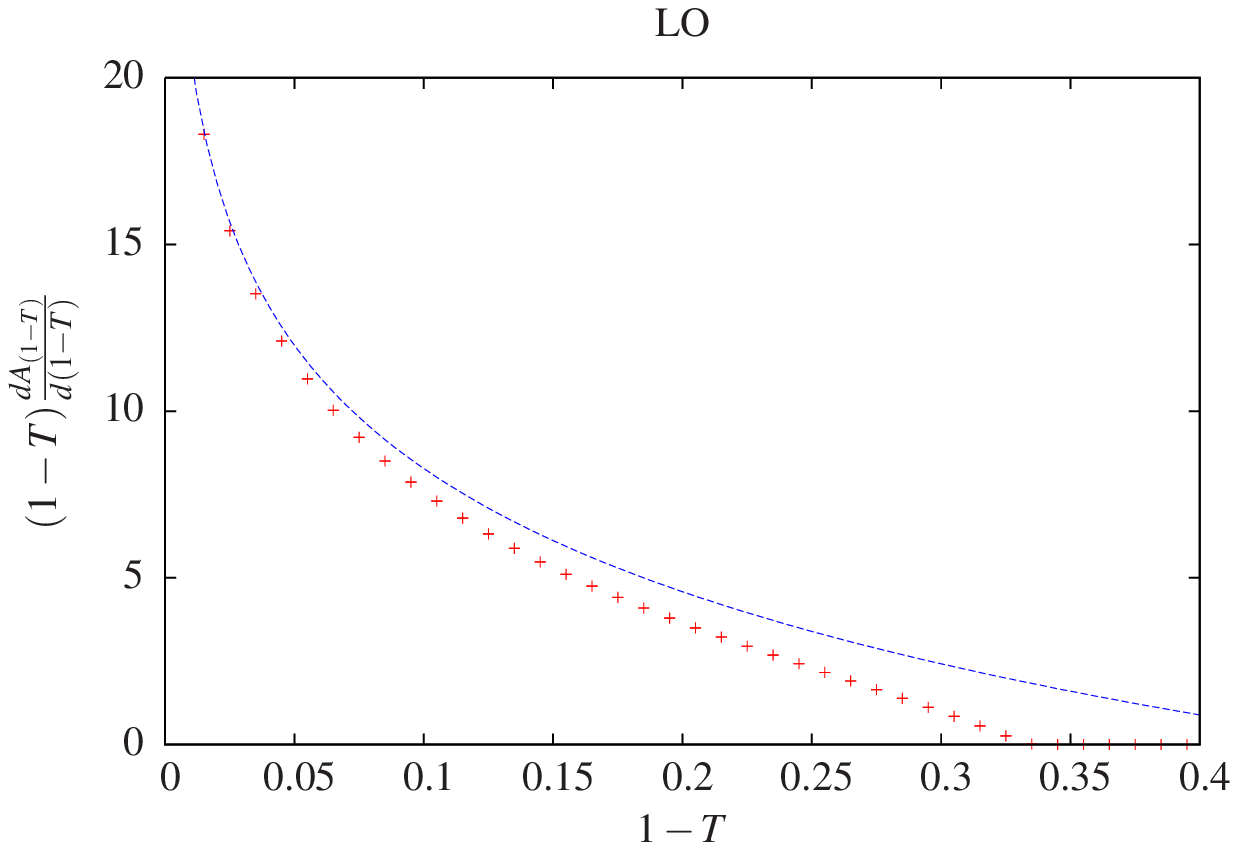}
\includegraphics[bb= 125 460 490 710,width=0.32\textwidth]{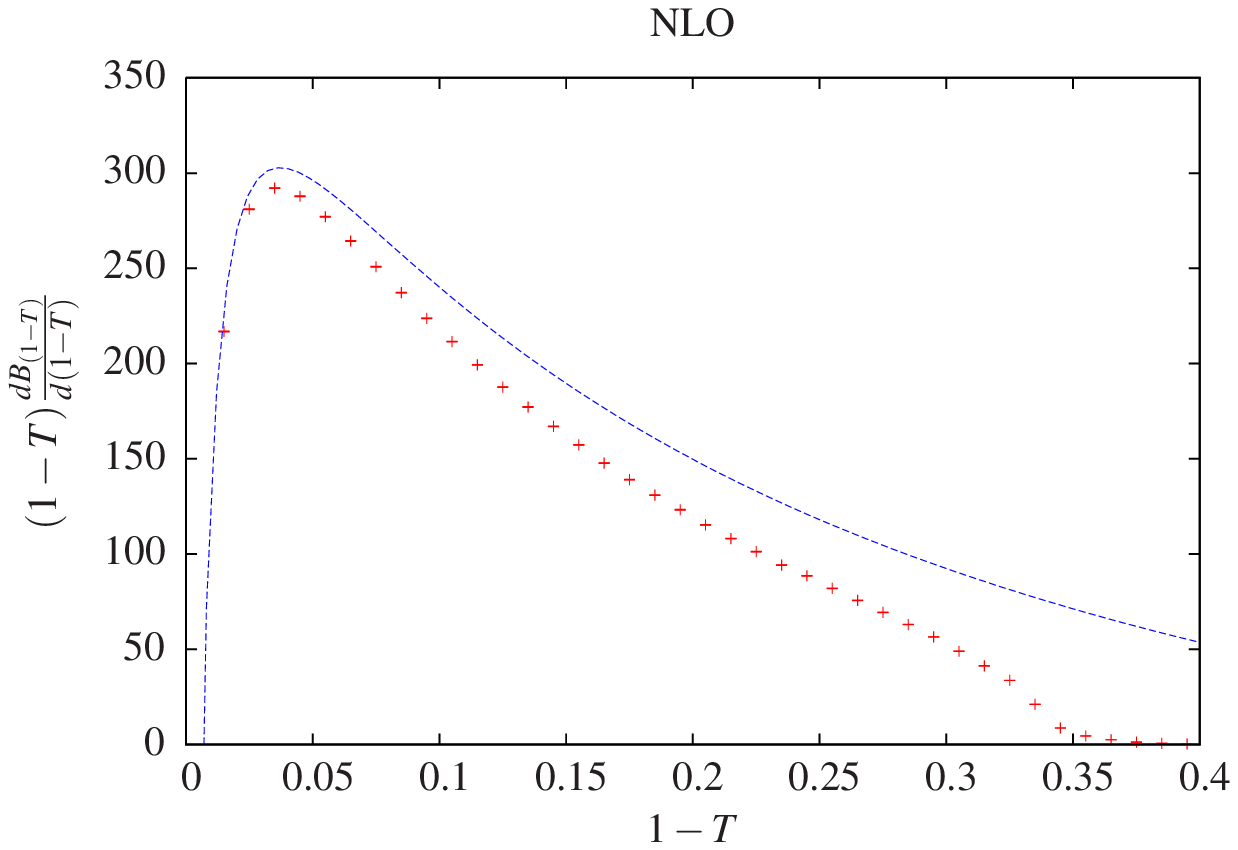}
\includegraphics[bb= 125 460 490 710,width=0.32\textwidth]{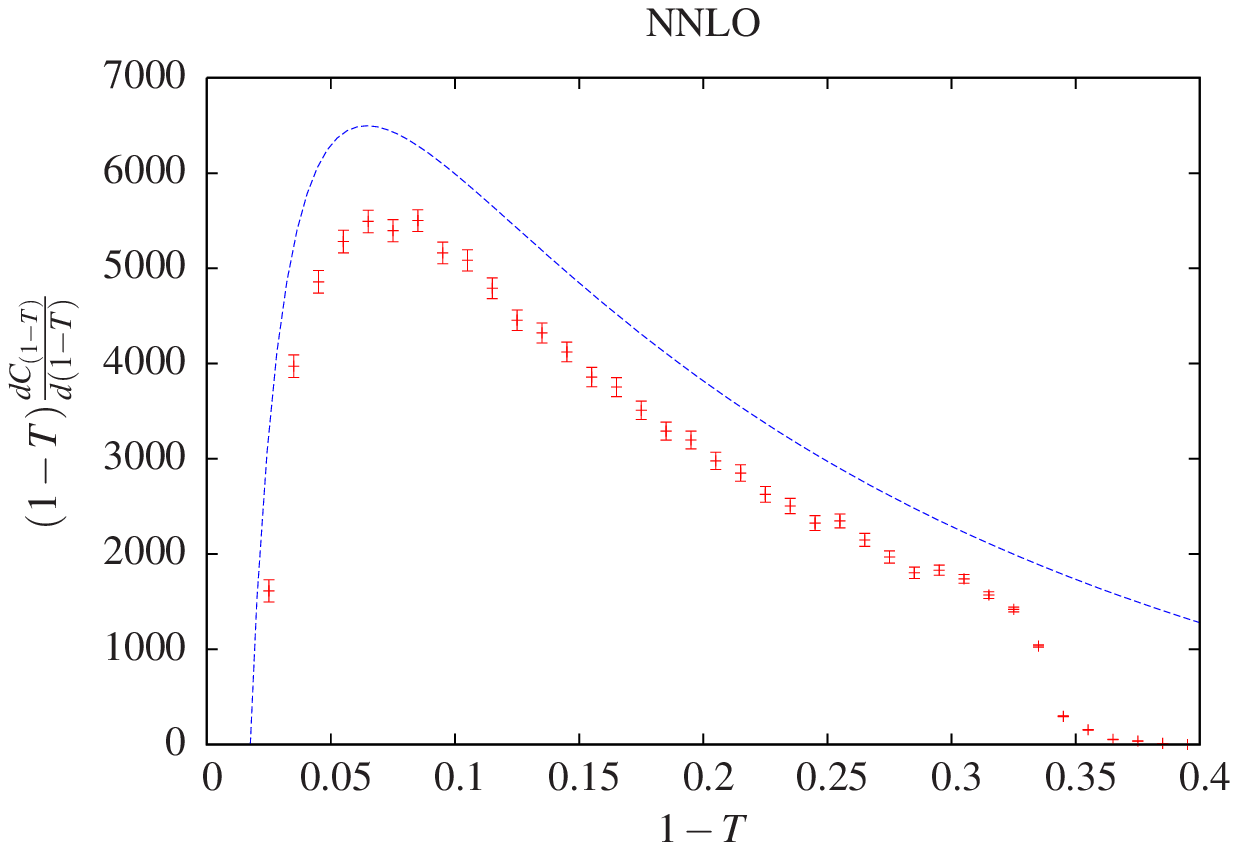}
\\
\includegraphics[bb= 125 460 490 710,width=0.32\textwidth]{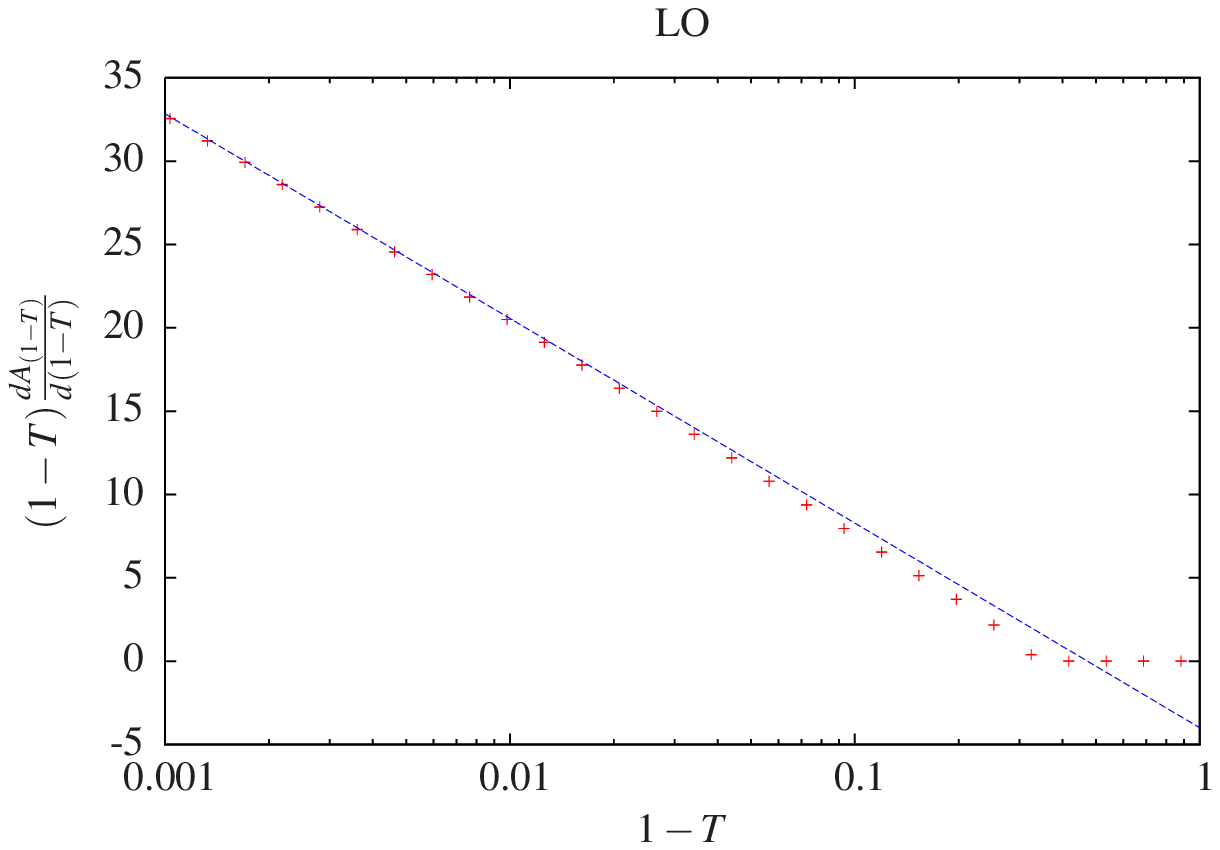}
\includegraphics[bb= 125 460 490 710,width=0.32\textwidth]{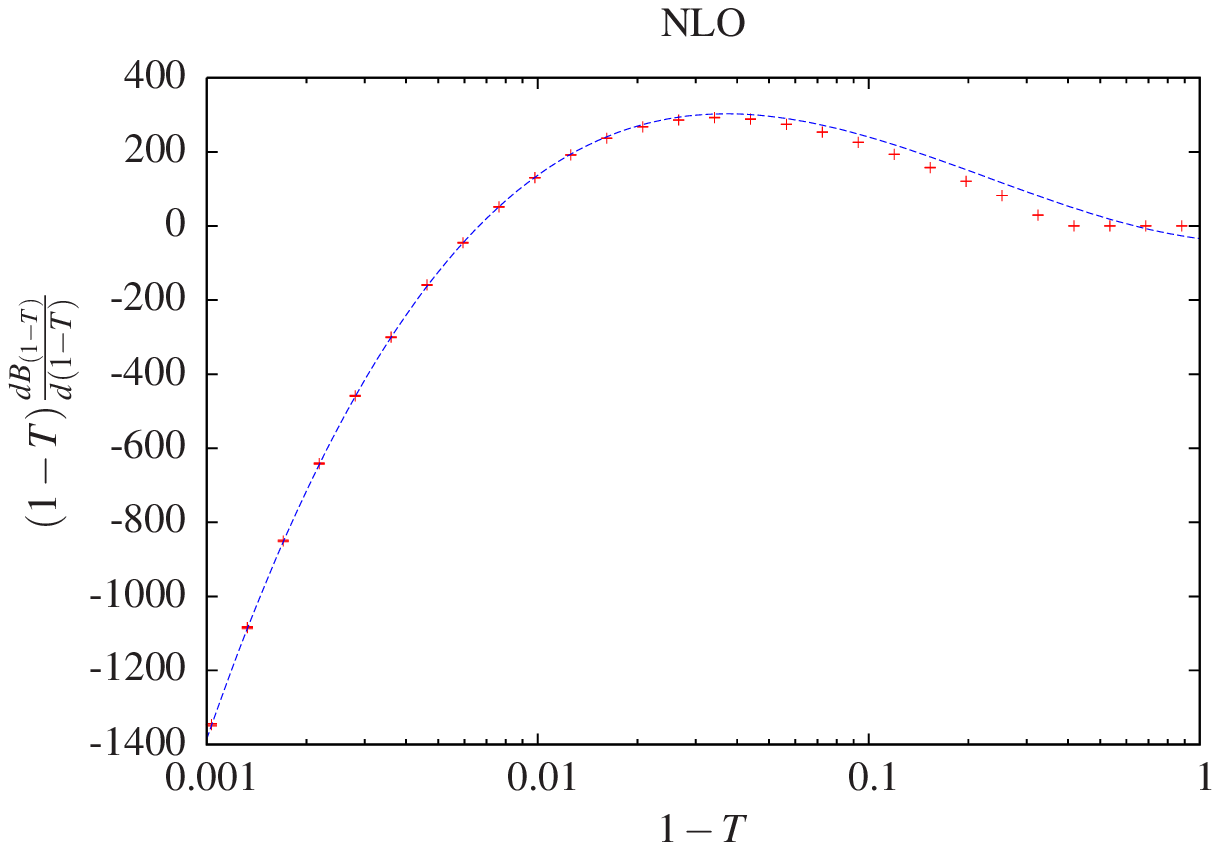}
\includegraphics[bb= 125 460 490 710,width=0.32\textwidth]{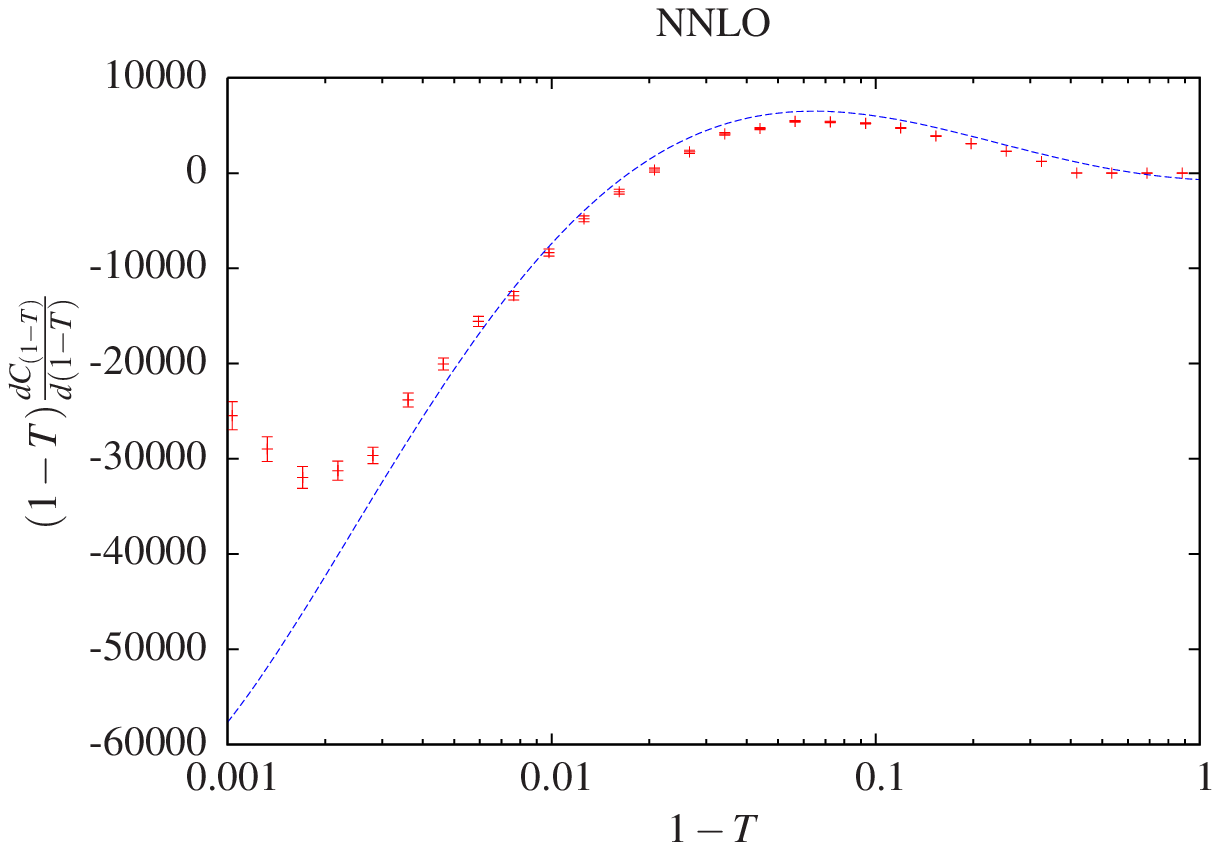}
\end{center}
\caption{Comparison of the results for the
coefficients of the leading-order ($A_{(1-T)}$, left), 
next-to-leading-order ($B_{(1-T)}$, middle)
and next-to-next-to-leading order ($C_{(1-T)}$, right)
contributions to the thrust distribution from perturbative QCD (red)
and SCET (blue line).
The upper row shows the distribution with a linear scale for $(1-T)$, the lower
row shows the distribution with a logarithmic scale for $(1-T)$.
}
\label{fig_thrustlog_ABC}
\end{figure}

\begin{figure}[p]
\begin{center}
\includegraphics[bb= 125 460 490 710,width=0.32\textwidth]{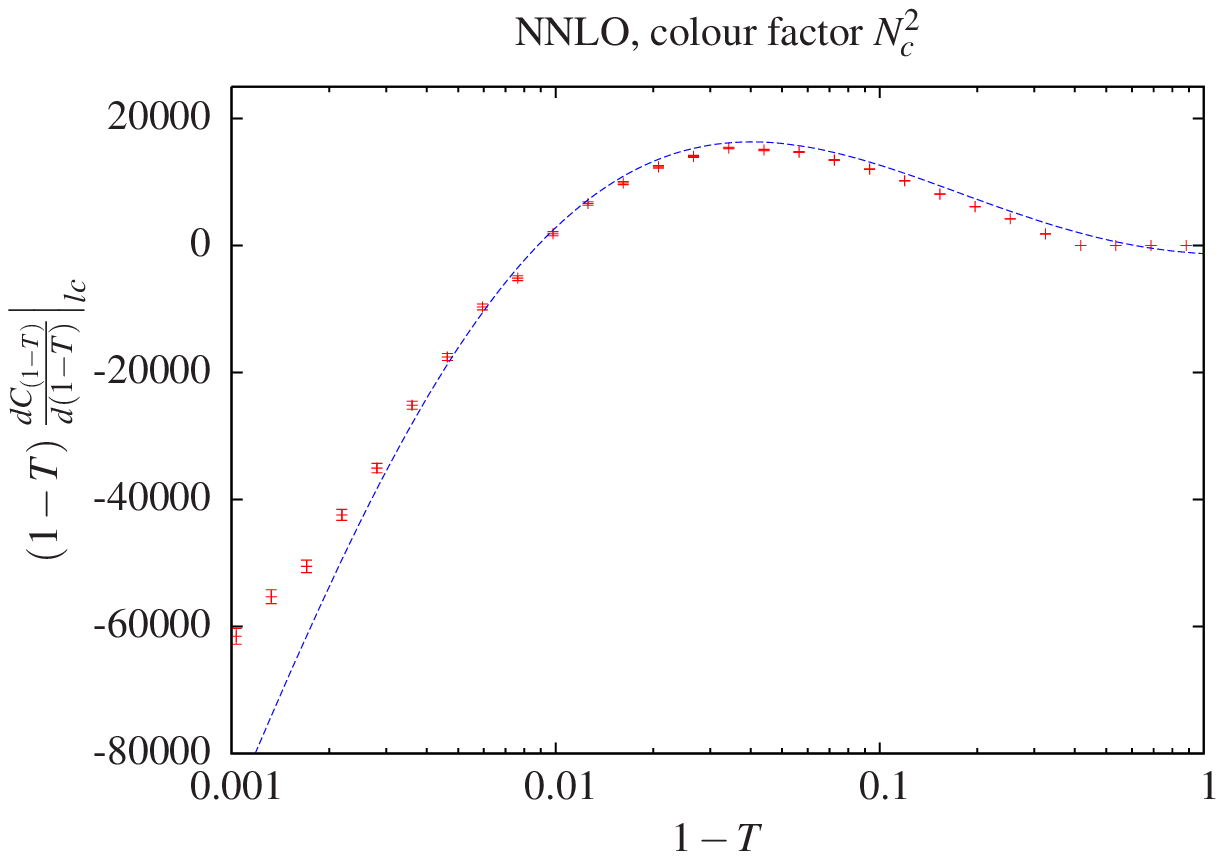}
\includegraphics[bb= 125 460 490 710,width=0.32\textwidth]{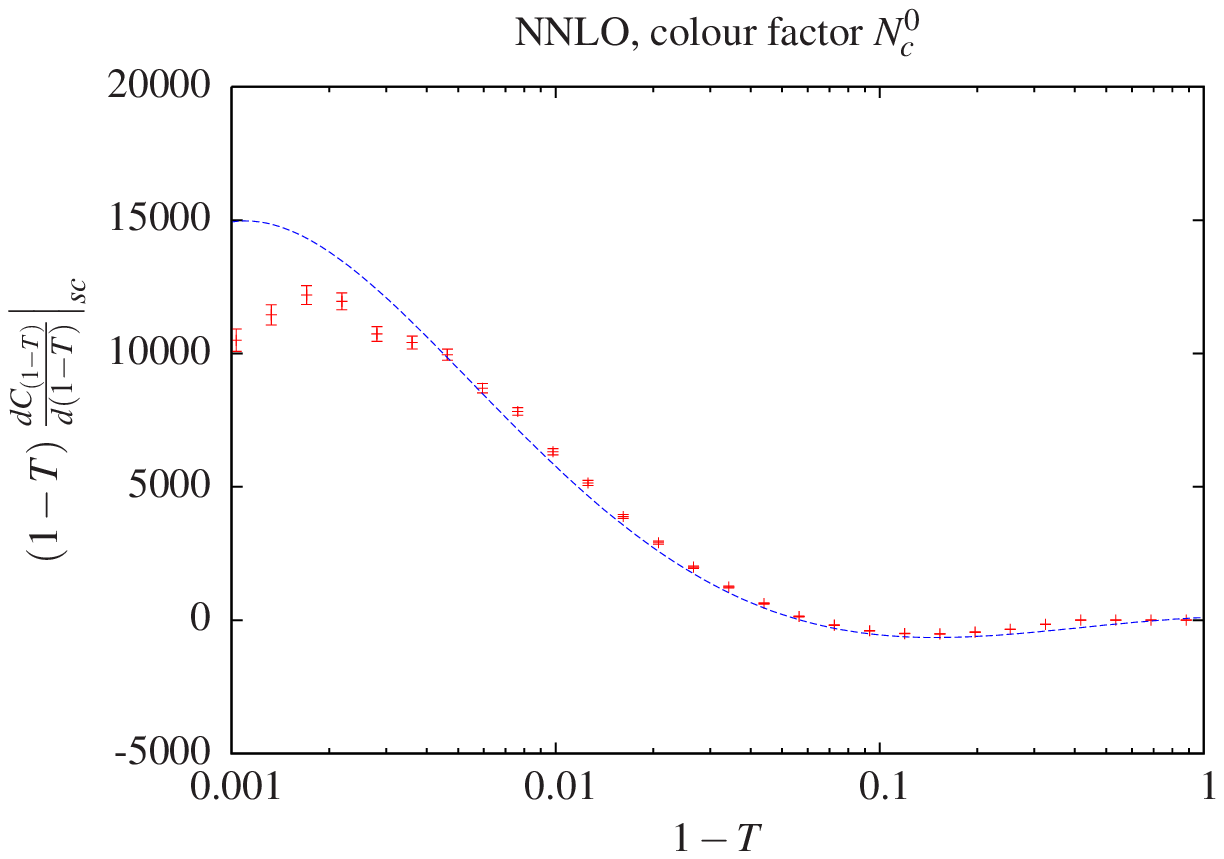}
\includegraphics[bb= 125 460 490 710,width=0.32\textwidth]{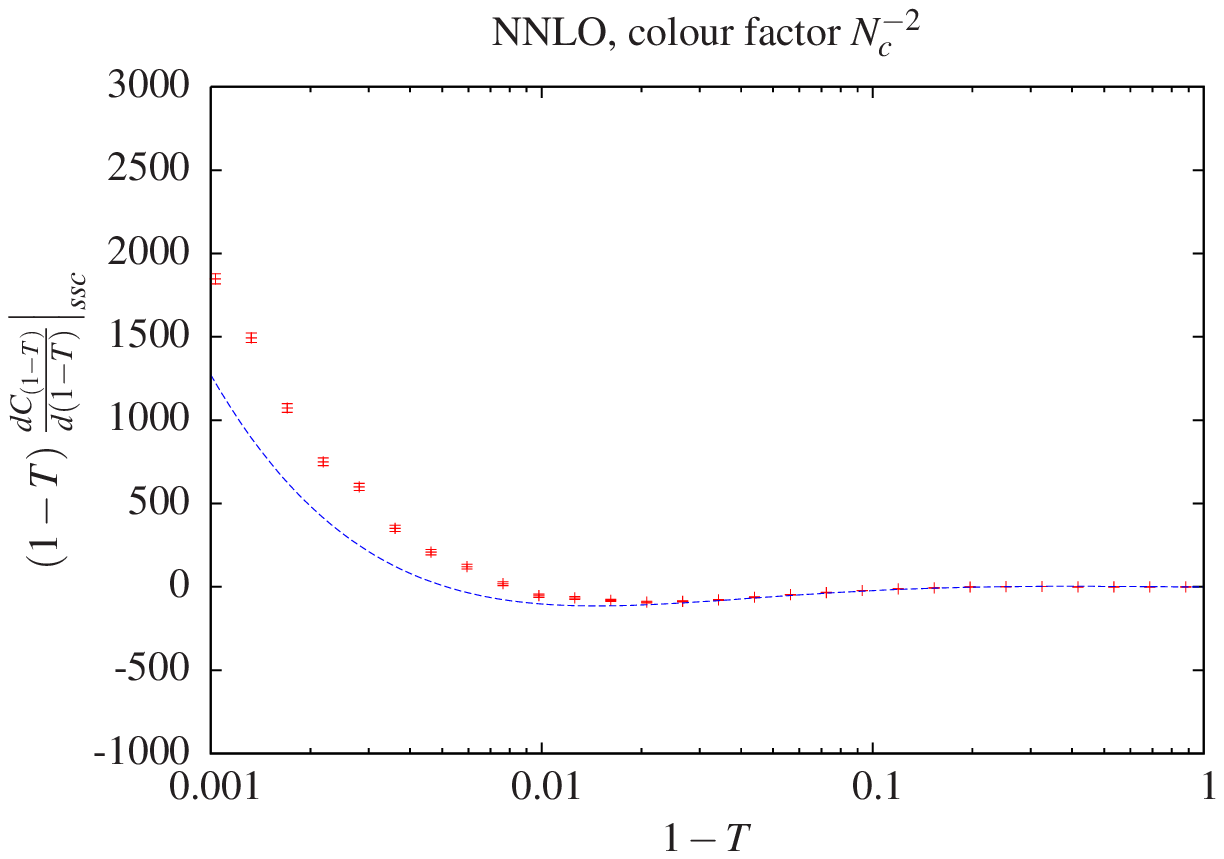}
\\
\includegraphics[bb= 125 460 490 710,width=0.32\textwidth]{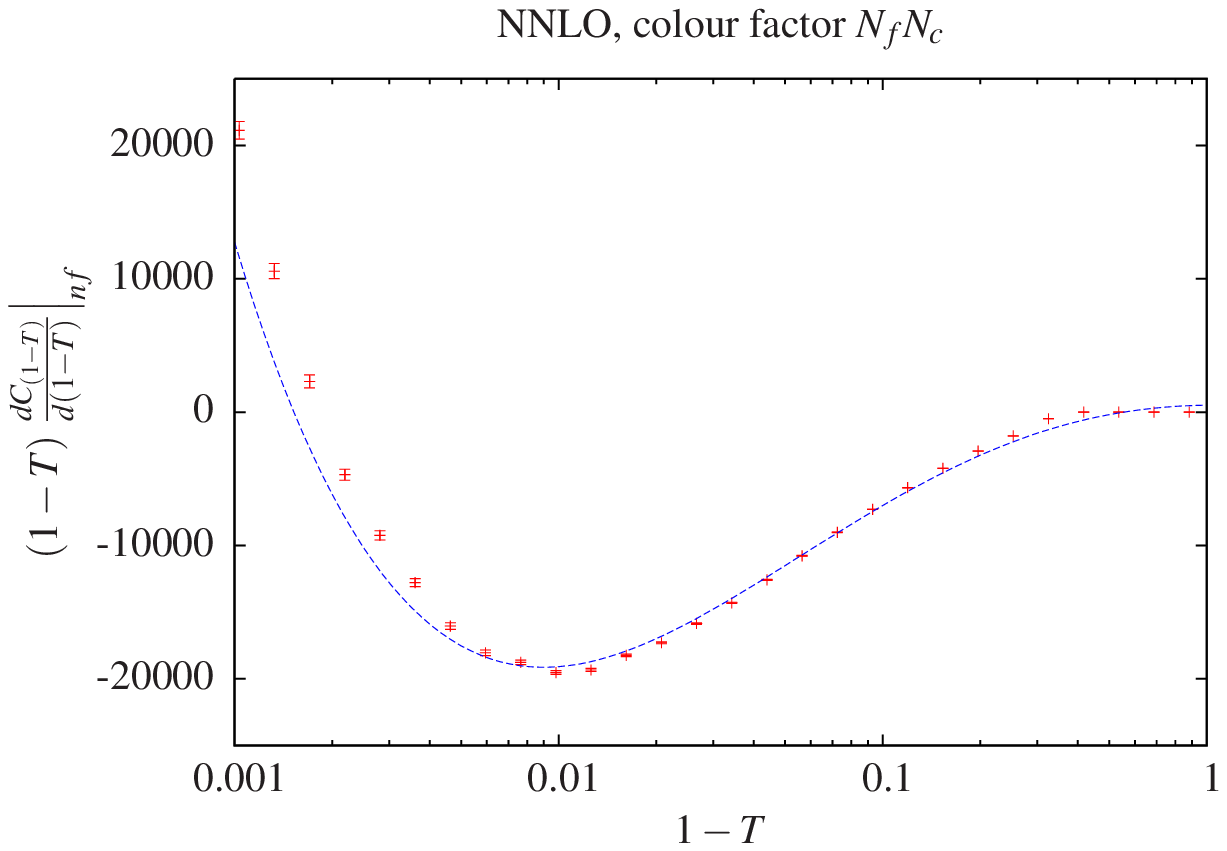}
\includegraphics[bb= 125 460 490 710,width=0.32\textwidth]{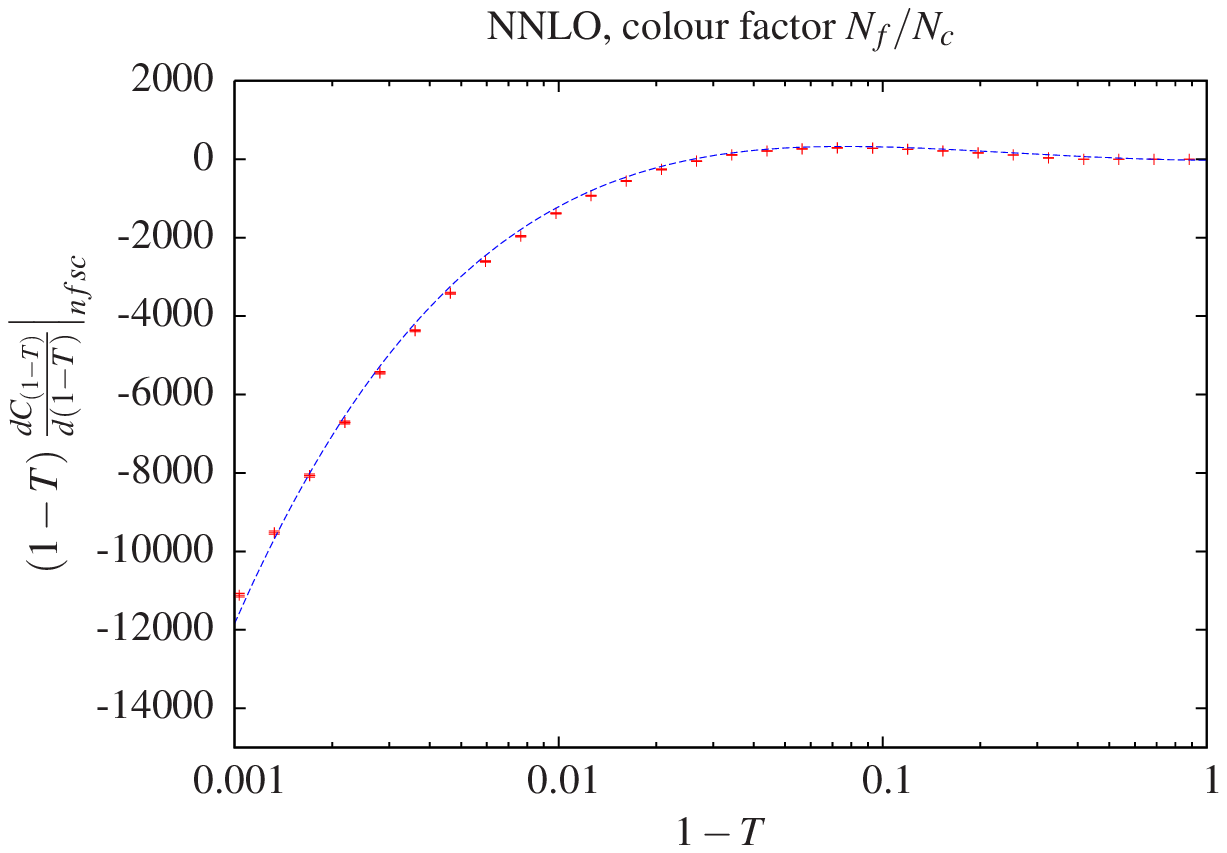}
\includegraphics[bb= 125 460 490 710,width=0.32\textwidth]{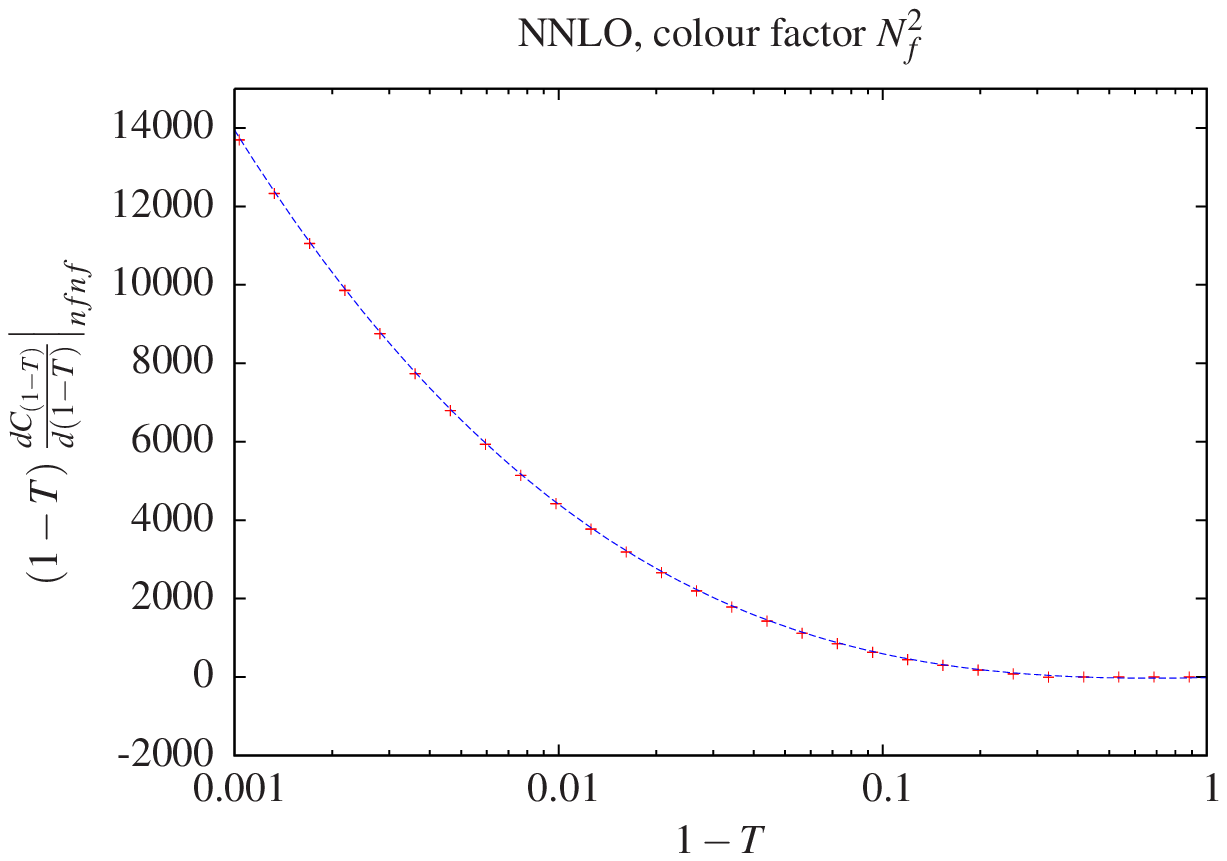}
\end{center}
\caption{
Comparison of the individual colour factors 
of the next-to-next-to-leading order coefficient $C_{(1-T)}$ for the thrust distribution
between perturbative QCD (red) and SCET (blue line).
The distributions are shown with a logarithmic scale for $(1-T)$.
The deviations of the Monte-Carlo results for small values of $(1-T)$ are due
to the slicing procedure.
}
\label{fig_thrustlog_C_col}
\end{figure}

\begin{figure}[p]
\begin{center}
\includegraphics[bb= 125 460 490 710,width=0.9\textwidth]{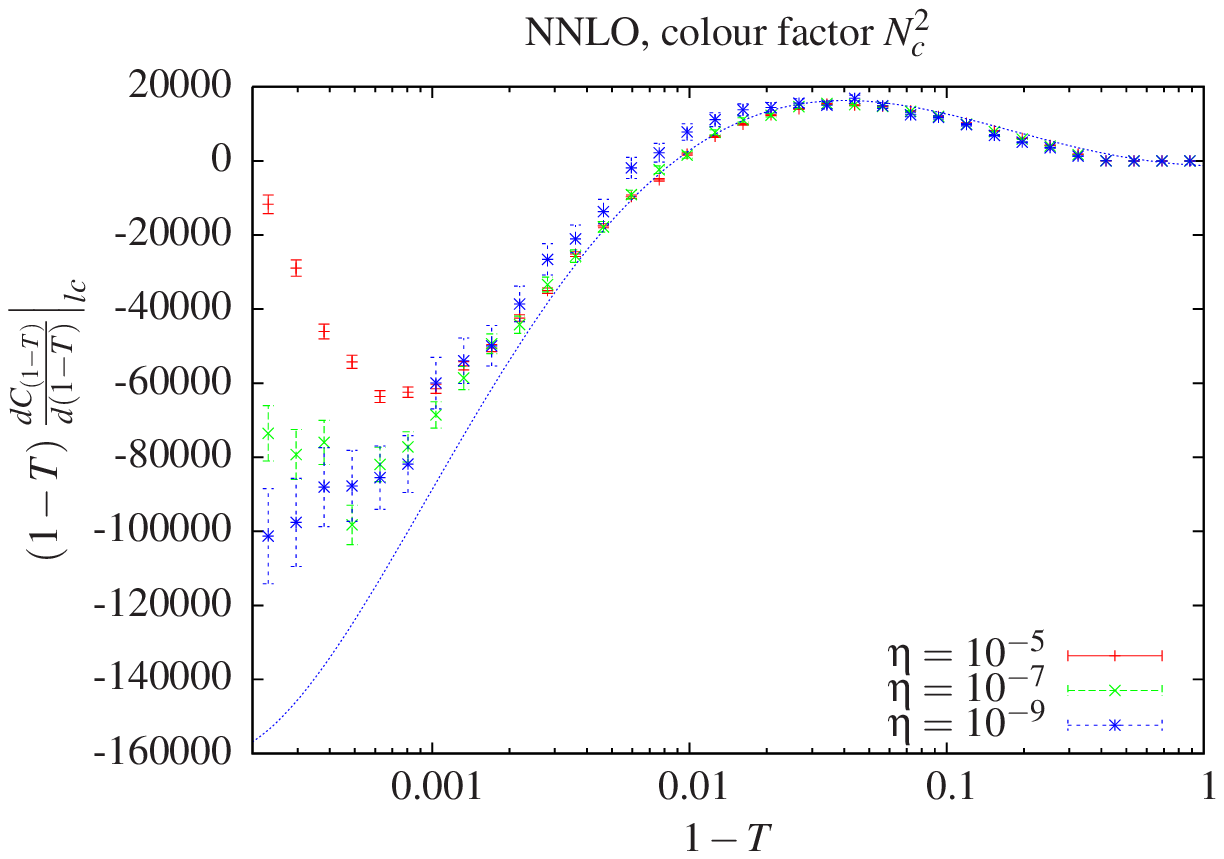}
\end{center}
\caption{Comparison of the next-to-next-to-leading order coefficient $C_{(1-T)}$ 
for the colour factor $N_c^2$ for the thrust distribution
between the numerical Monte-Carlo program for various values of $\eta$ and SCET (blue line).
}
\label{fig_thrustlog_C_col_eta}
\end{figure}

In fig.~\ref{fig_thrustlog_ABC}-\ref{fig_thrustlog_C_col_eta} 
the results for the thrust distribution are compared with the calculation of Becher and Schwartz
based on soft-collinear effective theory \cite{Becher:2008cf}. 
The leading-order, next-to-leading-order and next-to-next-to-leading order coefficients
$A_{1-T}$, $B_{1-T}$ and $C_{1-T}$ are compared in fig.~\ref{fig_thrustlog_ABC}.
The individual colour factors of the coefficient $C_{1-T}$ are compared in 
fig.~\ref{fig_thrustlog_C_col}.
The effective theory gives a good description of the thrust distribution for small values of $(1-T)$.
For values of $(1-T)$ close to $1$ the effective theory is not valid and deviations from the perturbative result
can be seen.
On the other hand the perturbative result obtained from Monte-Carlo integration has its limitation
through the slicing parameter introduced to handle the infrared singularities.
The deviations of the numerical Monte-Carlo results for the $C_{1-T}$ coefficient from the SCET
results for values of $(1-T) < 0.003$ are an artefact of the slicing parameter.
In figs.~\ref{fig_thrustlog_ABC} and \ref{fig_thrustlog_C_col} the value $\eta=10^{-5}$ was used for the slicing
parameter.
To study the situation in more detail, I show in fig.~\ref{fig_thrustlog_C_col_eta} the variation of the
numerical result for the leading colour factor of $C_{1-T}$ with the slicing parameter $\eta$.
The numerical results for $\eta=10^{-5}$, $\eta=10^{-7}$ and $\eta=10^{-9}$ are plotted.
For smaller values of $\eta$ the SCET result is approached.

\begin{figure}[p]
\begin{center}
\includegraphics[bb= 125 460 490 710,width=0.9\textwidth]{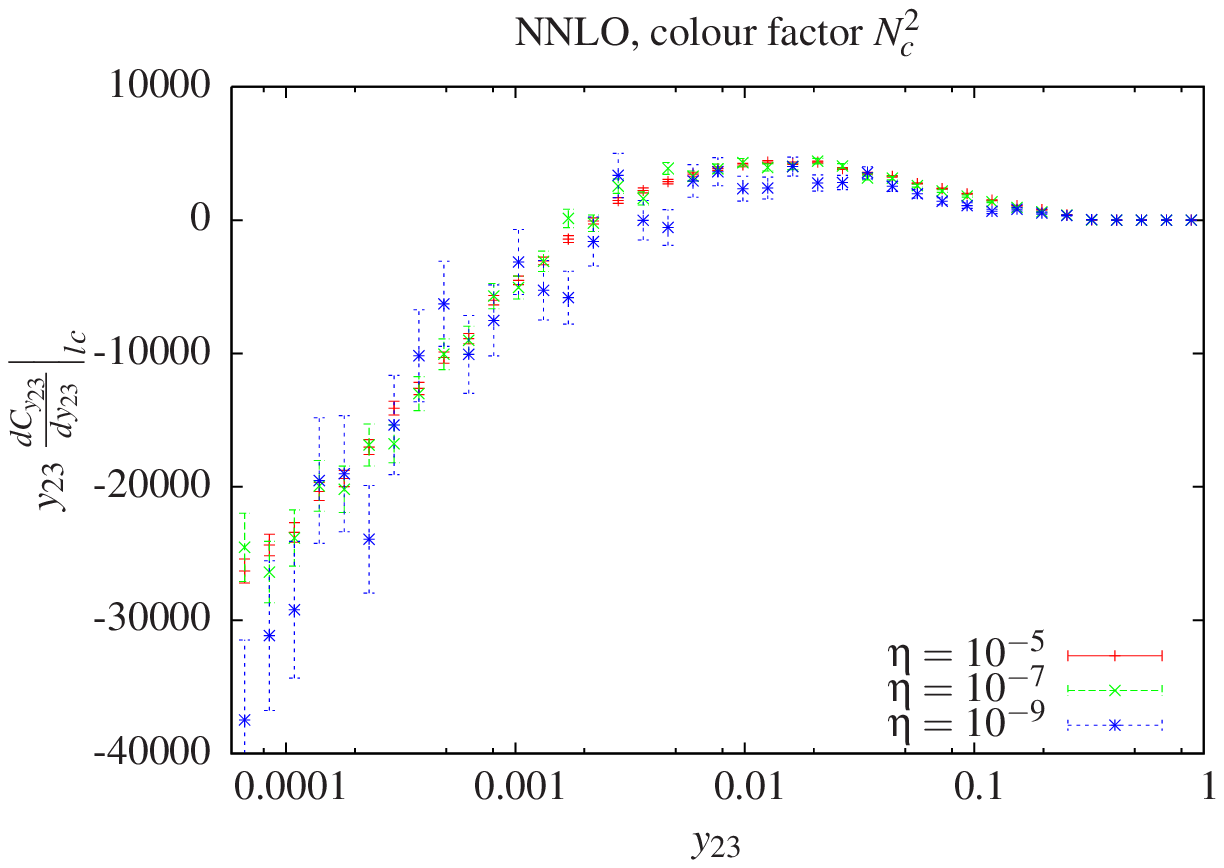}
\end{center}
\caption{Dependence of the numerical Monte-Carlo result for
the next-to-next-to-leading order coefficient $C_{y_{23}}$ 
for the colour factor $N_c^2$ for the three-to-two jet transition distribution 
on the slicing parameter $\eta$.
}
\label{fig_y23_C_col_eta}
\end{figure}

As a further example I also show the dependence on the slicing parameter $\eta$ for 
the colour factor $N_c^2$ for
the next-to-next-to-leading order coefficient $C_{y_{23}}$
for the three-to-two jet transition distribution in fig.~\ref{fig_y23_C_col_eta}.
In this plot the numerical results for $\eta=10^{-5}$, $\eta=10^{-7}$ and $\eta=10^{-9}$ are shown.

\begin{figure}[p]
\begin{center}
\includegraphics[bb= 125 460 490 710,width=0.32\textwidth]{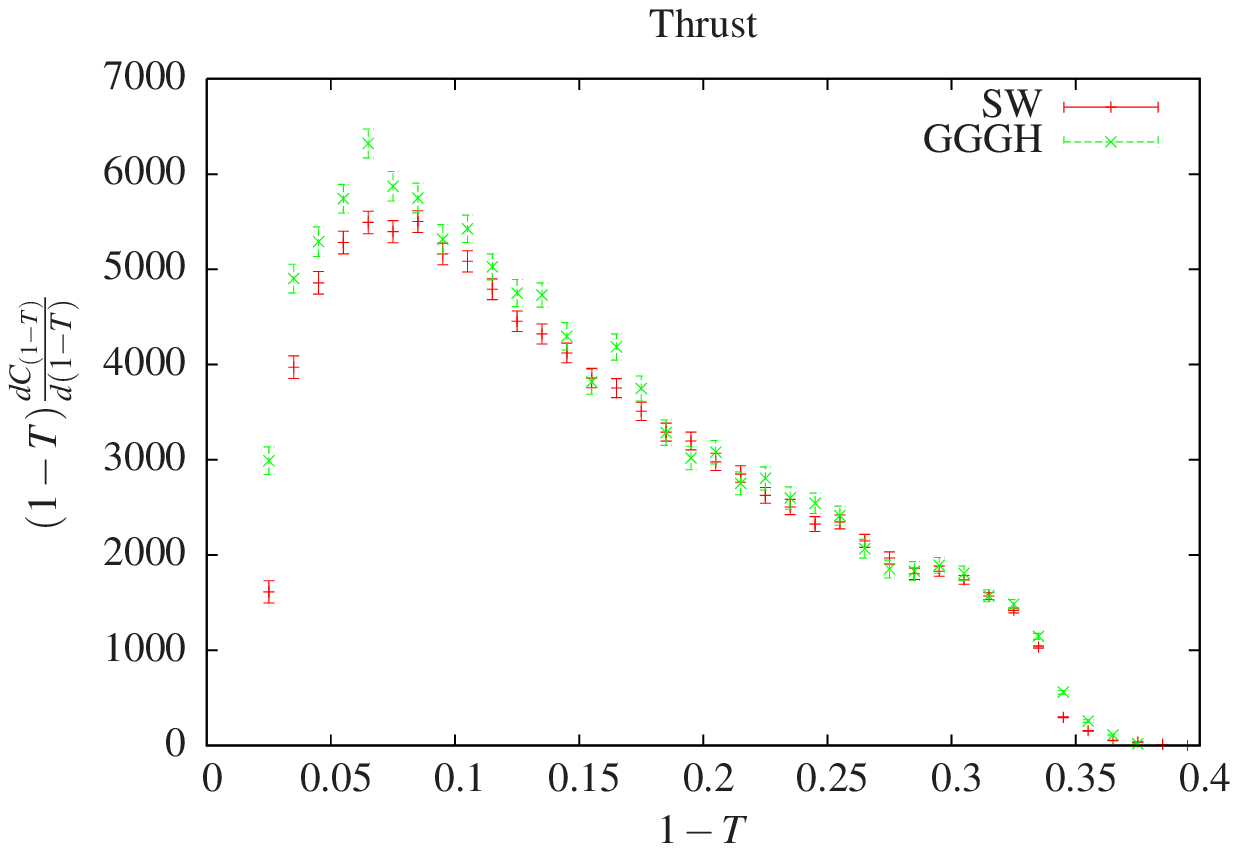}
\includegraphics[bb= 125 460 490 710,width=0.32\textwidth]{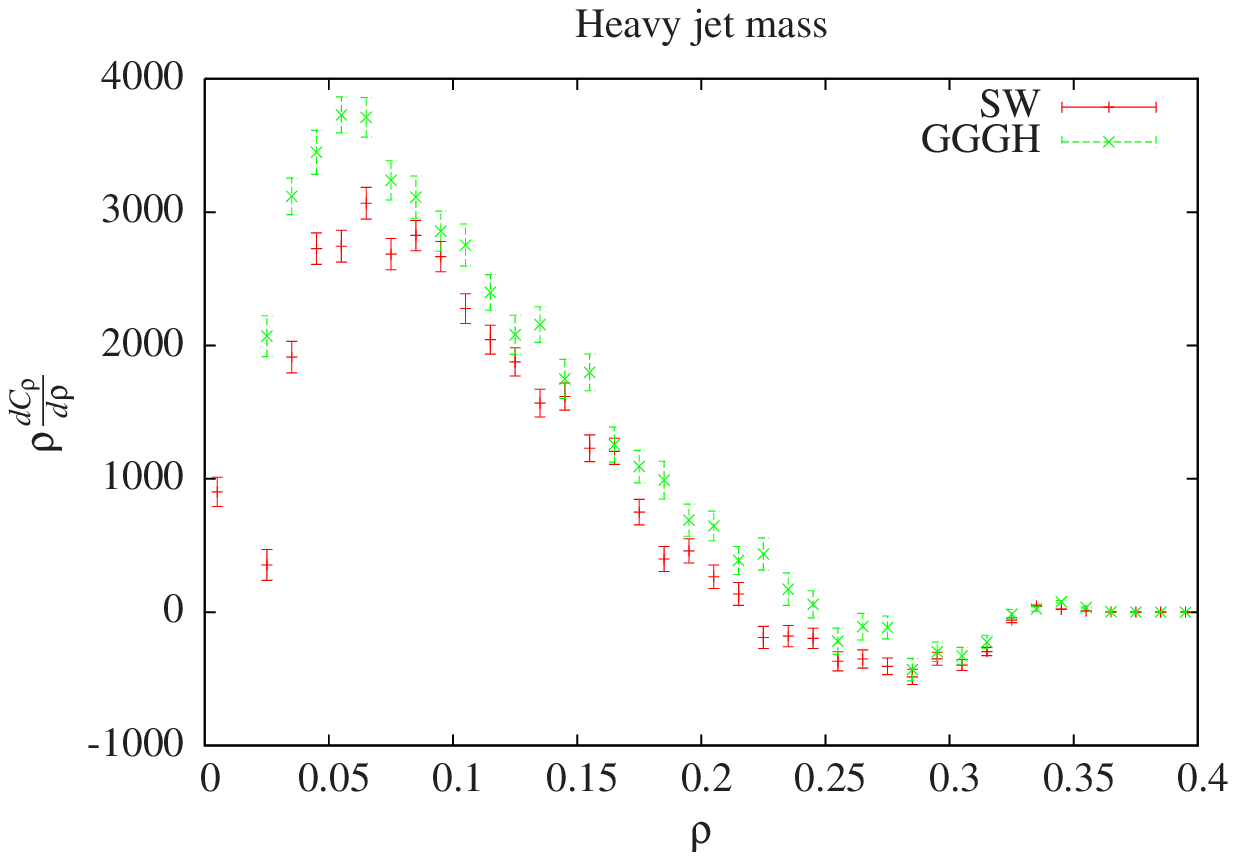}
\includegraphics[bb= 125 460 490 710,width=0.32\textwidth]{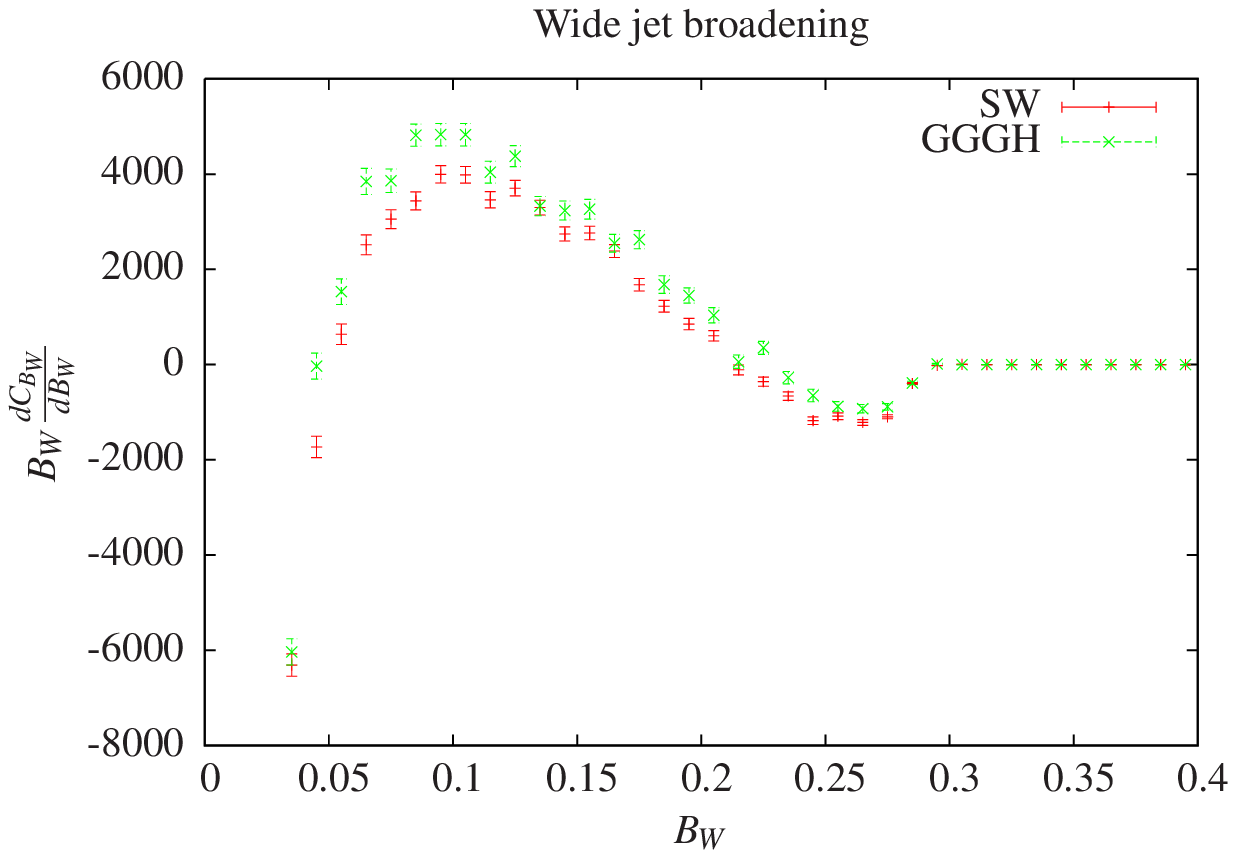}
\\
\includegraphics[bb= 125 460 490 710,width=0.32\textwidth]{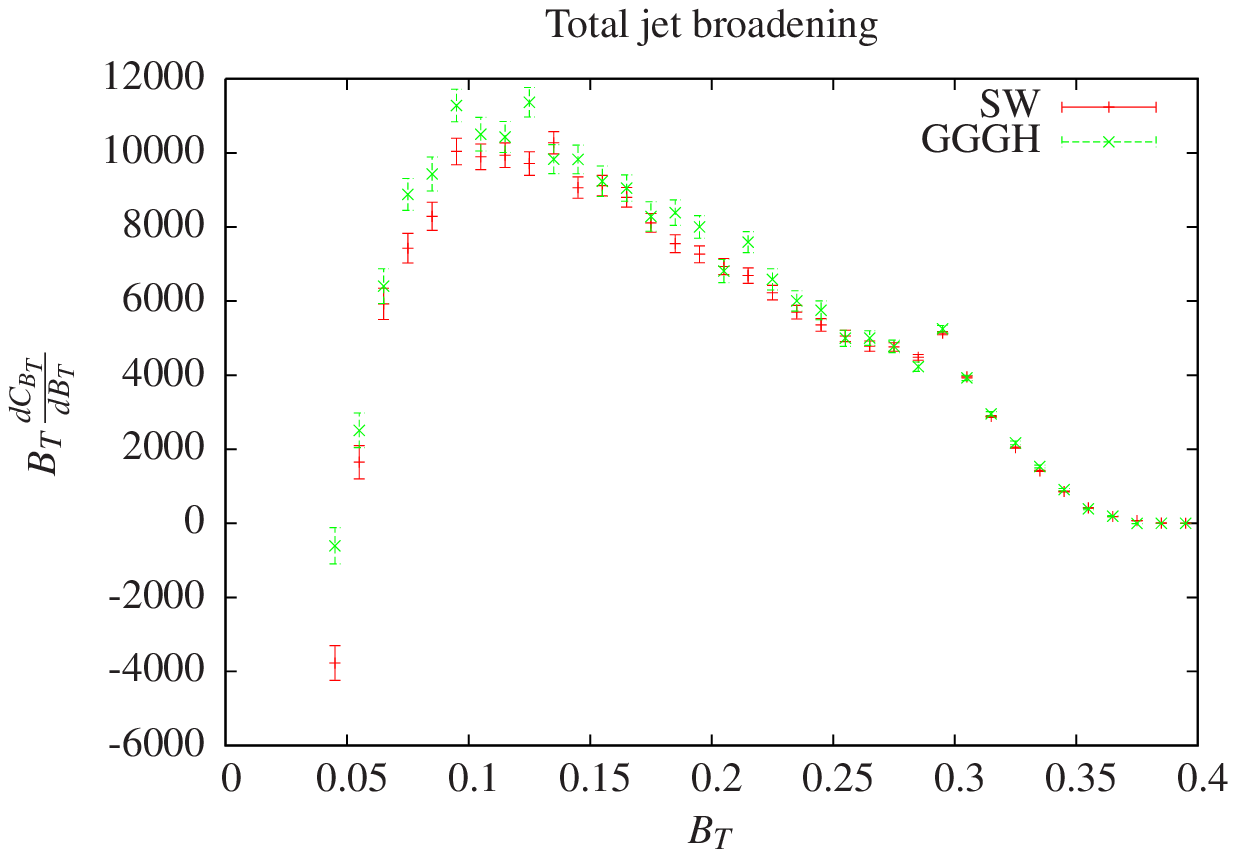}
\includegraphics[bb= 125 460 490 710,width=0.32\textwidth]{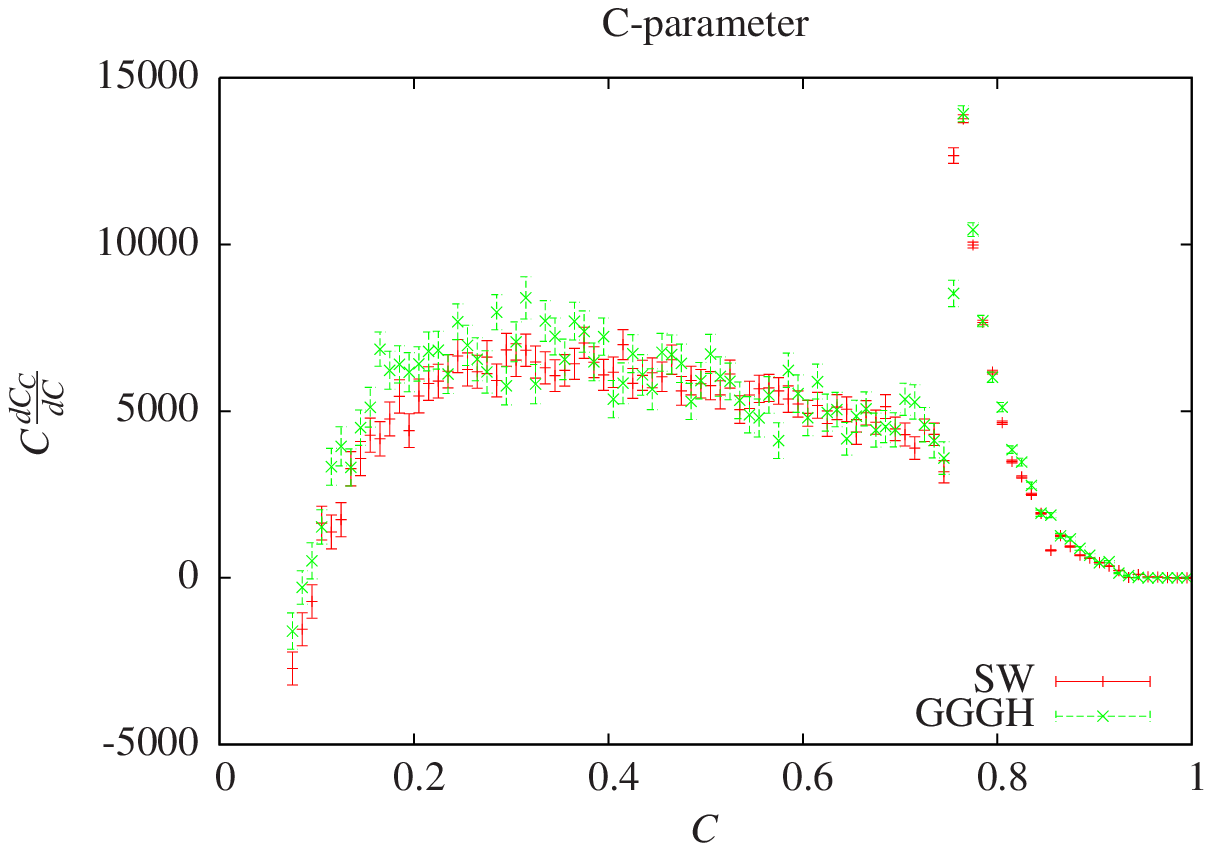}
\includegraphics[bb= 125 460 490 710,width=0.32\textwidth]{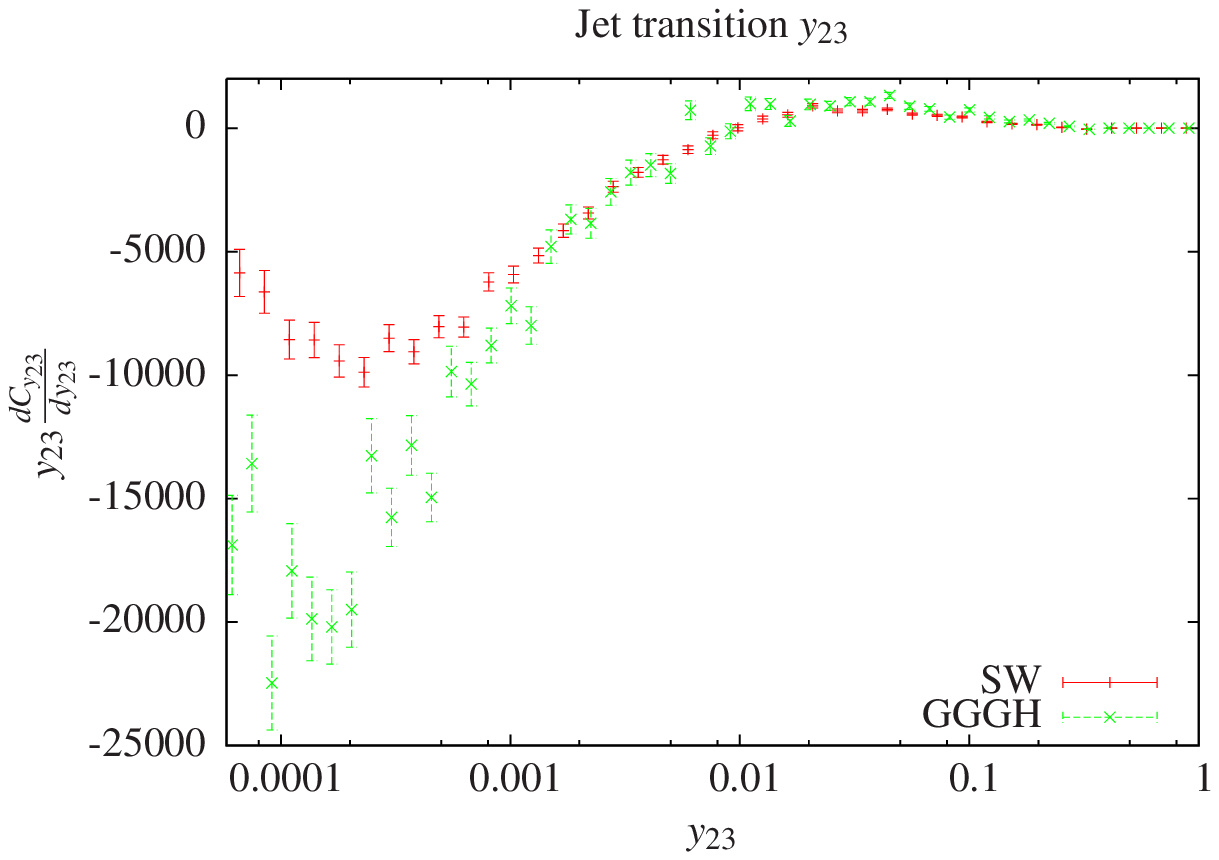}
\end{center}
\caption{
Comparison of the next-to-next-to-leading order coefficient $C_{\cal O}$
for the six distributions between ref.~\cite{GehrmannDeRidder:2007hr} and the present work.
}
\label{fig_comparison}
\end{figure}

Fig.~\ref{fig_comparison} compares the results of this calculation to the updated ones from
ref.~\cite{GehrmannDeRidder:2007hr}.
For the six event shape distributions one observes good agreement with the exception
of the three-to-two jet transition distribution at very small values of $y_{23}$.
The discrepancy at small values of $y_{23}$ can be traced back to the colour factor $N_c^2$
and could have an explanation in terms of a systematic error due to the slicing parameter.
The study of the $\eta$-dependence in fig.~\ref{fig_y23_C_col_eta} gives hints in this direction,
but does not give a conclusive answer. Although in fig.~\ref{fig_y23_C_col_eta} the three results
for $\eta=10^{-5}$, $\eta=10^{-7}$ and $\eta=10^{-9}$ are consistent with each other, using
$\eta=10^{-9}$ instead of $\eta=10^{-5}$ for the comparison with ref.~\cite{GehrmannDeRidder:2007hr}
would reduce the discrepancy significantly (but also enlarge the errorbars).
Given the complexity of the calculation the agreement of the two numerical programs
in all other distributions is remarkable.

\section{Conclusions}
\label{sec:conclusions}

In this article I reported on the NNLO calculation of observables associated to three-jet events
in electron-positron annihilation.
I provided NNLO results for the event shape variables
thrust, heavy jet mass, wide jet broadening,
total jet broadening, C parameter and the Durham three-to-two jet transition variable.
In addition the NNLO results for the jet rates defined in the schemes of Durham, Geneva,
Jade-E0 and Cambridge were given.
The results of this paper will be useful for an extraction of $\alpha_s$ from three-jet quantities.

\subsection*{Acknowledgements}

I would like to thank Th.~Gehrmann, G.~Heinrich, S. Kluth, Ch.~Pahl and J.~Schieck
for useful discussions.
In particular I would like to thank E.~Sch\"omer for pointing out the method for the determination of the thrust
axis to me.
The computer support from the Max-Planck-Institut for Physics is greatly acknowledged.


\begin{appendix}

\section{Calculation of the thrust for a single parton event}
\label{sec:thrust_algo}

For the event shape variable thrust and the ones related to it, it is necessary to determine
the maximum 
\bq
 \max\limits_{\vec{n}} \sum\limits_j \left| \vec{p}_j \cdot \vec{n} \right|
\eq
over all orientations of the unit vector $\vec{n}$.
The sum runs over all final state particles $j$.
For the NNLO calculation we have either $3$, $4$ or $5$ partons in the final state.
Of course there are numerical algorithms which can be used to find a local maximum.
However the computational cost for such a minimisation/maximisation is comparable
to the cost for the matrix elements and therefore not negligible.
Furthermore it is non-trivial to ensure that the found maximum is actually the global maximum.

There is a better way to calculate thrust and the thrust axis for a small number of final state particles.
This method finds the exact global maximum in $2^{N-1}-1$ steps for $N$ final state particles.
Let us define $N$ signs $s_j \in \{-1,1\}$ by
\bq
\label{def_signs_for_thrust}
 \vec{p}_j \cdot \vec{n} & = & s_j \left| \vec{p}_j \cdot \vec{n} \right|.
\eq
Of course the correct values for the signs $s_j$ are only known once the thrust axis $\vec{n}$ is known.
But we know in advance that the $N$-tuple $(s_1,s_2,...,s_N)$ will be one configuration out of the
$2^N$ possible ones.  
We can now step over all $2^N$ possibilities $(s_1,s_2,...,s_N)$.
For a given $N$-tuple $(s_1,s_2,...,s_N)$ we define
\bq
 \vec{P} & = & s_1 \vec{p}_1 + s_2 \vec{p_2} + ... + s_N \vec{p}_N.
\eq
We then have
\bq
 \max\limits_{\vec{n}} \sum\limits_j \left| \vec{p}_j \cdot \vec{n} \right|
 & = & 
 \max\limits_{\vec{n}} \sum\limits_j s_j \vec{p}_j \cdot \vec{n} 
 =
 \max\limits_{\vec{n}} \; \vec{P} \cdot \vec{n} 
\eq
This expression is maximised for $\vec{n}=\vec{P}/|\vec{P}|$.
Next, one checks if the $N$-tuple $(s_1,s_2,...,s_N)$ is actually allowed by verifying eq.~(\ref{def_signs_for_thrust})
for all $j=1,...,N$.
The maximum is then given by the maximum value obtained from all allowed $N$-tuples.
It is actually sufficient to restrict the search to $2^{N-1}-1$ possibilities, since thrust is invariant under
\bq
\vec{n} & \rightarrow & - \vec{n},
\eq
and the two cases where all signs are equal can be excluded due to momentum conservation.
With this algorithm we have to check for five-parton final states $15$ configurations to find the global maximum,
whereas for four-parton final states $7$ configurations have to be checked.
For the case of three massless particles in the final state the algorithm does not need to be used, since there
is a general formula for the thrust axis:
The thrust axis is given in this case by the direction of the most energetic particle \cite{Banfi:2000si}.

\end{appendix}


\bibliography{/home/stefanw/notes/biblio}
\bibliographystyle{/home/stefanw/latex-style/h-physrev3}

%
%

\clearpage

%
%
%


\clearpage

%
%
%

\clearpage

%
%
%

\clearpage

%
%
%

\clearpage

%
%
%

\clearpage

\clearpage

\clearpage

%
%
%
\begin{table}[p]
\begin{center}
{\scriptsize
\begin{tabular}{|r|c|c|c|}
\hline
 $(1-T)$ & $(1-T) \frac{dA_{(1-T)}}{d\left(1-T\right)}$ & $(1-T) \frac{dB_{(1-T)}}{d\left(1-T\right)}$ & $(1-T) \frac{dC_{(1-T)}}{d\left(1-T\right)}$ \\
\hline 
$0.005$ & $2.5797(2) \cdot 10^{1}$ & $-4.957(3) \cdot 10^{2}$ & $-9.0(1) \cdot 10^{3}$ \\
$0.015$ & $1.8299(1) \cdot 10^{1}$ & $2.166(3) \cdot 10^{2}$ & $-3.3(1) \cdot 10^{3}$ \\
$0.025$ & $1.5411(1) \cdot 10^{1}$ & $2.809(3) \cdot 10^{2}$ & $1.6(1) \cdot 10^{3}$ \\
$0.035$ & $1.35230(9) \cdot 10^{1}$ & $2.920(3) \cdot 10^{2}$ & $4.0(1) \cdot 10^{3}$ \\
$0.045$ & $1.21030(9) \cdot 10^{1}$ & $2.877(2) \cdot 10^{2}$ & $4.9(1) \cdot 10^{3}$ \\
$0.055$ & $1.09692(8) \cdot 10^{1}$ & $2.772(2) \cdot 10^{2}$ & $5.3(1) \cdot 10^{3}$ \\
$0.065$ & $1.00228(8) \cdot 10^{1}$ & $2.638(2) \cdot 10^{2}$ & $5.5(1) \cdot 10^{3}$ \\
\hline
\end{tabular}
}
\caption{\label{table_thrust}
Coefficients of the leading-order ($A_{(1-T)}$), 
next-to-leading-order ($B_{(1-T)}$)
and next-to-next-to-leading order ($C_{(1-T)}$)
contributions to the thrust distribution.
}
\end{center}
\end{table}
\begin{table}[p]
\begin{center}
{\scriptsize
\begin{tabular}{|r|c|c|c|}
\hline
 $\rho$ & $\rho \frac{dA_{\rho}}{d\rho}$ & $\rho \frac{dB_{\rho}}{d\rho}$ & $\rho \frac{dC_{\rho}}{d\rho}$ \\
\hline 
$0.005$ & $2.5797(2) \cdot 10^{1}$ & $-6.243(3) \cdot 10^{2}$ & $9(1) \cdot 10^{2}$ \\
$0.015$ & $1.8299(1) \cdot 10^{1}$ & $1.162(3) \cdot 10^{2}$ & $-3.0(1) \cdot 10^{3}$ \\
$0.025$ & $1.5411(1) \cdot 10^{1}$ & $1.917(3) \cdot 10^{2}$ & $4(1) \cdot 10^{2}$ \\
$0.035$ & $1.35230(9) \cdot 10^{1}$ & $2.088(3) \cdot 10^{2}$ & $1.9(1) \cdot 10^{3}$ \\
$0.045$ & $1.21030(9) \cdot 10^{1}$ & $2.092(2) \cdot 10^{2}$ & $2.7(1) \cdot 10^{3}$ \\
$0.055$ & $1.09692(8) \cdot 10^{1}$ & $2.020(2) \cdot 10^{2}$ & $2.7(1) \cdot 10^{3}$ \\
$0.065$ & $1.00228(8) \cdot 10^{1}$ & $1.918(2) \cdot 10^{2}$ & $3.1(1) \cdot 10^{3}$ \\
\hline
\end{tabular}
}
\caption{\label{table_heavyjetmass}
Coefficients of the leading-order ($A_{\rho}$), 
next-to-leading-order ($B_{\rho}$)
and next-to-next-to-leading order ($C_{\rho}$)
contributions to the heavy jet mass distribution.
}
\end{center}
\end{table}
\begin{table}[p]
\begin{center}
{\scriptsize
\begin{tabular}{|r|c|c|c|}
\hline
 $B_W$ & $B_W \frac{dA_{B_W}}{dB_W}$ & $B_W \frac{dB_{B_W}}{dB_W}$ & $B_W \frac{dC_{B_W}}{dB_W}$ \\
\hline 
$0.005$ & $5.146(1) \cdot 10^{1}$ & $-4.194(2) \cdot 10^{3}$ & $1.433(7) \cdot 10^{5}$ \\
$0.015$ & $3.6823(4) \cdot 10^{1}$ & $-6.385(7) \cdot 10^{2}$ & $-2.33(3) \cdot 10^{4}$ \\
$0.025$ & $3.1132(3) \cdot 10^{1}$ & $-1.015(5) \cdot 10^{2}$ & $-1.34(2) \cdot 10^{4}$ \\
$0.035$ & $2.7389(2) \cdot 10^{1}$ & $1.126(5) \cdot 10^{2}$ & $-6.3(2) \cdot 10^{3}$ \\
$0.045$ & $2.4563(2) \cdot 10^{1}$ & $2.136(4) \cdot 10^{2}$ & $-1.7(2) \cdot 10^{3}$ \\
$0.055$ & $2.2291(2) \cdot 10^{1}$ & $2.634(4) \cdot 10^{2}$ & $6(2) \cdot 10^{2}$ \\
$0.065$ & $2.0389(2) \cdot 10^{1}$ & $2.867(4) \cdot 10^{2}$ & $2.5(2) \cdot 10^{3}$ \\
\hline
\end{tabular}
}
\caption{\label{table_widejetbroadening}
Coefficients of the leading-order ($A_{B_W}$), 
next-to-leading-order ($B_{B_W}$)
and next-to-next-to-leading order ($C_{B_W}$)
contributions to the wide jet broadening distribution.
}
\end{center}
\end{table}
\begin{table}[p]
\begin{center}
{\scriptsize
\begin{tabular}{|r|c|c|c|}
\hline
 $B_T$ & $B_T \frac{dA_{B_T}}{dB_T}$ & $B_T \frac{dB_{B_T}}{dB_T}$ & $B_T \frac{dC_{B_T}}{dB_T}$ \\
\hline 
$0.005$ & $5.146(1) \cdot 10^{1}$ & $-3.776(2) \cdot 10^{3}$ & $9.16(8) \cdot 10^{4}$ \\
$0.015$ & $3.6823(4) \cdot 10^{1}$ & $-2.65(2) \cdot 10^{2}$ & $-5.59(5) \cdot 10^{4}$ \\
$0.025$ & $3.1132(3) \cdot 10^{1}$ & $2.40(2) \cdot 10^{2}$ & $-2.77(5) \cdot 10^{4}$ \\
$0.035$ & $2.7389(2) \cdot 10^{1}$ & $4.24(2) \cdot 10^{2}$ & $-1.24(5) \cdot 10^{4}$ \\
$0.045$ & $2.4563(2) \cdot 10^{1}$ & $5.11(1) \cdot 10^{2}$ & $-3.8(5) \cdot 10^{3}$ \\
$0.055$ & $2.2291(2) \cdot 10^{1}$ & $5.42(1) \cdot 10^{2}$ & $1.6(4) \cdot 10^{3}$ \\
$0.065$ & $2.0389(2) \cdot 10^{1}$ & $5.53(1) \cdot 10^{2}$ & $5.9(4) \cdot 10^{3}$ \\
\hline
\end{tabular}
}
\caption{\label{table_totaljetbroadening}
Coefficients of the leading-order ($A_{B_T}$), 
next-to-leading-order ($B_{B_T}$)
and next-to-next-to-leading order ($C_{B_T}$)
contributions to the total jet broadening distribution.
}
\end{center}
\end{table}
\begin{table}[p]
\begin{center}
{\scriptsize
\begin{tabular}{|r|c|c|c|}
\hline
 $C$ & $C \frac{dA_{C}}{dC}$ & $C \frac{dB_{C}}{dC}$ & $C \frac{dC_{C}}{dC}$ \\
\hline 
$0.005$ & $3.5327(6) \cdot 10^{1}$ & $-2.151(1) \cdot 10^{3}$ & $3.06(4) \cdot 10^{4}$ \\
$0.015$ & $2.7963(3) \cdot 10^{1}$ & $-4.75(1) \cdot 10^{2}$ & $-3.04(4) \cdot 10^{4}$ \\
$0.025$ & $2.5124(3) \cdot 10^{1}$ & $-1.45(1) \cdot 10^{2}$ & $-2.15(4) \cdot 10^{4}$ \\
$0.035$ & $2.3269(2) \cdot 10^{1}$ & $1.3(1) \cdot 10^{1}$ & $-1.48(4) \cdot 10^{4}$ \\
$0.045$ & $2.1887(2) \cdot 10^{1}$ & $1.09(1) \cdot 10^{2}$ & $-1.12(5) \cdot 10^{4}$ \\
$0.055$ & $2.0774(2) \cdot 10^{1}$ & $1.71(1) \cdot 10^{2}$ & $-7.6(5) \cdot 10^{3}$ \\
$0.065$ & $1.9831(2) \cdot 10^{1}$ & $2.15(1) \cdot 10^{2}$ & $-5.3(5) \cdot 10^{3}$ \\
\hline
\end{tabular}
}
\caption{\label{table_Cparameter_a}
Coefficients of the leading-order ($A_{C}$), 
next-to-leading-order ($B_{C}$)
and next-to-next-to-leading order ($C_{C}$)
contributions to the $C$-parameter distribution.
}
\end{center}
\end{table}
\begin{table}[p]
\begin{center}
{\scriptsize
\begin{tabular}{|r|c|c|c|}
\hline
 $C$ & $C \frac{dA_{C}}{dC}$ & $C \frac{dB_{C}}{dC}$ & $C \frac{dC_{C}}{dC}$ \\
\hline 
$0.505$ & $8.033(1) \cdot 10^{0}$ & $2.417(7) \cdot 10^{2}$ & $5.8(4) \cdot 10^{3}$ \\
$0.515$ & $7.918(1) \cdot 10^{0}$ & $2.376(7) \cdot 10^{2}$ & $5.5(4) \cdot 10^{3}$ \\
$0.525$ & $7.804(1) \cdot 10^{0}$ & $2.326(7) \cdot 10^{2}$ & $6.1(4) \cdot 10^{3}$ \\
$0.535$ & $7.691(1) \cdot 10^{0}$ & $2.312(7) \cdot 10^{2}$ & $5.0(4) \cdot 10^{3}$ \\
$0.545$ & $7.589(1) \cdot 10^{0}$ & $2.285(7) \cdot 10^{2}$ & $5.5(4) \cdot 10^{3}$ \\
$0.555$ & $7.476(1) \cdot 10^{0}$ & $2.247(7) \cdot 10^{2}$ & $5.7(4) \cdot 10^{3}$ \\
$0.565$ & $7.377(1) \cdot 10^{0}$ & $2.227(7) \cdot 10^{2}$ & $5.7(4) \cdot 10^{3}$ \\
\hline
\end{tabular}
}
\caption{\label{table_Cparameter_b}
Coefficients of the leading-order ($A_{C}$), 
next-to-leading-order ($B_{C}$)
and next-to-next-to-leading order ($C_{C}$)
contributions to the $C$-parameter distribution.
}
\end{center}
\end{table}
\begin{table}[p]
\begin{center}
{\scriptsize
\begin{tabular}{|r|c|c|c|}
\hline
 $\ln y_{23}$ & $y_{23} \frac{dA_{y_{23}}}{dy_{23}}$ & $y_{23} \frac{dB_{y_{23}}}{dy_{23}}$ & $y_{23} \frac{dC_{y_{23}}}{dy_{23}}$ \\
\hline 
$-9.875$ & $2.222(1) \cdot 10^{1}$ & $-1.133(3) \cdot 10^{3}$ & $-2(1) \cdot 10^{3}$ \\
$-9.625$ & $2.157(1) \cdot 10^{1}$ & $-1.004(3) \cdot 10^{3}$ & $-6(1) \cdot 10^{3}$ \\
$-9.375$ & $2.094(1) \cdot 10^{1}$ & $-8.81(3) \cdot 10^{2}$ & $-6.6(9) \cdot 10^{3}$ \\
$-9.125$ & $2.031(1) \cdot 10^{1}$ & $-7.63(2) \cdot 10^{2}$ & $-8.6(8) \cdot 10^{3}$ \\
$-8.875$ & $1.9650(8) \cdot 10^{1}$ & $-6.58(2) \cdot 10^{2}$ & $-8.6(7) \cdot 10^{3}$ \\
$-8.625$ & $1.8987(7) \cdot 10^{1}$ & $-5.63(2) \cdot 10^{2}$ & $-9.4(7) \cdot 10^{3}$ \\
$-8.375$ & $1.8361(6) \cdot 10^{1}$ & $-4.74(2) \cdot 10^{2}$ & $-9.9(6) \cdot 10^{3}$ \\
\hline
\end{tabular}
}
\caption{\label{table_y23}
Coefficients of the leading-order ($A_{y_{23}}$), 
next-to-leading-order ($B_{y_{23}}$)
and next-to-next-to-leading order ($C_{y_{23}}$)
contributions to the three-to-two jet transition distribution.
}
\end{center}
\end{table}
%
%
%
\clearpage
\begin{table}[p]
\begin{center}
{\scriptsize
\begin{tabular}{|r|c|c|c|}
\hline
 $y_{cut}$ & $A_{3-jet,Durham}$ & $B_{3-jet,Durham}$ & $C_{3-jet,Durham}$ \\
\hline 
$0.3$ & $2.2301(9) \cdot 10^{-2}$ & $1.35(2) \cdot 10^{-1}$ & $-6(3) \cdot 10^{0}$ \\
$0.15$ & $1.0028(3) \cdot 10^{0}$ & $1.582(3) \cdot 10^{1}$ & $5(1) \cdot 10^{1}$ \\
$0.1$ & $2.1150(6) \cdot 10^{0}$ & $3.415(7) \cdot 10^{1}$ & $1.7(2) \cdot 10^{2}$ \\
$0.06$ & $4.046(1) \cdot 10^{0}$ & $6.57(1) \cdot 10^{1}$ & $4.1(3) \cdot 10^{2}$ \\
$0.03$ & $7.627(2) \cdot 10^{0}$ & $1.141(2) \cdot 10^{2}$ & $5.0(6) \cdot 10^{2}$ \\
$0.015$ & $1.2359(3) \cdot 10^{1}$ & $1.495(3) \cdot 10^{2}$ & $-1.2(9) \cdot 10^{2}$ \\
$0.01$ & $1.5671(4) \cdot 10^{1}$ & $1.531(4) \cdot 10^{2}$ & $-1.3(1) \cdot 10^{3}$ \\
$0.006$ & $2.0430(5) \cdot 10^{1}$ & $1.238(5) \cdot 10^{2}$ & $-3.6(2) \cdot 10^{3}$ \\
$0.003$ & $2.7947(7) \cdot 10^{1}$ & $-5.5(7) \cdot 10^{0}$ & $-9.0(3) \cdot 10^{3}$ \\
$0.0015$ & $3.6694(8) \cdot 10^{1}$ & $-2.93(1) \cdot 10^{2}$ & $-1.63(5) \cdot 10^{4}$ \\
$0.001$ & $4.2386(9) \cdot 10^{1}$ & $-5.62(1) \cdot 10^{2}$ & $-2.16(6) \cdot 10^{4}$ \\
$0.0006$ & $5.016(1) \cdot 10^{1}$ & $-1.032(2) \cdot 10^{3}$ & $-2.84(8) \cdot 10^{4}$ \\
$0.0003$ & $6.181(1) \cdot 10^{1}$ & $-1.967(2) \cdot 10^{3}$ & $-3.1(1) \cdot 10^{4}$ \\
$0.00015$ & $7.471(2) \cdot 10^{1}$ & $-3.333(2) \cdot 10^{3}$ & $-1.8(3) \cdot 10^{4}$ \\
$0.0001$ & $8.285(2) \cdot 10^{1}$ & $-4.368(3) \cdot 10^{3}$ & $6(4) \cdot 10^{3}$ \\
\hline
\end{tabular}
}
\caption{\label{table_jet_rate_durham}
Coefficients of the leading-order ($A_{3-jet,Durham}$), 
next-to-leading-order ($B_{3-jet,Durham}$)
and next-to-next-to-leading order ($C_{3-jet,Durham}$)
contributions to the Durham three-jet rate for various values of $y_{cut}$.
}
\end{center}
\end{table}
\begin{table}[p]
\begin{center}
{\scriptsize
\begin{tabular}{|r|c|c|c|}
\hline
 $y_{cut}$ & $A_{3-jet,Geneva}$ & $B_{3-jet,Geneva}$ & $C_{3-jet,Geneva}$ \\
\hline 
$0.3$ & $1.7615(7) \cdot 10^{-1}$ & $2.10(1) \cdot 10^{0}$ & $-1.3(6) \cdot 10^{1}$ \\
$0.15$ & $4.7265(9) \cdot 10^{0}$ & $5.45(1) \cdot 10^{1}$ & $-4.4(6) \cdot 10^{2}$ \\
$0.1$ & $8.922(1) \cdot 10^{0}$ & $7.51(2) \cdot 10^{1}$ & $-1.3(1) \cdot 10^{3}$ \\
$0.06$ & $1.5520(3) \cdot 10^{1}$ & $3.87(4) \cdot 10^{1}$ & $-5.0(2) \cdot 10^{3}$ \\
$0.03$ & $2.6717(4) \cdot 10^{1}$ & $-2.403(7) \cdot 10^{2}$ & $-1.35(4) \cdot 10^{4}$ \\
$0.015$ & $4.0440(5) \cdot 10^{1}$ & $-9.725(7) \cdot 10^{2}$ & $-2.53(7) \cdot 10^{4}$ \\
$0.01$ & $4.9627(6) \cdot 10^{1}$ & $-1.709(2) \cdot 10^{3}$ & $-2.86(8) \cdot 10^{4}$ \\
$0.006$ & $6.2446(7) \cdot 10^{1}$ & $-3.072(3) \cdot 10^{3}$ & $-2.5(1) \cdot 10^{4}$ \\
$0.003$ & $8.204(1) \cdot 10^{1}$ & $-5.937(5) \cdot 10^{3}$ & $1.4(2) \cdot 10^{4}$ \\
$0.0015$ & $1.0418(1) \cdot 10^{2}$ & $-1.0300(6) \cdot 10^{4}$ & $1.53(4) \cdot 10^{5}$ \\
$0.001$ & $1.1833(1) \cdot 10^{2}$ & $-1.3736(9) \cdot 10^{4}$ & $2.81(5) \cdot 10^{5}$ \\
\hline
\end{tabular}
}
\caption{\label{table_jet_rate_geneva}
Coefficients of the leading-order ($A_{3-jet,Geneva}$), 
next-to-leading-order ($B_{3-jet,Geneva}$)
and next-to-next-to-leading order ($C_{3-jet,Geneva}$)
contributions to the Durham three-jet rate for various values of $y_{cut}$.
}
\end{center}
\end{table}
\begin{table}[p]
\begin{center}
{\scriptsize
\begin{tabular}{|r|c|c|c|}
\hline
 $y_{cut}$ & $A_{3-jet,Jade-E0}$ & $B_{3-jet,Jade-E0}$ & $C_{3-jet,Jade-E0}$ \\
\hline 
$0.3$ & $5.393(2) \cdot 10^{-2}$ & $1.119(5) \cdot 10^{0}$ & $9(8) \cdot 10^{0}$ \\
$0.15$ & $2.3086(5) \cdot 10^{0}$ & $5.293(6) \cdot 10^{1}$ & $5.3(4) \cdot 10^{2}$ \\
$0.1$ & $4.9170(9) \cdot 10^{0}$ & $1.118(1) \cdot 10^{2}$ & $9(1) \cdot 10^{2}$ \\
$0.06$ & $9.527(2) \cdot 10^{0}$ & $1.994(2) \cdot 10^{2}$ & $1.5(2) \cdot 10^{3}$ \\
$0.03$ & $1.8137(3) \cdot 10^{1}$ & $2.885(4) \cdot 10^{2}$ & $-8(3) \cdot 10^{2}$ \\
$0.015$ & $2.9425(4) \cdot 10^{1}$ & $2.397(8) \cdot 10^{2}$ & $-9.1(6) \cdot 10^{3}$ \\
$0.01$ & $3.7257(5) \cdot 10^{1}$ & $9.1(1) \cdot 10^{1}$ & $-1.88(6) \cdot 10^{4}$ \\
$0.006$ & $4.8389(6) \cdot 10^{1}$ & $-2.92(2) \cdot 10^{2}$ & $-2.9(1) \cdot 10^{4}$ \\
$0.003$ & $6.5768(9) \cdot 10^{1}$ & $-1.300(3) \cdot 10^{3}$ & $-4.6(2) \cdot 10^{4}$ \\
$0.0015$ & $8.572(1) \cdot 10^{1}$ & $-3.092(4) \cdot 10^{3}$ & $-4.6(2) \cdot 10^{4}$ \\
$0.001$ & $9.859(1) \cdot 10^{1}$ & $-4.603(5) \cdot 10^{3}$ & $-2.9(4) \cdot 10^{4}$ \\
$0.0006$ & $1.1607(1) \cdot 10^{2}$ & $-7.143(7) \cdot 10^{3}$ & $3.2(4) \cdot 10^{4}$ \\
$0.0003$ & $1.4199(2) \cdot 10^{2}$ & $-1.190(1) \cdot 10^{4}$ & $2.30(8) \cdot 10^{5}$ \\
\hline
\end{tabular}
}
\caption{\label{table_jet_rate_jadeE0}
Coefficients of the leading-order ($A_{3-jet,Jade-E0}$), 
next-to-leading-order ($B_{3-jet,Jade-E0}$)
and next-to-next-to-leading order ($C_{3-jet,Jade-E0}$)
contributions to the Durham three-jet rate for various values of $y_{cut}$.
}
\end{center}
\end{table}
\begin{table}[p]
\begin{center}
{\scriptsize
\begin{tabular}{|r|c|c|c|}
\hline
 $y_{cut}$ & $A_{3-jet,Cambridge}$ & $B_{3-jet,Cambridge}$ & $C_{3-jet,Cambridge}$ \\
\hline 
$0.3$ & $2.2301(9) \cdot 10^{-2}$ & $1.35(2) \cdot 10^{-1}$ & $-7(3) \cdot 10^{0}$ \\
$0.15$ & $1.0028(3) \cdot 10^{0}$ & $1.525(3) \cdot 10^{1}$ & $2(3) \cdot 10^{1}$ \\
$0.1$ & $2.1150(6) \cdot 10^{0}$ & $3.156(6) \cdot 10^{1}$ & $9(6) \cdot 10^{1}$ \\
$0.06$ & $4.046(1) \cdot 10^{0}$ & $5.72(1) \cdot 10^{1}$ & $4(9) \cdot 10^{1}$ \\
$0.03$ & $7.627(2) \cdot 10^{0}$ & $9.33(2) \cdot 10^{1}$ & $-2(2) \cdot 10^{2}$ \\
$0.015$ & $1.2359(3) \cdot 10^{1}$ & $1.127(3) \cdot 10^{2}$ & $-1.1(2) \cdot 10^{3}$ \\
$0.01$ & $1.5671(4) \cdot 10^{1}$ & $1.048(4) \cdot 10^{2}$ & $-2.3(3) \cdot 10^{3}$ \\
$0.006$ & $2.0430(5) \cdot 10^{1}$ & $5.96(6) \cdot 10^{1}$ & $-3.9(4) \cdot 10^{3}$ \\
$0.003$ & $2.7947(7) \cdot 10^{1}$ & $-9.66(7) \cdot 10^{1}$ & $-9.6(6) \cdot 10^{3}$ \\
$0.0015$ & $3.6694(8) \cdot 10^{1}$ & $-4.15(1) \cdot 10^{2}$ & $-1.41(8) \cdot 10^{4}$ \\
$0.001$ & $4.2386(9) \cdot 10^{1}$ & $-7.03(1) \cdot 10^{2}$ & $-1.7(1) \cdot 10^{4}$ \\
$0.0006$ & $5.016(1) \cdot 10^{1}$ & $-1.201(2) \cdot 10^{3}$ & $-1.8(1) \cdot 10^{4}$ \\
$0.0003$ & $6.181(1) \cdot 10^{1}$ & $-2.177(2) \cdot 10^{3}$ & $-1.3(2) \cdot 10^{4}$ \\
$0.00015$ & $7.471(2) \cdot 10^{1}$ & $-3.584(3) \cdot 10^{3}$ & $9(2) \cdot 10^{3}$ \\
$0.0001$ & $8.285(2) \cdot 10^{1}$ & $-4.656(4) \cdot 10^{3}$ & $3.3(3) \cdot 10^{4}$ \\
\hline
\end{tabular}
}
\caption{\label{table_jet_rate_cambridge}
Coefficients of the leading-order ($A_{3-jet,Cambridge}$), 
next-to-leading-order ($B_{3-jet,Cambridge}$)
and next-to-next-to-leading order ($C_{3-jet,Cambridge}$)
contributions to the Durham three-jet rate for various values of $y_{cut}$.
}
\end{center}
\end{table}
%
%
%
\clearpage
\begin{table}[p]
\begin{center}
{\scriptsize
\begin{tabular}{|r|c|c|c|c|c|c|}
\hline
 $(1-T)$ & $N_c^2$ & $N_c^0$ & $N_c^{-2}$ & $N_f N_c$ & $N_f/N_c$ & $N_f^2$ \\
\hline 
$0.005$ & $-1.67(1) \cdot 10^{4}$ & $8.14(2) \cdot 10^{3}$ & $5.73(2) \cdot 10^{2}$ & $-3.85(4) \cdot 10^{3}$ & $-5.412(5) \cdot 10^{3}$ & $8.3281(6) \cdot 10^{3}$ \\
$0.015$ & $8.5(1) \cdot 10^{3}$ & $4.35(3) \cdot 10^{3}$ & $-9.0(3) \cdot 10^{1}$ & $-1.880(4) \cdot 10^{4}$ & $-6.80(6) \cdot 10^{2}$ & $3.4298(3) \cdot 10^{3}$ \\
$0.025$ & $1.35(1) \cdot 10^{4}$ & $2.30(3) \cdot 10^{3}$ & $-9.2(3) \cdot 10^{1}$ & $-1.632(4) \cdot 10^{4}$ & $-1.02(6) \cdot 10^{2}$ & $2.3354(3) \cdot 10^{3}$ \\
$0.035$ & $1.52(1) \cdot 10^{4}$ & $1.24(3) \cdot 10^{3}$ & $-7.7(2) \cdot 10^{1}$ & $-1.425(4) \cdot 10^{4}$ & $1.12(6) \cdot 10^{2}$ & $1.7636(3) \cdot 10^{3}$ \\
$0.045$ & $1.52(1) \cdot 10^{4}$ & $5.4(3) \cdot 10^{2}$ & $-6.3(2) \cdot 10^{1}$ & $-1.243(4) \cdot 10^{4}$ & $2.12(6) \cdot 10^{2}$ & $1.4027(3) \cdot 10^{3}$ \\
$0.055$ & $1.47(1) \cdot 10^{4}$ & $2.0(2) \cdot 10^{2}$ & $-4.7(2) \cdot 10^{1}$ & $-1.101(4) \cdot 10^{4}$ & $2.65(6) \cdot 10^{2}$ & $1.1509(3) \cdot 10^{3}$ \\
$0.065$ & $1.41(1) \cdot 10^{4}$ & $-3(2) \cdot 10^{1}$ & $-4.0(2) \cdot 10^{1}$ & $-9.83(4) \cdot 10^{3}$ & $2.81(5) \cdot 10^{2}$ & $9.646(3) \cdot 10^{2}$ \\
\hline
\end{tabular}
}
\caption{\label{table_thrust_colour}
Individual contributions from the different colour factors to $(1-T) \frac{dC_{(1-T)}}{d\left(1-T\right)}$ for the thrust distribution.
}
\end{center}
\end{table}
\begin{table}[p]
\begin{center}
{\scriptsize
\begin{tabular}{|r|c|c|c|c|c|c|}
\hline
 $\rho$ & $N_c^2$ & $N_c^0$ & $N_c^{-2}$ & $N_f N_c$ & $N_f/N_c$ & $N_f^2$ \\
\hline 
$0.005$ & $-1.10(1) \cdot 10^{4}$ & $6.01(2) \cdot 10^{3}$ & $7.83(1) \cdot 10^{2}$ & $2.82(4) \cdot 10^{3}$ & $-6.094(5) \cdot 10^{3}$ & $8.3694(7) \cdot 10^{3}$ \\
$0.015$ & $5.3(1) \cdot 10^{3}$ & $3.64(3) \cdot 10^{3}$ & $3.2(1) \cdot 10^{1}$ & $-1.435(4) \cdot 10^{4}$ & $-1.137(4) \cdot 10^{3}$ & $3.4802(5) \cdot 10^{3}$ \\
$0.025$ & $9.2(1) \cdot 10^{3}$ & $2.04(2) \cdot 10^{3}$ & $-1.0(1) \cdot 10^{1}$ & $-1.278(4) \cdot 10^{4}$ & $-4.76(4) \cdot 10^{2}$ & $2.3914(4) \cdot 10^{3}$ \\
$0.035$ & $1.03(1) \cdot 10^{4}$ & $1.18(2) \cdot 10^{3}$ & $-1.5(1) \cdot 10^{1}$ & $-1.119(4) \cdot 10^{4}$ & $-2.05(4) \cdot 10^{2}$ & $1.8219(4) \cdot 10^{3}$ \\
$0.045$ & $1.04(1) \cdot 10^{4}$ & $6.9(2) \cdot 10^{2}$ & $-1.6(1) \cdot 10^{1}$ & $-9.71(4) \cdot 10^{3}$ & $-7.6(4) \cdot 10^{1}$ & $1.4625(3) \cdot 10^{3}$ \\
$0.055$ & $9.7(1) \cdot 10^{3}$ & $3.6(2) \cdot 10^{2}$ & $-1.5(1) \cdot 10^{1}$ & $-8.51(4) \cdot 10^{3}$ & $4(4) \cdot 10^{0}$ & $1.2114(3) \cdot 10^{3}$ \\
$0.065$ & $9.3(1) \cdot 10^{3}$ & $1.7(2) \cdot 10^{2}$ & $-1.0(1) \cdot 10^{1}$ & $-7.48(4) \cdot 10^{3}$ & $5.3(4) \cdot 10^{1}$ & $1.0256(3) \cdot 10^{3}$ \\
\hline
\end{tabular}
}
\caption{\label{table_heavyjetmass_colour}
Individual contributions from the different colour factors to $\rho \frac{dC_{\rho}}{d\rho}$ for the heavy jet mass distribution.
}
\end{center}
\end{table}
\begin{table}[p]
\begin{center}
{\scriptsize
\begin{tabular}{|r|c|c|c|c|c|c|}
\hline
 $B_W$ & $N_c^2$ & $N_c^0$ & $N_c^{-2}$ & $N_f N_c$ & $N_f/N_c$ & $N_f^2$ \\
\hline 
$0.005$ & $3.30(6) \cdot 10^{4}$ & $-8.6(1) \cdot 10^{3}$ & $4.845(5) \cdot 10^{3}$ & $1.192(2) \cdot 10^{5}$ & $-2.799(2) \cdot 10^{4}$ & $2.2782(2) \cdot 10^{4}$ \\
$0.015$ & $-3.05(2) \cdot 10^{4}$ & $1.169(6) \cdot 10^{4}$ & $7.19(4) \cdot 10^{2}$ & $-8.55(9) \cdot 10^{3}$ & $-6.85(1) \cdot 10^{3}$ & $1.0195(1) \cdot 10^{4}$ \\
$0.025$ & $-7.5(2) \cdot 10^{3}$ & $8.19(6) \cdot 10^{3}$ & $2.20(3) \cdot 10^{2}$ & $-1.788(8) \cdot 10^{4}$ & $-3.356(9) \cdot 10^{3}$ & $6.8578(7) \cdot 10^{3}$ \\
$0.035$ & $3.8(2) \cdot 10^{3}$ & $5.78(5) \cdot 10^{3}$ & $7.1(3) \cdot 10^{1}$ & $-1.922(7) \cdot 10^{4}$ & $-1.908(8) \cdot 10^{3}$ & $5.1322(6) \cdot 10^{3}$ \\
$0.045$ & $9.9(2) \cdot 10^{3}$ & $3.95(4) \cdot 10^{3}$ & $1.1(2) \cdot 10^{1}$ & $-1.853(6) \cdot 10^{4}$ & $-1.145(7) \cdot 10^{3}$ & $4.0453(6) \cdot 10^{3}$ \\
$0.055$ & $1.25(2) \cdot 10^{4}$ & $2.84(4) \cdot 10^{3}$ & $0(2) \cdot 10^{-1}$ & $-1.731(6) \cdot 10^{4}$ & $-7.03(6) \cdot 10^{2}$ & $3.2925(5) \cdot 10^{3}$ \\
$0.065$ & $1.40(2) \cdot 10^{4}$ & $1.96(4) \cdot 10^{3}$ & $-1.4(2) \cdot 10^{1}$ & $-1.571(6) \cdot 10^{4}$ & $-4.33(6) \cdot 10^{2}$ & $2.7390(5) \cdot 10^{3}$ \\
\hline
\end{tabular}
}
\caption{\label{table_widejetbroadening_colour}
Individual contributions from the different colour factors to $B_W \frac{dC_{B_W}}{dB_W}$ for the wide jet broadening distribution.
}
\end{center}
\end{table}
\begin{table}[p]
\begin{center}
{\scriptsize
\begin{tabular}{|r|c|c|c|c|c|c|}
\hline
 $B_T$ & $N_c^2$ & $N_c^0$ & $N_c^{-2}$ & $N_f N_c$ & $N_f/N_c$ & $N_f^2$ \\
\hline 
$0.005$ & $-2.9(7) \cdot 10^{3}$ & $-3.2(1) \cdot 10^{3}$ & $4.480(8) \cdot 10^{3}$ & $9.64(2) \cdot 10^{4}$ & $-2.592(2) \cdot 10^{4}$ & $2.2742(2) \cdot 10^{4}$ \\
$0.015$ & $-5.33(5) \cdot 10^{4}$ & $1.97(1) \cdot 10^{4}$ & $-1.1(1) \cdot 10^{2}$ & $-2.75(2) \cdot 10^{4}$ & $-4.92(2) \cdot 10^{3}$ & $1.01489(9) \cdot 10^{4}$ \\
$0.025$ & $-1.23(5) \cdot 10^{4}$ & $1.32(1) \cdot 10^{4}$ & $-3.8(1) \cdot 10^{2}$ & $-3.34(2) \cdot 10^{4}$ & $-1.69(2) \cdot 10^{3}$ & $6.8099(7) \cdot 10^{3}$ \\
$0.035$ & $7.1(4) \cdot 10^{3}$ & $9.3(1) \cdot 10^{3}$ & $-3.83(9) \cdot 10^{2}$ & $-3.31(2) \cdot 10^{4}$ & $-4.4(2) \cdot 10^{2}$ & $5.0819(6) \cdot 10^{3}$ \\
$0.045$ & $1.70(4) \cdot 10^{4}$ & $6.2(1) \cdot 10^{3}$ & $-3.46(8) \cdot 10^{2}$ & $-3.08(2) \cdot 10^{4}$ & $1.6(2) \cdot 10^{2}$ & $3.9942(5) \cdot 10^{3}$ \\
$0.055$ & $2.26(4) \cdot 10^{4}$ & $4.1(1) \cdot 10^{3}$ & $-2.95(7) \cdot 10^{2}$ & $-2.85(1) \cdot 10^{4}$ & $4.6(2) \cdot 10^{2}$ & $3.2413(5) \cdot 10^{3}$ \\
$0.065$ & $2.58(4) \cdot 10^{4}$ & $2.92(9) \cdot 10^{3}$ & $-2.57(6) \cdot 10^{2}$ & $-2.59(1) \cdot 10^{4}$ & $6.5(2) \cdot 10^{2}$ & $2.6872(5) \cdot 10^{3}$ \\
\hline
\end{tabular}
}
\caption{\label{table_totaljetbroadening_colour}
Individual contributions from the different colour factors to $B_T \frac{dC_{B_T}}{dB_T}$ for the total jet broadening distribution.
}
\end{center}
\end{table}
\begin{table}[p]
\begin{center}
{\scriptsize
\begin{tabular}{|r|c|c|c|c|c|c|}
\hline
 $C$ & $N_c^2$ & $N_c^0$ & $N_c^{-2}$ & $N_f N_c$ & $N_f/N_c$ & $N_f^2$ \\
\hline 
$0.005$ & $-2.82(3) \cdot 10^{4}$ & $4.28(9) \cdot 10^{3}$ & $2.583(6) \cdot 10^{3}$ & $5.10(1) \cdot 10^{4}$ & $-1.578(1) \cdot 10^{4}$ & $1.6764(2) \cdot 10^{4}$ \\
$0.015$ & $-3.60(3) \cdot 10^{4}$ & $1.22(1) \cdot 10^{4}$ & $4.7(1) \cdot 10^{2}$ & $-1.04(2) \cdot 10^{4}$ & $-5.45(2) \cdot 10^{3}$ & $8.808(1) \cdot 10^{3}$ \\
$0.025$ & $-1.76(4) \cdot 10^{4}$ & $9.8(1) \cdot 10^{3}$ & $1.6(1) \cdot 10^{2}$ & $-1.73(2) \cdot 10^{4}$ & $-3.27(2) \cdot 10^{3}$ & $6.697(1) \cdot 10^{3}$ \\
$0.035$ & $-7.3(4) \cdot 10^{3}$ & $8.6(1) \cdot 10^{3}$ & $2(1) \cdot 10^{1}$ & $-1.94(2) \cdot 10^{4}$ & $-2.23(2) \cdot 10^{3}$ & $5.522(1) \cdot 10^{3}$ \\
$0.045$ & $-1.0(4) \cdot 10^{3}$ & $6.6(1) \cdot 10^{3}$ & $-6(1) \cdot 10^{1}$ & $-1.99(2) \cdot 10^{4}$ & $-1.59(2) \cdot 10^{3}$ & $4.7388(9) \cdot 10^{3}$ \\
$0.055$ & $4.2(4) \cdot 10^{3}$ & $5.8(1) \cdot 10^{3}$ & $-6(1) \cdot 10^{1}$ & $-2.05(2) \cdot 10^{4}$ & $-1.16(2) \cdot 10^{3}$ & $4.1679(9) \cdot 10^{3}$ \\
$0.065$ & $6.8(4) \cdot 10^{3}$ & $5.2(1) \cdot 10^{3}$ & $-9(1) \cdot 10^{1}$ & $-2.01(2) \cdot 10^{4}$ & $-8.3(2) \cdot 10^{2}$ & $3.7300(9) \cdot 10^{3}$ \\
\hline
\end{tabular}
}
\caption{\label{table_Cparameter_colour_a}
Individual contributions from the different colour factors to $\rho \frac{dC_{C}}{dC}$ for the C parameter distribution.
}
\end{center}
\end{table}
\begin{table}[p]
\begin{center}
{\scriptsize
\begin{tabular}{|r|c|c|c|c|c|c|}
\hline
 $C$ & $N_c^2$ & $N_c^0$ & $N_c^{-2}$ & $N_f N_c$ & $N_f/N_c$ & $N_f^2$ \\
\hline 
$0.505$ & $1.24(4) \cdot 10^{4}$ & $-8.6(7) \cdot 10^{2}$ & $-4(5) \cdot 10^{0}$ & $-6.6(1) \cdot 10^{3}$ & $3.6(1) \cdot 10^{2}$ & $5.051(8) \cdot 10^{2}$ \\
$0.515$ & $1.20(4) \cdot 10^{4}$ & $-6.7(6) \cdot 10^{2}$ & $-9(4) \cdot 10^{0}$ & $-6.7(1) \cdot 10^{3}$ & $3.8(1) \cdot 10^{2}$ & $4.909(8) \cdot 10^{2}$ \\
$0.525$ & $1.24(4) \cdot 10^{4}$ & $-7.0(6) \cdot 10^{2}$ & $-5(4) \cdot 10^{0}$ & $-6.4(1) \cdot 10^{3}$ & $3.4(1) \cdot 10^{2}$ & $4.762(8) \cdot 10^{2}$ \\
$0.535$ & $1.13(4) \cdot 10^{4}$ & $-8.5(6) \cdot 10^{2}$ & $-1.5(4) \cdot 10^{1}$ & $-6.2(1) \cdot 10^{3}$ & $3.5(1) \cdot 10^{2}$ & $4.619(8) \cdot 10^{2}$ \\
$0.545$ & $1.15(4) \cdot 10^{4}$ & $-7.3(6) \cdot 10^{2}$ & $-6(4) \cdot 10^{0}$ & $-6.1(1) \cdot 10^{3}$ & $3.3(1) \cdot 10^{2}$ & $4.473(8) \cdot 10^{2}$ \\
$0.555$ & $1.16(4) \cdot 10^{4}$ & $-7.9(6) \cdot 10^{2}$ & $-4(4) \cdot 10^{0}$ & $-5.9(1) \cdot 10^{3}$ & $3.2(1) \cdot 10^{2}$ & $4.370(8) \cdot 10^{2}$ \\
$0.565$ & $1.16(4) \cdot 10^{4}$ & $-7.6(6) \cdot 10^{2}$ & $-7(4) \cdot 10^{0}$ & $-5.9(1) \cdot 10^{3}$ & $3.5(1) \cdot 10^{2}$ & $4.225(8) \cdot 10^{2}$ \\
\hline
\end{tabular}
}
\caption{\label{table_Cparameter_colour_b}
Individual contributions from the different colour factors to $\rho \frac{dC_{C}}{dC}$ for the C parameter distribution.
}
\end{center}
\end{table}
\begin{table}[p]
\begin{center}
{\scriptsize
\begin{tabular}{|r|c|c|c|c|c|c|}
\hline
 $\ln y_{23}$ & $N_c^2$ & $N_c^0$ & $N_c^{-2}$ & $N_f N_c$ & $N_f/N_c$ & $N_f^2$ \\
\hline 
$-9.875$ & $-2.6(1) \cdot 10^{4}$ & $3.1(2) \cdot 10^{3}$ & $1.426(9) \cdot 10^{3}$ & $1.89(3) \cdot 10^{4}$ & $-8.30(2) \cdot 10^{3}$ & $8.762(3) \cdot 10^{3}$ \\
$-9.625$ & $-2.63(9) \cdot 10^{4}$ & $3.7(2) \cdot 10^{3}$ & $1.208(8) \cdot 10^{3}$ & $1.49(3) \cdot 10^{4}$ & $-7.51(2) \cdot 10^{3}$ & $8.164(2) \cdot 10^{3}$ \\
$-9.375$ & $-2.44(8) \cdot 10^{4}$ & $3.9(1) \cdot 10^{3}$ & $1.046(7) \cdot 10^{3}$ & $1.19(3) \cdot 10^{4}$ & $-6.74(2) \cdot 10^{3}$ & $7.594(2) \cdot 10^{3}$ \\
$-9.125$ & $-2.34(7) \cdot 10^{4}$ & $4.7(1) \cdot 10^{3}$ & $9.18(7) \cdot 10^{2}$ & $8.2(2) \cdot 10^{3}$ & $-6.04(1) \cdot 10^{3}$ & $7.048(2) \cdot 10^{3}$ \\
$-8.875$ & $-2.04(7) \cdot 10^{4}$ & $4.7(1) \cdot 10^{3}$ & $7.41(6) \cdot 10^{2}$ & $5.2(2) \cdot 10^{3}$ & $-5.40(1) \cdot 10^{3}$ & $6.535(2) \cdot 10^{3}$ \\
$-8.625$ & $-1.94(6) \cdot 10^{4}$ & $4.9(1) \cdot 10^{3}$ & $6.49(5) \cdot 10^{2}$ & $3.2(2) \cdot 10^{3}$ & $-4.79(1) \cdot 10^{3}$ & $6.040(2) \cdot 10^{3}$ \\
$-8.375$ & $-1.70(6) \cdot 10^{4}$ & $4.7(1) \cdot 10^{3}$ & $5.33(5) \cdot 10^{2}$ & $5(2) \cdot 10^{2}$ & $-4.25(1) \cdot 10^{3}$ & $5.578(1) \cdot 10^{3}$ \\
\hline
\end{tabular}
}
\caption{\label{table_y23_colour}
Individual contributions from the different colour factors to $y_{23} \frac{dC_{y_{23}}}{dy_{23}}$ for the three-to-two jet transition distribution.
}
\end{center}
\end{table}
%
%
%
\clearpage
\begin{table}[p]
\begin{center}
{\scriptsize
\begin{tabular}{|r|c|c|c|c|c|c|}
\hline
 $y_{cut}$ & $N_c^2$ & $N_c^0$ & $N_c^{-2}$ & $N_f N_c$ & $N_f/N_c$ & $N_f^2$ \\
\hline 
$0.3$ & $-2(3) \cdot 10^{0}$ & $4(2) \cdot 10^{-1}$ & $-7(4) \cdot 10^{-3}$ & $-5.9(7) \cdot 10^{0}$ & $2(4) \cdot 10^{-2}$ & $1.48(3) \cdot 10^{0}$ \\
$0.15$ & $4.4(1) \cdot 10^{2}$ & $-1.82(8) \cdot 10^{1}$ & $-2.7(3) \cdot 10^{-1}$ & $-4.47(4) \cdot 10^{2}$ & $1.13(2) \cdot 10^{1}$ & $5.93(2) \cdot 10^{1}$ \\
$0.1$ & $1.05(2) \cdot 10^{3}$ & $-3.1(1) \cdot 10^{1}$ & $-4.9(3) \cdot 10^{-1}$ & $-1.000(6) \cdot 10^{3}$ & $2.19(3) \cdot 10^{1}$ & $1.336(3) \cdot 10^{2}$ \\
$0.06$ & $2.16(3) \cdot 10^{3}$ & $-2.0(2) \cdot 10^{1}$ & $-1.20(6) \cdot 10^{0}$ & $-2.048(8) \cdot 10^{3}$ & $3.19(4) \cdot 10^{1}$ & $2.843(4) \cdot 10^{2}$ \\
$0.03$ & $3.84(5) \cdot 10^{3}$ & $1.23(3) \cdot 10^{2}$ & $-2.71(6) \cdot 10^{0}$ & $-4.09(2) \cdot 10^{3}$ & $1.08(8) \cdot 10^{1}$ & $6.238(9) \cdot 10^{2}$ \\
$0.015$ & $4.83(8) \cdot 10^{3}$ & $5.78(5) \cdot 10^{2}$ & $-3.1(2) \cdot 10^{0}$ & $-6.59(3) \cdot 10^{3}$ & $-1.10(1) \cdot 10^{2}$ & $1.183(2) \cdot 10^{3}$ \\
$0.01$ & $4.4(1) \cdot 10^{3}$ & $1.060(7) \cdot 10^{3}$ & $2(2) \cdot 10^{-1}$ & $-8.12(3) \cdot 10^{3}$ & $-2.66(2) \cdot 10^{2}$ & $1.646(2) \cdot 10^{3}$ \\
$0.006$ & $2.4(2) \cdot 10^{3}$ & $1.96(1) \cdot 10^{3}$ & $1.35(4) \cdot 10^{1}$ & $-9.68(5) \cdot 10^{3}$ & $-6.22(3) \cdot 10^{2}$ & $2.409(3) \cdot 10^{3}$ \\
$0.003$ & $-4.7(3) \cdot 10^{3}$ & $3.76(2) \cdot 10^{3}$ & $6.54(4) \cdot 10^{1}$ & $-1.050(9) \cdot 10^{4}$ & $-1.499(5) \cdot 10^{3}$ & $3.862(5) \cdot 10^{3}$ \\
$0.0015$ & $-1.74(4) \cdot 10^{4}$ & $6.13(3) \cdot 10^{3}$ & $1.95(2) \cdot 10^{2}$ & $-8.1(1) \cdot 10^{3}$ & $-3.064(9) \cdot 10^{3}$ & $5.899(9) \cdot 10^{3}$ \\
$0.001$ & $-2.84(6) \cdot 10^{4}$ & $7.61(4) \cdot 10^{3}$ & $3.38(1) \cdot 10^{2}$ & $-4.1(2) \cdot 10^{3}$ & $-4.43(1) \cdot 10^{3}$ & $7.41(1) \cdot 10^{3}$ \\
$0.0006$ & $-4.60(8) \cdot 10^{4}$ & $9.38(6) \cdot 10^{3}$ & $6.17(3) \cdot 10^{2}$ & $4.7(2) \cdot 10^{3}$ & $-6.79(2) \cdot 10^{3}$ & $9.74(2) \cdot 10^{3}$ \\
$0.0003$ & $-7.1(1) \cdot 10^{4}$ & $1.070(9) \cdot 10^{4}$ & $1.273(3) \cdot 10^{3}$ & $2.58(3) \cdot 10^{4}$ & $-1.142(3) \cdot 10^{4}$ & $1.380(3) \cdot 10^{4}$ \\
$0.00015$ & $-9.2(3) \cdot 10^{4}$ & $9.0(2) \cdot 10^{3}$ & $2.406(8) \cdot 10^{3}$ & $6.22(6) \cdot 10^{4}$ & $-1.821(4) \cdot 10^{4}$ & $1.906(5) \cdot 10^{4}$ \\
$0.0001$ & $-9.4(4) \cdot 10^{4}$ & $5.4(2) \cdot 10^{3}$ & $3.413(7) \cdot 10^{3}$ & $9.13(7) \cdot 10^{4}$ & $-2.341(6) \cdot 10^{4}$ & $2.271(5) \cdot 10^{4}$ \\
\hline
\end{tabular}
}
\caption{\label{table_jet_rate_durham_colour}
Individual contributions from the different colour factors to the NNLO coefficient $C_{3-jet,Durham}$ 
for the three-jet rate with the Durham jet algorithm 
for various values of $y_{cut}$.
}
\end{center}
\end{table}
\begin{table}[p]
\begin{center}
{\scriptsize
\begin{tabular}{|r|c|c|c|c|c|c|}
\hline
 $y_{cut}$ & $N_c^2$ & $N_c^0$ & $N_c^{-2}$ & $N_f N_c$ & $N_f/N_c$ & $N_f^2$ \\
\hline 
$0.3$ & $4.3(6) \cdot 10^{1}$ & $-3(2) \cdot 10^{0}$ & $-6(4) \cdot 10^{-2}$ & $-6.5(2) \cdot 10^{1}$ & $1.68(9) \cdot 10^{0}$ & $1.091(1) \cdot 10^{1}$ \\
$0.15$ & $1.17(5) \cdot 10^{3}$ & $6(1) \cdot 10^{1}$ & $-4(2) \cdot 10^{-1}$ & $-2.05(3) \cdot 10^{3}$ & $2(9) \cdot 10^{-1}$ & $3.835(2) \cdot 10^{2}$ \\
$0.1$ & $1.33(9) \cdot 10^{3}$ & $3.5(2) \cdot 10^{2}$ & $2.0(4) \cdot 10^{0}$ & $-3.74(4) \cdot 10^{3}$ & $-9.8(2) \cdot 10^{1}$ & $8.788(4) \cdot 10^{2}$ \\
$0.06$ & $-2.0(2) \cdot 10^{3}$ & $1.16(3) \cdot 10^{3}$ & $1.82(8) \cdot 10^{1}$ & $-5.66(8) \cdot 10^{3}$ & $-4.95(5) \cdot 10^{2}$ & $1.9342(9) \cdot 10^{3}$ \\
$0.03$ & $-1.49(3) \cdot 10^{4}$ & $3.07(5) \cdot 10^{3}$ & $1.20(2) \cdot 10^{2}$ & $-4.2(2) \cdot 10^{3}$ & $-1.94(1) \cdot 10^{3}$ & $4.461(2) \cdot 10^{3}$ \\
$0.015$ & $-4.24(7) \cdot 10^{4}$ & $5.2(1) \cdot 10^{3}$ & $4.60(2) \cdot 10^{2}$ & $7.8(3) \cdot 10^{3}$ & $-5.17(3) \cdot 10^{3}$ & $8.808(3) \cdot 10^{3}$ \\
$0.01$ & $-6.44(7) \cdot 10^{4}$ & $6.1(1) \cdot 10^{3}$ & $8.78(3) \cdot 10^{2}$ & $2.47(4) \cdot 10^{4}$ & $-8.34(4) \cdot 10^{3}$ & $1.2472(5) \cdot 10^{4}$ \\
$0.006$ & $-9.7(1) \cdot 10^{4}$ & $4.3(2) \cdot 10^{3}$ & $1.783(5) \cdot 10^{3}$ & $6.09(6) \cdot 10^{4}$ & $-1.427(6) \cdot 10^{4}$ & $1.8585(8) \cdot 10^{4}$ \\
$0.003$ & $-1.33(2) \cdot 10^{5}$ & $-6.1(3) \cdot 10^{3}$ & $4.055(9) \cdot 10^{3}$ & $1.46(1) \cdot 10^{5}$ & $-2.68(1) \cdot 10^{4}$ & $3.015(1) \cdot 10^{4}$ \\
$0.0015$ & $-1.19(3) \cdot 10^{5}$ & $-3.48(5) \cdot 10^{4}$ & $8.18(2) \cdot 10^{3}$ & $2.98(1) \cdot 10^{5}$ & $-4.63(2) \cdot 10^{4}$ & $4.655(2) \cdot 10^{4}$ \\
$0.001$ & $-8.7(5) \cdot 10^{4}$ & $-6.47(6) \cdot 10^{4}$ & $1.169(2) \cdot 10^{4}$ & $4.24(2) \cdot 10^{5}$ & $-6.16(3) \cdot 10^{4}$ & $5.875(3) \cdot 10^{4}$ \\
\hline
\end{tabular}
}
\caption{\label{table_jet_rate_geneva_colour}
Individual contributions from the different colour factors to the NNLO coefficient $C_{3-jet,Geneva}$ 
for the three-jet rate with the the Geneva jet algorithm 
for various values of $y_{cut}$.
}
\end{center}
\end{table}
\begin{table}[p]
\begin{center}
{\scriptsize
\begin{tabular}{|r|c|c|c|c|c|c|}
\hline
 $y_{cut}$ & $N_c^2$ & $N_c^0$ & $N_c^{-2}$ & $N_f N_c$ & $N_f/N_c$ & $N_f^2$ \\
\hline 
$0.3$ & $3.5(8) \cdot 10^{1}$ & $-3(1) \cdot 10^{0}$ & $-5(3) \cdot 10^{-2}$ & $-2.7(1) \cdot 10^{1}$ & $1.0(1) \cdot 10^{0}$ & $2.770(5) \cdot 10^{0}$ \\
$0.15$ & $1.85(3) \cdot 10^{3}$ & $-8.9(9) \cdot 10^{1}$ & $-2.4(3) \cdot 10^{0}$ & $-1.41(1) \cdot 10^{3}$ & $5.4(1) \cdot 10^{1}$ & $1.3607(9) \cdot 10^{2}$ \\
$0.1$ & $3.86(9) \cdot 10^{3}$ & $-1.1(2) \cdot 10^{2}$ & $-8.9(5) \cdot 10^{0}$ & $-3.33(2) \cdot 10^{3}$ & $1.10(2) \cdot 10^{2}$ & $3.349(2) \cdot 10^{2}$ \\
$0.06$ & $7.2(2) \cdot 10^{3}$ & $1.4(3) \cdot 10^{2}$ & $-2.7(1) \cdot 10^{1}$ & $-6.83(5) \cdot 10^{3}$ & $1.54(4) \cdot 10^{2}$ & $7.945(4) \cdot 10^{2}$ \\
$0.03$ & $9.2(3) \cdot 10^{3}$ & $1.48(6) \cdot 10^{3}$ & $-5.8(2) \cdot 10^{1}$ & $-1.33(1) \cdot 10^{4}$ & $-3(1) \cdot 10^{1}$ & $1.9819(9) \cdot 10^{3}$ \\
$0.015$ & $2.6(5) \cdot 10^{3}$ & $4.7(2) \cdot 10^{3}$ & $-7.1(3) \cdot 10^{1}$ & $-1.95(2) \cdot 10^{4}$ & $-9.4(2) \cdot 10^{2}$ & $4.116(2) \cdot 10^{3}$ \\
$0.01$ & $-8.9(5) \cdot 10^{3}$ & $7.7(1) \cdot 10^{3}$ & $-2.1(5) \cdot 10^{1}$ & $-2.13(3) \cdot 10^{4}$ & $-2.12(3) \cdot 10^{3}$ & $5.947(2) \cdot 10^{3}$ \\
$0.006$ & $-2.74(9) \cdot 10^{4}$ & $1.18(3) \cdot 10^{4}$ & $1.88(7) \cdot 10^{2}$ & $-1.77(4) \cdot 10^{4}$ & $-4.55(8) \cdot 10^{3}$ & $9.025(4) \cdot 10^{3}$ \\
$0.003$ & $-6.8(1) \cdot 10^{4}$ & $1.67(6) \cdot 10^{4}$ & $1.02(1) \cdot 10^{3}$ & $3(6) \cdot 10^{2}$ & $-1.08(1) \cdot 10^{4}$ & $1.4875(6) \cdot 10^{4}$ \\
$0.0015$ & $-1.16(2) \cdot 10^{5}$ & $1.81(6) \cdot 10^{4}$ & $2.99(2) \cdot 10^{3}$ & $4.6(1) \cdot 10^{4}$ & $-2.02(3) \cdot 10^{4}$ & $2.311(1) \cdot 10^{4}$ \\
$0.001$ & $-1.38(4) \cdot 10^{5}$ & $1.41(7) \cdot 10^{4}$ & $4.84(2) \cdot 10^{3}$ & $9.1(1) \cdot 10^{4}$ & $-2.93(3) \cdot 10^{4}$ & $2.923(1) \cdot 10^{4}$ \\
$0.0006$ & $-1.45(4) \cdot 10^{5}$ & $0(1) \cdot 10^{2}$ & $8.03(2) \cdot 10^{3}$ & $1.74(2) \cdot 10^{5}$ & $-4.36(3) \cdot 10^{4}$ & $3.855(2) \cdot 10^{4}$ \\
$0.0003$ & $-7.1(7) \cdot 10^{4}$ & $-4.5(1) \cdot 10^{4}$ & $1.414(4) \cdot 10^{4}$ & $3.47(3) \cdot 10^{5}$ & $-6.93(5) \cdot 10^{4}$ & $5.455(3) \cdot 10^{4}$ \\
\hline
\end{tabular}
}
\caption{\label{table_jet_rate_jadeE0_colour}
Individual contributions from the different colour factors to the NNLO coefficient $C_{3-jet,Jade-E0}$ 
for the three-jet rate with the the Jade-E0 jet algorithm 
for various values of $y_{cut}$.
}
\end{center}
\end{table}
\begin{table}[p]
\begin{center}
{\scriptsize
\begin{tabular}{|r|c|c|c|c|c|c|}
\hline
 $y_{cut}$ & $N_c^2$ & $N_c^0$ & $N_c^{-2}$ & $N_f N_c$ & $N_f/N_c$ & $N_f^2$ \\
\hline 
$0.3$ & $-2(3) \cdot 10^{0}$ & $3(6) \cdot 10^{-1}$ & $-5(2) \cdot 10^{-2}$ & $-6.5(6) \cdot 10^{0}$ & $-1.5(4) \cdot 10^{-1}$ & $1.475(3) \cdot 10^{0}$ \\
$0.15$ & $4.3(3) \cdot 10^{2}$ & $-2.5(4) \cdot 10^{1}$ & $0(9) \cdot 10^{-3}$ & $-4.48(7) \cdot 10^{2}$ & $1.01(5) \cdot 10^{1}$ & $5.986(4) \cdot 10^{1}$ \\
$0.1$ & $9.3(6) \cdot 10^{2}$ & $-2.1(6) \cdot 10^{1}$ & $-4(2) \cdot 10^{-1}$ & $-9.7(1) \cdot 10^{2}$ & $1.67(7) \cdot 10^{1}$ & $1.364(1) \cdot 10^{2}$ \\
$0.06$ & $1.62(9) \cdot 10^{3}$ & $1(1) \cdot 10^{1}$ & $-4(4) \cdot 10^{-1}$ & $-1.90(2) \cdot 10^{3}$ & $1.4(1) \cdot 10^{1}$ & $2.915(2) \cdot 10^{2}$ \\
$0.03$ & $2.6(2) \cdot 10^{3}$ & $2.2(2) \cdot 10^{2}$ & $2(6) \cdot 10^{-1}$ & $-3.70(5) \cdot 10^{3}$ & $-3.8(2) \cdot 10^{1}$ & $6.431(4) \cdot 10^{2}$ \\
$0.015$ & $2.9(2) \cdot 10^{3}$ & $5.9(3) \cdot 10^{2}$ & $6(1) \cdot 10^{0}$ & $-5.61(9) \cdot 10^{3}$ & $-1.98(4) \cdot 10^{2}$ & $1.2207(6) \cdot 10^{3}$ \\
$0.01$ & $2.1(2) \cdot 10^{3}$ & $1.12(3) \cdot 10^{3}$ & $1.31(9) \cdot 10^{1}$ & $-6.8(1) \cdot 10^{3}$ & $-3.82(6) \cdot 10^{2}$ & $1.6975(9) \cdot 10^{3}$ \\
$0.006$ & $2(4) \cdot 10^{2}$ & $1.91(6) \cdot 10^{3}$ & $3.6(1) \cdot 10^{1}$ & $-7.8(2) \cdot 10^{3}$ & $-7.93(8) \cdot 10^{2}$ & $2.485(1) \cdot 10^{3}$ \\
$0.003$ & $-8.1(5) \cdot 10^{3}$ & $3.53(7) \cdot 10^{3}$ & $1.12(3) \cdot 10^{2}$ & $-7.4(2) \cdot 10^{3}$ & $-1.76(2) \cdot 10^{3}$ & $3.960(2) \cdot 10^{3}$ \\
$0.0015$ & $-1.89(7) \cdot 10^{4}$ & $5.5(2) \cdot 10^{3}$ & $2.78(4) \cdot 10^{2}$ & $-3.5(3) \cdot 10^{3}$ & $-3.42(2) \cdot 10^{3}$ & $6.034(3) \cdot 10^{3}$ \\
$0.001$ & $-2.79(9) \cdot 10^{4}$ & $6.6(1) \cdot 10^{3}$ & $4.47(3) \cdot 10^{2}$ & $1.3(3) \cdot 10^{3}$ & $-4.89(3) \cdot 10^{3}$ & $7.590(3) \cdot 10^{3}$ \\
$0.0006$ & $-4.0(1) \cdot 10^{4}$ & $7.9(1) \cdot 10^{3}$ & $7.69(5) \cdot 10^{2}$ & $1.06(4) \cdot 10^{4}$ & $-7.27(6) \cdot 10^{3}$ & $9.959(5) \cdot 10^{3}$ \\
$0.0003$ & $-5.9(1) \cdot 10^{4}$ & $7.8(3) \cdot 10^{3}$ & $1.505(7) \cdot 10^{3}$ & $3.44(7) \cdot 10^{4}$ & $-1.207(8) \cdot 10^{4}$ & $1.4070(6) \cdot 10^{4}$ \\
$0.00015$ & $-7.3(2) \cdot 10^{4}$ & $4.6(4) \cdot 10^{3}$ & $2.76(1) \cdot 10^{3}$ & $7.46(8) \cdot 10^{4}$ & $-1.92(1) \cdot 10^{4}$ & $1.9362(9) \cdot 10^{4}$ \\
$0.0001$ & $-7.5(2) \cdot 10^{4}$ & $-1(4) \cdot 10^{2}$ & $3.81(1) \cdot 10^{3}$ & $1.06(1) \cdot 10^{5}$ & $-2.45(1) \cdot 10^{4}$ & $2.312(1) \cdot 10^{4}$ \\
\hline
\end{tabular}
}
\caption{\label{table_jet_rate_cambridge_colour}
Individual contributions from the different colour factors to the NNLO coefficient $C_{3-jet,Cambridge}$ 
for the three-jet rate with the the Cambridge jet algorithm 
for various values of $y_{cut}$.
}
\end{center}
\end{table}


\end{document}